\documentclass[a4paper,11pt]{article}

\usepackage[dvips]{graphicx}
\usepackage{amsfonts, color, float,  amsmath}
\usepackage{amssymb}\usepackage{subfigure}
\usepackage{url}
\usepackage{array}

\usepackage{slashed}
\usepackage[svgnames]{xcolor} 

\usepackage[small,bf,sf]{caption} 
\usepackage{microtype}
\usepackage{cite}
\usepackage{authblk}
\usepackage{datetime}
\usepackage[bottom]{footmisc} 
\usepackage{booktabs}  

\allowdisplaybreaks

\newcommand{\wh}{\widehat}

\newcommand{\so}{\sin\theta_1}
\newcommand{\st}{\sin\theta_2}
\newcommand{\co}{\cos\theta_1}
\newcommand{\ct}{\cos\theta_2}
\newcommand{\sosq}{\sin^2\theta_1}
\newcommand{\stsq}{\sin^2\theta_2}
\newcommand{\cosq}{\cos^2\theta_1}
\newcommand{\ctsq}{\cos^2\theta_2}

\newcommand{\sgn}{{\rm sgn}}
\newcommand{\Eqs}[2]{Eqs.~(\ref{#1}) and~(\ref{#2})}

\def\aem{\alpha_{\rm em}}
\def\SM {\circ}
\def\beq {\begin{equation}}
\def\eeq {\end{equation}}
\def\bea {\begin{eqnarray}}
\def\eea {\end{eqnarray}}

\def\nn {\nonumber}

\def\HZll {H \to Z \lplm}
\def\Hllll {H \to Z(\to \lplm) \lplm}
\def\eeHZ {e^+ e^- \to H Z}
\def\eeHll {e^+ e^- \to H Z(\to  \lplm)}
\def\calO{\mathcal{O}}
\def\mn{\mu\nu}

\def\haV{\widehat\alpha^V_{\Phi \ell}}
\def\haA{\widehat\alpha^A_{\Phi \ell}}
\def\aV{\alpha^V_{\Phi \ell}}

\def\aVA{\alpha^{V,A}_{\Phi \ell}}
\def\haVA{\widehat\alpha^{V,A}_{\Phi \ell}}
\def\ha{\widehat \alpha}
\def\haAZ{\widehat \alpha_{AZ}}
\def\haZZ{\widehat \alpha_{ZZ}}
\def\haone{\widehat\alpha_{1}^{\rm eff}}
\def\hatwo{\widehat\alpha_{2}^{\rm eff}}
\def \wh{\widehat}
\def\barg{\bar g}
\def \gAgVsq {g_A^2+g_V^2}

\def\bp {\mbox{\boldmath $p$}}

\def\AsymThree {\mathcal{A}_\phi^{(3)}}
\def\AsymCoCt {\mathcal{A}_{{\rm c}\theta_1,{\rm c}\theta_2}}

\def\lplm{\ell^+\ell^-}

\AtBeginDocument{}

\title{\LARGE {\bf \sffamily \boldmath  Anomalous Higgs couplings in 
angular asymmetries of  $H\to Z \,  \lplm$ and $e^+e^- \to H Z$}}

\author[a]{Martin Beneke}
\author[a]{Diogo Boito}
\author[a,b]{Yu-Ming Wang}
\affil[a]{\sl Physik Department T31, Technische Universit\"at M\"unchen,
James-Franck-Stra\ss e~1, D-85748 Garching, Germany}

\affil[b]{\sl Institut f\"ur Theoretische Teilchenphysik und Kosmologie, 
RWTH Aachen University, 
D-52056 Aachen, Germany}

\date{}

\begin{document}
\begin{flushright}
{\small 
TUM-HEP-949/14,  TTK-14-11,   SFB/CPP-14-28\\ 
12 November 2014} 
\end{flushright}

\vspace*{-0.7cm}
\begingroup
\let\newpage\relax
\maketitle
\endgroup
\date{}

\begin{abstract}
\noindent
We study in detail the impact of anomalous Higgs couplings in angular
asymmetries of the crossing-symmetric processes $\HZll$ and
$\eeHZ$. Beyond Standard Model physics is parametrized in terms of the
$SU(3)\times SU(2)_L\times U(1)_Y$ dimension-six effective Lagrangian.
In the light of present bounds on $d=6$ interactions we study how
angular asymmetries can reveal non-standard CP-even and CP-odd
couplings.  We provide approximate expressions to all observables of
interest making transparent their dominant dependence on anomalous
couplings.  We show that some asymmetries may reveal BSM effects that
are hidden in other observables. In particular, CP-even and CP-odd
$d=6$ $HZ\gamma$ couplings as well as (to a lesser extent) $HZ\ell^+
\ell^-$ contact interactions can generate asymmetries at the several
percent level, while having small or no effects on the di-lepton
invariant mass spectrum of $\HZll$. Finally, the higher di-lepton
invariant mass probed in $\eeHZ$ leads to interesting differences in
the asymmetries with respect to those of $\HZll$ that may lead to
complementary anomalous coupling searches at the LHC and $e^+ e^-$
colliders.
\end{abstract}

\thispagestyle{empty}


\section{Introduction}

The discovery of a light boson $H$ with mass around 125~GeV in the
first run of LHC~\cite{Atlas_Higgs,CMS_Higgs} opened a new window to
physics beyond the Standard Model (BSM). At present, the Standard
Model (SM) Higgs quantum numbers, $J^P=0^+$, are favoured by the
data~--- all other tested hypotheses have been excluded at confidence
levels above~95\%~\cite{Atlas_HiggsQN,CMS_HiggsQN}.  Furthermore, from
the study of the signal strengths of the new state, all Higgs
couplings to SM particles are compatible with SM predictions (see
e.g. Ref.~\cite{EY13} and references therein). In particular, those to
the $W$ and $Z$ bosons are constrained to be within 10~\% of their SM
values~\cite{FRU13}. In the absence of evidence for any other new
state, the SM seems to be a good effective field theory (EFT) 
above the electroweak scale, at least
up to the scales currently probed by the LHC.

In the spirit of an EFT --- assuming the characteristic scale
$\Lambda$ of BSM physics to be much larger than the electroweak scale
--- the SM should be supplemented with all operators compatible with
its symmetries. The $SU(2)_L\times U(1)_Y$ symmetry can be linearly or
non-linearly realized. The non-linear realization gives rise to a
theory in close analogy to 
ChPT~\cite{FF93,KC93,GT07,Alonso_etal13,BCK13,Brivio_etal13}, with an additional 
light, SM-singlet scalar.  In the linear
realization, that we adopt in this work, one must add to the SM all
dimension-six operators constructed from the SM 
fields~\cite{BW86,GIMR10}. These operators are
suppressed by the large scale $\Lambda$ and generate anomalous
couplings of the Higgs boson.

The present status of the search for BSM physics in the Higgs sector
is somewhat similar to that of the flavour sector of the SM, in which
evidence for BSM physics has proven to be much more elusive than
naively expected. In the search for new flavour-changing neutral currents, 
dedicated observables constructed from the angular
distribution of $B\to K^* \lplm$ have been constructed to 
unveil BSM effects in weak interactions more 
efficiently~\cite{KSSS00,Egede:2008uy}. The
angular distribution of the analogous decay $H\to Z \lplm$
offers similar possibilities.

The study of $H\to Z \lplm$, with the on-shell $Z$ also decaying into
$\lplm$, has a long history.  Angular distributions were used in the
determination of the Higgs quantum numbers~\cite{Atlas_HiggsQN,CMS_HiggsQN} as
suggested years ago (see
e.g.~Refs.~\cite{GMM07,CMMZ2003,Bea12,DeRujula_etal10}). More
recently, the di-lepton-mass distribution has been proposed as a way to
reveal effects that would otherwise be hidden in the total decay
width~\cite{IMT13,IT13,GMP13}. The full angular distribution of
the final state leptons has been revisited recently~\cite{BCD13} 
in the framework of the EFT parametrization of anomalous couplings, 
and it was shown that angular asymmetries can be
constructed in order to reveal effects of anomalous Higgs couplings
that would remain hidden even in the di-lepton invariant mass
distribution.

In the present work we perform an extended study of the 
angular asymmetries of $H\to Z(\to \lplm)\lplm$ and of the
crossing-symmetric reaction $e^+e^-\to HZ(\to \lplm)$. The latter
process should be measured with high precision at a high-energy $e^+ e^-$ 
collider (such as the ILC~\cite{Baeretal13}) and provide
a clean way to extract the Higgs couplings~\cite{Barger:1993wt,KKZ96,Hagiwara:1993sw}.  In the massless lepton limit,  the two processes are
described by the same set of six form factors, albeit in different
kinematic regimes, related by analyticity.  The form factors can be
written in terms of the couplings of the general $d=6$
Lagrangian.  Ignoring loop corrections and neglecting the lepton masses, 
the processes are described by six independent angular functions of 
the three independent angles among the four leptons, which can be expressed 
in terms of the six form factors. Our focus is on these asymmetries,
their sensitivity to anomalous Higgs couplings, and the interplay
between the asymmetries and the di-lepton mass distributions.

In $\HZll$, we show that the most promising anomalous coupling that
could generate sizeable asymmetries is the $d=6$ $HZ\gamma$ interaction. The
contact interactions $HZ\ell^+\ell^-$, whose effects were recently
investigated in Ref.~\cite{BCD13}, also have a more prominent impact in the
asymmetries than in the decay rate. However, the present constraints
on these couplings make the magnitude of the asymmetries rather small. Next,
we perform a study of the total cross section at intermediate energies
for the reaction $e^+e^-\to HZ(\to \lplm)$ and of angular
asymmetries akin to those of $\HZll$. We fully exploit the
crossing-symmetric nature of these processes to make the relation 
between them transparent.  Although described by the same form
factors, the sensitivity to specific BSM couplings differs in the
asymmetries of these two reactions due to the different
characteristic energy scale involved. In $\HZll$ the $d=6$
corrections scale as $m_H^2/\Lambda^2$ while in $\eeHZ$ they scale as
$q^2/\Lambda^2$, where $q^2$, the  
center-of-mass (CM) energy of the $e^+e^-$ pair, must be 
larger than $(217\,\mbox{GeV})^2$ to produce 
the $HZ$ final state~on-shell.

At present, the experimental study of the di-lepton mass distribution
and of the angular asymmetries in $\Hllll$ is not feasible due to low
statistics --- ATLAS~\cite{ATLAS_HVV} and CMS~\cite{CMS_HiggsQN}
observed only around 30 $H\to Z \,Z^*\to 4\ell$ events each.
However,  higher luminosities will permit these studies in the future. 
With an integrated luminosity of 350~fb$^{-1}$ at $14$~TeV, which
could be reached by 2021~\cite{ATLAS_Upgrade,CMS_Upgrade}, the number
of observed events could attain 1000. 
With the high-luminosity up-grade (integrated luminosity 3000~fb$^{-1}$), 
this number was estimated by a recent study of the sensitivity to 
anomalous Higgs-gauge boson interactions  to be of the order of 
6000~\cite{Andersonetal14}. The number of reconstructed events of
$\eeHll$ at an $e^+ e^-$ collider would be approximately $2000$ at
$\sqrt{q^2}=250$~GeV with an integrated luminosity of
250~fb$^{-1}$~\cite{Andersonetal14}.

The paper is organized as follows. In Sec.~\ref{sec:Dyn} we discuss
the relevant operators in the linear realization of the $d=6$
Lagrangian and the relevant Higgs anomalous couplings. In
Sec.~\ref{sec:HZll} we study the angular distribution of $H\to Z
\lplm$ and we show some of the promising angular asymmetries. In
Sec.~\ref{sec:eeHZ} we perform a similar study of the reaction
$e^+e^-\to HZ(\to \lplm)$. Our calculations are done at tree level.
However, in Sec.~\ref{sec:SM-loop} we discuss briefly the 
generic effect of SM loops. We summarize in Sec.~\ref{sec:Conc}. We 
relegate to App.~\ref{app:Kine} the kinematics and definitions of 
angular distributions, while the explicit expressions of the angular
coefficient functions are given in App.~\ref{app:JFunctions}.

\section{Effective Lagrangian and couplings}
\label{sec:Dyn}

In order to parametrize BSM effects in a general way, we resort to the 
linear realization of the $SU(2)_L\times U(1)_Y$ SM electroweak symmetry. 
Assuming the new physics sector to be characterized by a
scale $\Lambda$, larger than the electroweak scale, the
SM is supplemented with 59 independent $d=6$ operators~\cite{BW86,GIMR10}. 
This Lagrangian can be schematically cast~as
\beq
\mathcal{L}_{\rm eff} = \mathcal{L}_{\rm SM}^{(4)} + 
\frac{1}{\Lambda^2} \sum_{k=1}^{59} \alpha_{k}\calO_k,
\label{leff}
\eeq
where the $\alpha_k$ is the coupling of operator $\calO_{k}$. 
The effective Lagrangian implies a parametrization of anomalous 
Higgs interactions (contained in $\calO_k$) constrained by the SM gauge 
symmetry.  In our expressions, we often employ the 
notation~$\widehat\alpha_k$ defined as
\beq
\widehat\alpha_k = \frac{v^2}{\Lambda^2} \alpha_k,
\label{ahat}
\eeq
where  $v$ is the classical Higgs vacuum expectation value.
The dimensionless coefficients $\widehat\alpha_k$ should be smaller 
than $\mathcal{O}(1)$ for the EFT description to be applicable.

\bgroup
\def\arraystretch{1.3}%
\begin{table}[t]
\begin{center}{
\caption{The subset of $d=6$ operators that contribute to $\HZll$ and 
$\eeHZ$ in the basis defined in Ref.~\cite{GIMR10}. The four-lepton operator 
given in Eq.~(\ref{eq:O4L}) gives an indirect contribution solely through 
the redefinition of $\delta_{G_F}$ and is not listed in this table.}
\vspace{-2mm}
\begin{tabular}{l l l}
\toprule
 $\Phi^4\, D^2$& $X^2\, \Phi^2$ & $\psi^2\, \Phi^2 \,D$\\
\midrule
$\calO_{\Phi\Box}=(\Phi^\dagger\Phi)\Box(\Phi^\dagger \Phi)$ & 
$\calO_{\Phi W}=(\Phi^\dagger\Phi) W^{I}_{\mn} W^{I\mn}$ & 
$\calO^{(1)}_{\Phi\, \ell}=(\Phi^\dagger i 
\overset{\leftrightarrow}{D}_\mu \Phi)(\bar\ell\gamma^\mu \ell)$ \\
$\calO_{\Phi D}=(\Phi^\dagger D^\mu \Phi)^*(\Phi^\dagger D_\mu \Phi)$& 
$\calO_{\Phi B} = (\Phi^\dagger \Phi)B_{\mn}B^{\mn}$ & 
$\calO^{(3)}_{\Phi\, \ell}=(\Phi^\dagger i 
\overset{\leftrightarrow}{D} \ \!\!\!^{\, I}_\mu \Phi)
(\bar\ell\gamma^\mu\tau^I \ell)$  \\
&$\calO_{\Phi W\! B} =(\Phi^\dagger \tau^I\Phi) W^{I}_{\mn} B^{\mn} $& 
$\calO_{\Phi e}=(\Phi^\dagger i 
\overset{\leftrightarrow}{D}_\mu \Phi)(\bar e \gamma^\mu e)$\\
& $\calO_{\Phi \widetilde W} =(\Phi^\dagger\Phi) \widetilde W^{I}_{\mn} 
W^{I\mn}$  &   \\
& $\calO_{\Phi \widetilde B} =(\Phi^\dagger\Phi) \widetilde B_{\mn}B^{\mn}$& 
\\
& $\calO_{\Phi \widetilde W\! B} =(\Phi^\dagger\tau ^I\Phi) 
\widetilde W_{\mn}^IB^{\mn}$  &  \\[0.1cm]
\bottomrule
\end{tabular}
\label{tab:Operators}
}\end{center}
\end{table}
\egroup

Different choices for the operator basis are possible and in use. Here we 
stick to the basis defined in Ref.~\cite{GIMR10}.
In practice we only need to work with a  subset of the 59
operators, since not all of them contribute at tree level to the
processes of interest. Furthermore, assuming
minimal-flavour violation to avoid tree-level flavour-changing neutral
currents, flavour matrices of operators that involve a left-handed
doublet and a right-handed singlet are fixed to be the same as in the
SM Yukawa couplings. Within this approximation, these operators are
proportional to lepton masses and are henceforward
neglected.\footnote{In $\HZll$ the lepton mass corrections are at most
  of the order of $m_\tau^2/m_H^2\approx 2\times 10^{-4}$. The typical
  contribution from a $d=6$ operator scales as $m_H^2/\Lambda^2$ which
  is 5-10 times larger for $\Lambda$ of a few~TeV. } 

The operators
considered in this work are listed in Tab.~\ref{tab:Operators}. The
notation and conventions follow those of Ref.~\cite{Heinemeyer_etal13}. 
The Higgs doublet is denoted $\Phi$. The field strength tensors 
for the $SU(2)_L\times U(1)_Y$ gauge group are
\bea
W^I_{\mn} &=& \partial_\mu W_\nu^I - \partial_\nu W_\mu^I - 
g\epsilon^{IJK} W_\mu^J W^K_\nu,\qquad I=1,2,3, \nn \\
B_{\mn} &=& \partial_\mu B_\nu -\partial_\nu B_\mu,
\eea
with the gauge couplings $g$ and $g'$, respectively. With a tilde we denote
the dual field strength tensors
\beq
\widetilde X_{\mn} = \frac{1}{2}\epsilon_{\mn \rho\sigma} X^{\rho\sigma}
\eeq
where $X=W^I,B$ and $\epsilon_{0123}=+1$.
When acting on  $SU(2)$ doublets, the covariant derivative is written
\beq
D_\mu = \partial_\mu + ig\frac{\tau^I}{2}W_\mu^I + ig'Y B_\mu,
\eeq
where $Y$ is the hypercharge and $\tau^I$ are the Pauli matrices.
In
Tab.~\ref{tab:Operators} the left-handed lepton doublets and the
right-handed charged leptons are written $\ell_p^i$ and $e_p$, where
$i=1,2$ and $p=1,2,3$ are weak-isospin and flavour indices,
respectively. We make the simplifying assumption, stronger
than minimal flavour violation, that the coefficients 
$\alpha_{\Phi \ell}^{(1)}$,
$\alpha_{\Phi \ell}^{(3)}$,  and $\alpha_{\Phi e}$ are flavour independent, 
and define the fermion-bilinear operators with flavour indices contracted.
Because of this the operators are hermitian and all couplings $\alpha_k$
are real. A few comments on the operators in Tab.~\ref{tab:Operators} are in
order.

\begin{itemize}
\item  The $\psi^2\, \Phi^2 \,D$ type operators yield contact 
$HZ\ell  \ell$ interactions as well
as modifications of the gauge-boson couplings to leptons.
\item  The $X^2\Phi^2$ operators generate anomalous couplings of the Higgs to $ZZ$, $\gamma Z$, and $WW$.
After performing field redefinitions of the gauge
fields, the SM Higgs couplings to gauge bosons are not modified 
due to cancellations against the redefinitions of input parameters 
(see Ref.~\cite{Hagiwara:1996kf} and Eq.~(\ref{eq:d6Zstandard}) below).
\item  Operators of the type $\Phi^4\, D^2$ modify the Higgs-gauge
couplings and entail a redefinition of the Higgs field to
preserve canonically normalized kinetic terms.
\end{itemize}


The $d=4$ couplings of the electroweak sector of the SM Lagrangian are 
the gauge couplings $g$, $g'$, the Higgs self-coupling $\lambda$, and 
the classical Higgs vacuum expectation value $v$. We trade these 
couplings for the experimental observables $G_F$ (the Fermi 
constant as measured in $\mu \to e \nu_\mu \bar \nu_e$ decay), 
the $Z$ mass $m_Z$, the electromagnetic coupling $\alpha_{\rm em}$, and 
the Higgs mass $m_H$. In the presence of $d=6$ operators,
the first three of these quantities are given by
\beq
m_Z= m_{ Z\SM}\left(1+\delta_Z \right), \qquad  
G_F =G_{ F\SM}\left(1+\delta_{G_F}\right), \qquad
\alpha_{\rm em} = \alpha_{\rm em\SM}\left(1+\delta_A\right),
\label{eq:d6Zstandard}
\eeq
where $X_\SM$ denotes the quantity $X$ in the absence of $d=6$ operators, 
expressed in terms of the Lagrangian parameters $g$, $g'$, and $v$. 
The above relations are then inverted to express the $g$, $g'$, and $v$ 
in terms of $m_Z$, $G_F$ and $\alpha_{\rm em}$ and the $d=6$ 
couplings. The explicit expressions for the $d=6$ contributions 
to Eq.~(\ref{eq:d6Zstandard}) in our basis 
read~\cite{Heinemeyer_etal13,BCRM13,AJMT13} 
\beq
\delta_Z = \widehat\alpha_{ZZ} +\frac{1}{4}\widehat\alpha_{\Phi D},\qquad 
\delta_{G_F} = -\widehat\alpha_{4L} +2 \widehat\alpha_{\Phi \ell}^{(3)}, 
\qquad \delta_A = 2\widehat\alpha_{AA}.
\eeq
The combinations of coupling coefficients $\alpha_{ZZ}$ and $\alpha_{AA}$ are 
defined in Eq.~(\ref{eq:effHVVCouplingsAlpha}) below. In $\delta_{G_F}$ a 
four-lepton operator (not listed in the Tab.~1) intervenes
\beq
\mathcal{O}_{4L}^{prst} = 
(\bar \ell_p \gamma_\mu \ell_r)(\bar \ell_s \gamma^\mu \ell_t),
\label{eq:O4L}
\eeq
with $p$, $r$, $s$, and $t$ denoting flavour indices. We  
assume that the coefficients of  $\mathcal{O}_{4L}^{prst}$ are flavour 
independent.  In the expressions below we will also use the Weinberg angle 
\beq
\sin^2 \theta_W \equiv s_W^2 = 
\frac{1}{2}\left( 1 - \sqrt{1 - \frac{2 \sqrt{2}\pi\aem}{m_Z^2G_F}}\,\right), 
\qquad  \cos^2\theta_W \equiv c_W^2=1-s_W^2.
\label{eq:swZstandard} 
\eeq
It should be understood as an abbreviation for the combination 
of input parameters as given, which appears after eliminating the 
$d=4$ Lagrangian couplings as described above. 


Apart from the SM tree contributions we only consider effects of order
$1/\Lambda^2$ on the decay amplitude. In the broken-symmetry phase the 
effective Lagrangian Eq.~(\ref{leff}) generates the terms   
\bea
\mathcal{L_{\rm eff}} &\!\!\supset\!\!& c_{ZZ}^{(1)} \, H Z_\mu Z^\mu + 
c_{ZZ}^{(2)} H\, Z_{\mn}Z^{\mn} + 
c_{Z\widetilde Z}H\, Z_{\mn}\widetilde Z^{\mn} + 
c_{AZ}H \, Z_{\mn} A^{\mn} + 
c_{A\widetilde Z} H Z_{\mn}\widetilde A^{\mn} \nn \\[0.2cm]
&& + \,H Z_\mu \bar \ell \gamma^\mu \left( c_V + c_A\gamma_5   \right) \ell
+Z_\mu \bar \ell \gamma^\mu (g_V - g_A \gamma_5)\ell -
g_{\rm em}Q_\ell A_\mu \bar \ell \gamma^\mu \ell,
\label{eq:effLag}
\eea
which include the relevant tree-level SM terms. 
We omit the $H\gamma\gamma$ vertex, since it does not
contribute to the processes studied here within our approximations.
The effective couplings of this Lagrangian are related to 
the coefficients $\alpha_k$ of the fundamental $d=6$ operators 
as given explicitly in Tab.~\ref{tab:Operators}.  We define the
following combinations of coupling coefficients:
\bea
\alpha_{ZZ}^{(1)} &=&  \alpha_{\Phi \Box} -\frac{1}{2}\delta_{G_F}+ 
\frac{1 }{4}  \alpha_{\Phi D}, \nn\\
\alpha_{ZZ} &=& c_W^2 \alpha_{\Phi W}+s_W^2 \alpha_{\Phi B} 
+ s_W\,c_W \alpha_{\Phi W\!B},\nn\\[0.14cm]
\alpha_{AZ} &=& 2s_W\,c_W ( \alpha_{\Phi W}-\alpha_{\Phi B}) + (s_W^2-c_W^2)\alpha_{\Phi W\!B},\nn\\
\alpha_{AA} &=&  s_W^2 \alpha_{\Phi W}+ c_W^2\alpha_{\Phi B}- s_W\,c_W \alpha_{\Phi W\!B}.
\label{eq:effHVVCouplingsAlpha}
\eea
with analogous expressions for $\alpha_{ Z\widetilde  Z}$ and
$\alpha_{ A\widetilde  Z}$, where the couplings on the r.h.s. are replaced
by their tilde counterparts.    
The Higgs-gauge couplings of Eq.~(\ref{eq:effLag}) are then given by
\bea
c_{ZZ}^{(1)}  &=& m_Z^2(\sqrt{2}G_F)^{1/2}\,\left(1 + 
\widehat\alpha_{ZZ}^{(1)}  \right), \nn \\
c_{ZZ}^{(2)} &=& (\sqrt{2}G_F)^{1/2}\,\widehat\alpha_{ZZ},\nn \\
c_{Z\widetilde Z} &=&(\sqrt{2}G_F)^{1/2}\, 
\widehat \alpha_{Z\widetilde Z},\nn \\
c_{AZ} &=&  (\sqrt{2}G_F)^{1/2}\,\widehat \alpha_{AZ},   \nn \\
c_{A\widetilde Z} &=&(\sqrt{2}G_F)^{1/2}\, \widehat \alpha_{A\widetilde Z}.
\label{eq:effHVVCouplings}
\eea
The contact $HZ\ell \ell$ couplings can be written as
\bea
c_V &=& \sqrt{2}G_F\,m_Z \,\widehat\alpha^V_{\Phi \ell}, \nn\\
c_A &=& \sqrt{2}G_F \,m_Z\,\widehat\alpha^A_{\Phi \ell}.
\label{eq:effCVandCA}
\end{eqnarray}
with
\bea
\haV  &=& \widehat\alpha_{\Phi e} +
\left(\widehat\alpha_{\Phi \ell}^{(1)}+
\widehat\alpha_{\Phi \ell}^{(3)}\right),\nn\\
\haA   &=& \widehat\alpha_{\Phi e} -  
\left(\widehat\alpha_{\Phi \ell}^{(1)}+
\widehat\alpha_{\Phi \ell}^{(3)}\right).
\label{eq:alphaVandA}
\eea
Note that in the operator basis employed here $\haV$ and $\haA$ are, 
in general, different. Therefore, in the effective Lagrangian we have both
left-handed and right-handed $HZ\ell\ell$ couplings. This contrasts
with the so-called SILH basis~\cite{SILH}, where the absence of the
operators $\mathcal{O}_{\Phi \ell}^{(1,3)}$ implies $\haV=\haA$ and
hence only right-handed $HZ\ell\ell$ couplings enter the effective 
Lagrangian directly~\cite{PR13}.

The same combinations $\haVA$ also participate in the $d=6$ corrections 
to the  $Z$ couplings to fermions in Eq.~(\ref{eq:effLag}), that can be 
cast as
\bea
g_V&=& \frac{m_Z}{2} (\sqrt{2} G_F)^{1/2}
\left[  \left(1- 4\, s_W^2\right) -  \delta g_V\right],\nn\\
g_A&=& \frac{m_Z}{2} (\sqrt{2} G_F)^{1/2} 
\left(1 + \delta g_A\right).
\label{eq:Zll}
\eea
The corrections $\delta g_{V,A}$ from the $d=6$ operators are given by
\bea
\delta g_V &=& -\haV +\frac{\widehat\alpha_{\Phi D}}{4} +\frac{\delta_{G_F}}{2} + 
\frac{4 s_W^2}{c_{2W}}\left[ \frac{\widehat\alpha_{\Phi D}}{4} +
\frac{c_W}{s_W} \,\widehat\alpha_{\Phi WB} +\frac{\delta_{G_F}}{2}   \right], 
\nn \\
\delta g_A &=&  - \haA  -\frac{\widehat\alpha_{\Phi D}}{4} -
\frac{\delta_{G_F}}{2},
\label{eq:cA6}
\eea
with $c_{2W}\equiv \cos 2\theta_W$. The contributions from 
$\widehat\alpha_{\Phi D}$, $\widehat\alpha_{\Phi WB}$,
and $\delta_{G_F}$ arise from the redefinition of SM fields and
the rewriting of Lagrangian parameters in terms of input parameters  
in the presence of $d=6$ operators.


In this work, we are interested in the effects of Higgs anomalous
couplings in certain angular asymmetries of $\HZll$ and $\eeHZ$. 
In order to estimate the maximal effect that is still possible, we
have to incorporate the constraints on the anomalous couplings 
from all existing data, not necessarily related to Higgs observables. 
However, to the best of our knowledge, a full
analysis along the lines of Refs.~\cite{PR13,Corbettetal} is not 
available in the operator basis employed here. Although different bases are 
related by a linear transformation, the results cannot be straightforwardly 
translated, since the correlations are not known (to us). We are 
particularly interested in the coefficients $\haVA$ of the contact 
interactions. In the absence of a complete analysis,
we perform here an order-of-magnitude estimate of the 
present constraints on~$\haVA$, which is sufficient for the discussion 
of the angular asymmetries.

In order to use data for $g_{V,A}$ to constrain $\haVA$ one needs to
estimate the allowed range for the other three combinations of Wilson
coefficients that enter Eq.~(\ref{eq:cA6}), namely,
$\widehat\alpha_{\Phi D}$, $\widehat\alpha_{\Phi WB}$, and
$\delta_{G_F}$. One therefore needs  five observables. 
Apart from the $Z$ coupling to leptons, $g_V$ and $ g_A$, we employ 
the electroweak precision observables $S$ and $T$, and the $W$ mass.
The operators $\mathcal{O}_{\Phi WB}$ and
$\mathcal{O}_{\Phi D}$ give tree-level contributions to $S$ and $T$, 
respectively (for the explicit expressions in our basis, see~\cite{AJMT13}). 
Experimental values for these two parameters~\cite{EWfit12} constrain 
$\widehat\alpha_{\Phi WB}$ and $\widehat\alpha_{\Phi D}$ to be at the 
permille level. With these bounds as input,
$m_W$ can be utilized to constrain $\delta_{G_F}$, since
\beq
m_W = m_Z(1-s_W^2)^{1/2}\left[1 -\frac{1}{2 c_{2W}}
\left(\frac{c_W^2}{2}\ha_{\Phi D} +s_W^2 
\delta_{G_F} + s_{2W}\, \ha_{\Phi WB}   \right)    \right],
\label{eq:mWd6}
\eeq
with $  s_{2W} \equiv  \sin2\theta_W  $.
This constrains $\delta_{G_F}$ to be at the level of a few $10^{-3}$.
Finally, using these results and the tight constraints on $\delta
g_{V,A}$~\cite{Schael_etal06,Pich13}, we find that $\haA$ and $\haV$ 
cannot exceed a few times $10^{-3}$.\footnote{The triple-gauge boson
  coupling $\Delta g_1^{Z}$~\cite{MS13,PR13} could also be used to
  constrain $\delta_{G_F}$ but leads to less stringent bounds than the
  value of $m_W$ does. Data for the decay $Z\to \bar \nu \nu$ help
  disentangling $\ha_{\Phi \ell}^{(1)}$ and $\ha_{\Phi \ell}^{(3)}$
  but this is immaterial to the present work.}
This agrees with the conclusion of Ref.~\cite{PR13} 
that the bounds on $\haVA$ are at the permille level,
though this refers to the SILH basis in which $\haA=\haV$. 

Since we do not perform a full fit, we allow $\haA$ and $\haV$ to vary 
within slightly weaker bounds than those arising from the five observables 
described above. We therefore employ the  conservative interval
\beq
\haVA \in [-5,5] \times 10^{-3}.
\label{contactrange}
\eeq
According to Eq.~(\ref{ahat}), for $\aVA=1$, this corresponds to the 
BSM physics scale of $\Lambda\approx 3.5\,$TeV.
Note that the maximally allowed value of $\haV$
is smaller by a factor of about 4 compared to the one used
in Ref.~\cite{BCD13}.

The $d=6$ anomalous $H Z\gamma$ vertex also plays an important role 
in our analysis.  This coupling is constrained by Higgs measurements, 
especially the direct searches for $H\to Z \gamma$ decays, which 
presently limit the branching fraction to about ten times the SM 
expectation~\cite{ATLAS_HZgamma,CMS_HZgamma}. This leads to a weaker 
bound than those on $\haVA$ and here we employ the result of Ref.~\cite{PR13}, 
which with our definitions reads
\beq
\ha_{AZ} \in [ -1.3,2.6] \times 10^{-2}.
\label{hzarange}
\eeq

One-loop corrections to the SM amplitude give contributions to the 
$\HZll$ and $\eeHZ$ processes studied in this paper that can be of the same 
order of $d=6$ terms. They have been computed in the
past~\cite{Kniehl90,Kniehl91,Kniehl94,BDDW06} and should eventually 
be included in a quantitative extraction of the anomalous couplings from 
data.  Since this data is not yet available, and our purpose is to 
determine the sensitivity of angular observables to $d=6$ operators, 
we neglect loop corrections here. In Sec.~\ref{sec:SM-loop} we provide 
a rough estimate in order to ascertain whether or not loop effects 
affect the main conclusions of this work. Loop corrections to amplitudes 
with $d=6$ operator insertions are evidently negligible.

In the following we discuss the effects of anomalous
Higgs couplings of \Eqs{eq:effHVVCouplings}{eq:effCVandCA} in the
differential decay width of $\Hllll$, in the total cross section
of $\eeHll$, as well as on angular asymmetries of
these two crossing-symmetric processes.
Two main scenarios will be investigated in detail. In the first
we allow for non-vanishing $\haVA$, which gives rise to the
$HZ\ell\ell$ contact interaction of Eq.~(\ref{eq:effCVandCA}), 
and in the second for non-vanishing
$\wh\alpha_{AZ}$. In each scenario we set the other couplings 
 to zero. In scenarios with non-vanishing
$\haVA$ their contribution to $\delta g_{V,A}$ is  taken into
account.


\section{\boldmath Angular asymmetries of $H\to Z(\to \lplm) \lplm$}
\label{sec:HZll}

\begin{figure}[t]
\begin{center}
\includegraphics[width=0.95\columnwidth]{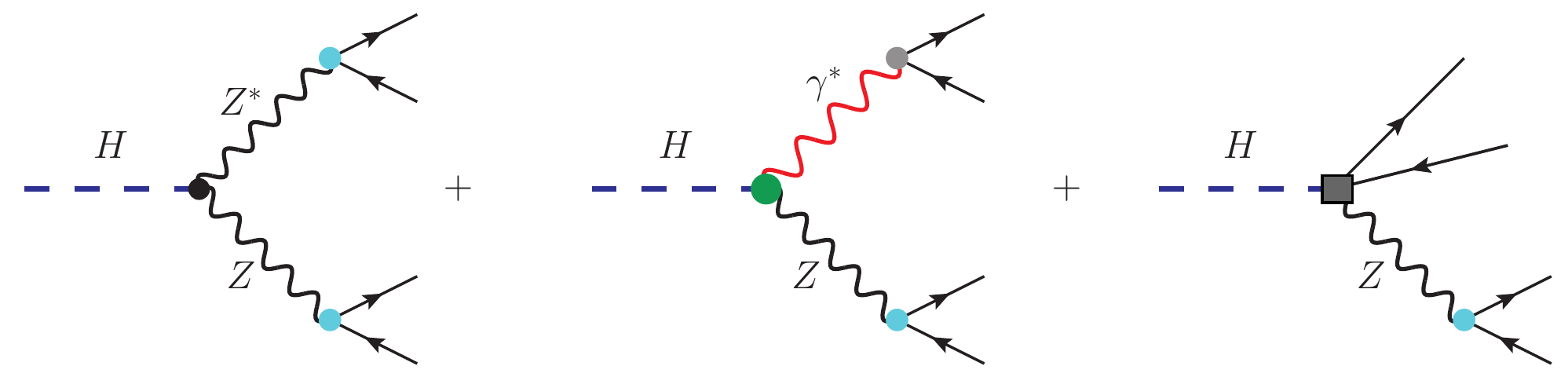}
\vspace*{0.1cm}
\caption{ Feynman diagrams for the decay $\Hllll$.
}
\label{fig:Feynman}
\end{center}
\end{figure}

The decay of the on-shell Higgs boson to four leptons with an intermediate
on-shell $Z$ boson can proceed with an off-shell $Z$ through the $H\to ZZ$
interaction, as in the SM, but with $d=6$ operators added to the 
Lagrangian it can also proceed through a
$HZ\gamma$ coupling or the contact interaction
$HZ\ell \ell$. The three types of diagrams are depicted in
Fig.~\ref{fig:Feynman}.

\subsection{Form factors and angular distribution}

The  amplitude for the decay $H(p_H) \to Z(p) (\to \ell^- (p_1)
\ell^+(p_2)) \ell^- (p_3)\ell^+(p_4)$ can be written as 
\beq
\mathcal{M}(\Hllll) = \mathcal{M}_{HZ\ell\ell}^\mu\,
\frac{1}{p^2-m_Z^2+i\,m_Z\,\Gamma_Z}\,\mathcal{M}_{Z\ell\ell,\mu},
\label{eq:totM}
\eeq
where $\mathcal{M}^\mu_X$ denotes the matrix element of process $X$ 
with the polarization vector of the on-shell $Z$ boson stripped off. 
As already mentioned, we neglect lepton mass effects. When squaring 
the amplitude we employ the narrow-width approximation for the 
intermediate $Z$ boson, but include spin correlations. 
Summing over spins of the
final-state leptons, the four-fold differential decay width for the
process $\Hllll$ in the massless lepton limit can be written as a
function of the di-lepton invariant mass squared $q^2 = (p_3+p_4)^2$
and of three angles (see App.~\ref{app:KineHZll} for their 
definitions). The expression reads
\begin{eqnarray}
\frac{d^4 \Gamma }{  dq^2 d \cos \theta_1 d \cos \theta_2 d \phi} &=& 
\frac{1 }{ 2^{10} (2 \pi)^5}\,
\frac{1 }{m_H^3} \,\frac {1 }{ m_Z \Gamma_Z} \,
\lambda^{1/2}(m_H^2,m_Z^2,q^2) \,   \sum_{\rm spins} \,
|\mathcal{M}_{HZ\ell\ell}^\mu\,
\mathcal{M}_{Z\ell\ell,\mu} |^2 \, \nonumber \\
&=& \frac{1}{m_H}\, \mathcal{N} (q^2)\,\mathcal{J}(q^2, \theta_1, \theta_2, \phi).
\label{eq:DiffDecayRate}
\end{eqnarray}
In the last equation we introduced the dimensionless function
\beq
\mathcal{J}(q^2, \theta_1, \theta_2, \phi) = \frac{1}{m_H^2}  
\sum_{\rm spins} \,
|\mathcal{M}_{HZ\ell\ell}^\mu\,
\mathcal{M}_{Z\ell\ell,\mu} |^2,
\label{eq:calJ}
\eeq
and the normalization
\begin{eqnarray}
\mathcal{N}(q^2)=\,\frac{1 }{ 2^{10} (2 \pi)^5}  \frac{1}{\sqrt{r}\, \gamma_Z}
\lambda^{1/2}(1, r,s),\label{eq:Norm}
\end{eqnarray}
written in terms of the  dimensionless variables
\beq
s= \frac{q^2}{m_H^2},\qquad r= \frac{m_Z^2}{m_H^2}\approx 0.53, \qquad 
\qquad \gamma_Z = \frac{\Gamma_Z}{m_H}\approx 0.020,
\label{eq:AdVar}
\eeq
and  the function $\lambda(a,b,c) = a^2 +b^2 +c^2 -2ab -2ac -2bc$.
The maximum value of $q^2 $ is
$q_{\rm max}^2 = (m_H-m_Z)^2 \approx (34.4~\mbox{GeV})^2$ which gives
\beq
0\le  s \le \frac{(m_H-m_Z)^2}{m_H^2} \approx 0.075.
\eeq

The decay of the on-shell $Z$ boson is described by
\beq
 {\cal M}_{Z\ell\ell}^{\mu}
=  \bar u(k_1,s_1)  \left [ \gamma^\mu \left(g_V-g_A\gamma_5   \right)
 \right ]v(k_2,s_2) ,\label{eq:MZll}
\eeq
with the couplings given in Eq.~(\ref{eq:Zll}).  It is important to
observe that $g_A$ largely dominates the interaction $Z\to \ell^+
\ell^-$ due to the partial cancellation in the factor $(1- 4\,
s_W^2)$ in $g_V$. This fact plays an important role in the interpretation of 
the numerical results for the angular asymmetries.

Neglecting the lepton masses the general expression for the amplitude 
of $H\to Z(p) \ell^-(p_3)
\ell^+(p_4)$ at $\mathcal{O}\left(1/\Lambda^2\right)$ in the
$d=6$ Lagrangian   can be written in terms of six form
factors~\cite{GMM07,CMMZ2003,IMT13,GMP13,BCD13}. 
Denoting them by $H_{i,V/A}$ ($i=1,2,3$), we adopt the parametrization 
\begin{eqnarray}
{\cal M}_{HZ\ell\ell}^\mu &=& \frac{1}{m_H} \,
\bar u(p_3,s_3) \bigg[ \gamma^\mu\left(H_{1,V}+H_{1,A}\,\gamma_5\right) + 
\frac{ q^\mu \slashed{p}}{m_H^2} \left(H_{2,V}+H_{2,A}\,\gamma_5\right)  
\nonumber \\
&&+\, \frac{\epsilon^{\mu\nu\sigma\rho}p_\nu q_\sigma}{m_H^2} \, 
\gamma_\rho \,\left(H_{3,V}+H_{3,A}\,\gamma_5\right)\bigg] 
v(p_4,s_4),\label{eq:MHZll}
\end{eqnarray}
where $\epsilon_{0123}=+1$ and $q=p_3+p_4$. 
The form factors $H_{2,V/A}$ and $H_{3,V/A}$  vanish
in the SM at tree level.  The expressions for $H_{i,V/A}$
at $\mathcal{O}(1/\Lambda^2)$  are 
\begin{eqnarray}
H_{1,V}&=& - \frac{2m_H (\sqrt{2} G_F)^{1/2} \,r}{r-s} \,g_V
\left(1 + \wh\alpha_1^{\rm eff}-\frac{\kappa}{r}\, \wh\alpha_{ZZ}
- \frac{\kappa}{2r}\frac{Q_\ell\, g_{em} \, (r-s)}{s\, g_V} \,\wh\alpha_{AZ} 
  \right) ,  \nonumber \\
H_{1,A}&=&\frac{2m_H (\sqrt{2} G_F)^{1/2} \,r}{r-s} \,g_A
\left(1 + \wh\alpha_2^{\rm eff}-\frac{\kappa}{r}\, 
\wh\alpha_{ZZ}\right) ,\nn \\
H_{2,V}&=& -\frac{2m_H (\sqrt{2} G_F)^{1/2}}{r-s} \,g_V\,
\bigg [ 2  \, 
\wh\alpha_{ZZ}   + \frac{Q_\ell\, g_{em} \, (r-s)}{s\, g_V} \,\wh\alpha_{AZ} 
\bigg]  \,,  \nonumber \\
H_{2,A}&=&  \frac{4m_H (\sqrt{2} G_F)^{1/2}}{r-s} \, g_A  \, 
\wh\alpha_{ZZ} \,,  \nonumber \\
H_{3,V}&=& -\frac{2m_H (\sqrt{2} G_F)^{1/2}}{r-s}\,g_V \,
\bigg [  
2   \, \wh\alpha_{Z\widetilde{Z}} +\frac{Q_\ell \, g_{em} \, (r-s)}{s\, g_V} \,  
\wh\alpha_{A\widetilde{Z}}\bigg] \,,  \nonumber \\
H_{3,A}&=& \frac{4m_H (\sqrt{2} G_F)^{1/2}}{r-s} \, g_A   
\,\wh \alpha_{Z\widetilde{Z}},
\label{eq:ExplicitH}
\end{eqnarray}
where $Q_\ell=-1$. 
The couplings $g_A$ and $g_V$ are those of Eq.~(\ref{eq:Zll}) including  
the $d=6$ corrections. (Of course, within our approximations, 
this matters only when $g_{V,A}$ multiply the 
SM ``1'' in the bracket in $H_{1,V/A}$.) We defined the combinations
\bea
\wh \alpha_{1}^{\rm eff} &\equiv& \wh\alpha_{ZZ}^{(1)} \,
- \frac{ m_H (\sqrt{2} G_F)^{1/2} \, (r-s)} {2\sqrt{r}  } \, 
\frac{\wh\alpha_{\Phi l}^V}{g_V}  , \nn \\
\wh\alpha_{2}^{\rm eff}&\equiv&\wh \alpha_{ZZ}^{(1)} \,
+ \frac{ m_H (\sqrt{2} G_F)^{1/2} \, (r-s)} {2\sqrt{r}  } \, 
\frac{\wh\alpha_{\Phi l}^A}{g_A} ,
\label{eq:alpha1alpha2}
\eea
where the couplings $\wh\alpha_{ZZ}^{(1)}$ and $\wh\alpha_{\Phi l}^{V/A}$ 
are defined in  Eqs.~(\ref{eq:effHVVCouplingsAlpha})
and~(\ref{eq:alphaVandA}), respectively.
Last, we introduced 
\begin{equation}
 \kappa=1-r-s.
\end{equation}
At order $1/\Lambda^2$, ignoring loop-suppressed contributions and lepton 
masses, the form factors of Eq.~(\ref{eq:ExplicitH}) are real. 
Note that the absence of $i$ in front of the epsilon-symbol in 
Eq.~(\ref{eq:MHZll}) implies that with this definition 
real $H_{3,V/A}$ are CP-odd form factors 
as can also be seen  from their expressions in 
Eq.~(\ref{eq:ExplicitH}).

Computing $\mathcal{J}(q^2, \theta_1, \theta_2, \phi)$ explicitly, 
we find nine independent angular structures with coefficient 
functions  $J_1$,...,$J_{9}$, which we write 
as\footnote{\label{foot:deltaJ} To make 
contact with Ref.~\cite{GMM07}, we remark that final-state interactions,  
which would generate (loop-suppressed) imaginary parts
in the form factors, lead to six new angular structures. Denoting these new
structures by $\delta \mathcal J$, the expression
\begin{eqnarray*}
\delta\mathcal{J} &=&\left(J_{10} \sin2\theta_1\st + J_{11} 
\so \sin2\theta_2\right)\sin\phi 
 +\, \left(J_{12} \sin2\theta_1\st + 
J_{13} \so \sin2\theta_2\right)\cos\phi
\nonumber \\
&& +\,J_{14}\ct(1+\cosq) +J_{15}\co(1+\ctsq)
\end{eqnarray*}
has to be added to Eq.~(\ref{eq:FullJ}). The new angular functions 
depend on the imaginary parts of the form factors $H_{i,V/A}$.}
\begin{eqnarray}
\mathcal{J}(q^2, \theta_1, \theta_2, \phi) &=& 
J_1(1+ \cosq\ctsq {+\cosq +\ctsq}) \nn \\ 
&& +\, J_2 \sosq\stsq +J_3\co\ct  \nn \\
&& +\, \left(J_4 \so\st +J_5\sin2\theta_1\sin2\theta_2\right)\sin\phi
\nn \\
&& +\, \left(J_6 \so\st +J_7\sin2\theta_1\sin2\theta_2\right)\cos\phi
\nn \\
&& +\,J_8\sin^2\theta_1\,\sin^2\theta_2\,\sin2\phi
+J_9\sin^2\theta_1\,\sin^2\theta_2\,\cos2\phi  .
\label{eq:FullJ}
\end{eqnarray}
The  expressions for the non-vanishing $J$ functions at
$\mathcal{O}(1/\Lambda^2)$ in the limit $m_\ell\to 0$ in terms of the form
factors of Eq.~(\ref{eq:MHZll}) are\footnote{Our expression agrees with 
Ref.~\cite{BCD13} with adjustments for the different
definitions of angles and form factors.
In particular, we have $\theta_1 \to \pi -\alpha$ and $\theta_2\to \beta$, 
$H_{1,V}\to  2F_1 G_V$, $H_{1,A}\to - 2F_1 G_A$, $H_{2,V}\to H_{V}$, 
$H_{2,A}\to- H_{A}$, $H_{3,V}\to- K_{V}$, and $H_{3,A}\to K_{A}$. 
Note that the definitions of $J_1$ and $J_2$ are different.}
\bea
&&J_1=2 \, r \,s\, \left(g_A^2+g_V^2\right)
\left(|H_{1,V}|^{2}+|H_{1,A}|^{2}\right), \nn \\[0.1cm]
&&J_2=\kappa\, \left(g_A^2+g_V^2\right) \left[\kappa \,
\left(|H_{1,V}|^{2 }+|H_{1,A}|^{2}\right)+\lambda
\,  {\rm Re} \left( H_{1,V} H^{\ast}_{2,V} + H_{1,A} H^{\ast}_{2,A}\right)
\right],\nn \\[0.1cm]
&&J_3=32 \,r\, s\, g_A\, g_V\, {\rm Re} \left ( H_{1,V} \, H_{1,A}^{\ast}
\right ), \nn \\
&&J_4=4\kappa\, \sqrt{r\,s\,\lambda} \,g_A\,g_V \, {\rm Re}
\left(H_{1,V} H_{3,A}^{\ast}+H_{1,A}H_{3,V}^{\ast}\right),\nn \\
&&J_5=\frac{1}{2}\kappa\, \sqrt{r\,s\,\lambda} \,
\left(g_A^2+g_V^2\right) {\rm Re} \left(H_{1,V} H_{3,V}^{\ast}+H_{1,A}\,
H_{3,A}^{\ast}\right),\nn \\
&&J_6=4 \sqrt{r\,s}\, g_A\, g_V\, \left[4 \kappa  \, {\rm Re}
\left (H_{1,V}H_{1,A}^{\ast} \right )
+\lambda  \, {\rm Re} \left (H_{1,V}H_{2,A}^{\ast} + H_{1,A}H_{2,V}^{\ast} 
\right)
\right] ,\nn \\
&&J_7=\frac{1}{2} \sqrt{r\, s} \left(g_A^2+g_V^2\right) \left[2 \kappa \, 
\left(|H_{1,V}|^{2}+|H_{1,A}|^{2}\right)
+ \lambda \, {\rm Re} \left(H_{1,V} H_{2,V}^{\ast} + H_{1,A} H_{2,A}^{\ast} 
\right)
\right] ,\nn \\
&&J_8=2\, r\, s\, \sqrt{\lambda } \left(g_A^2+g_V^2\right) {\rm Re}
\left( H_{1,V}H_{3,V}^{\ast}+ H_{1,A}H_{3,A}^{\ast}\right),\nn \\[0.1cm]
&&J_9=2 \,r \,s \,\left(g_A^2+g_V^2\right)
\left(|H_{1,V}|^{2}+|H_{1,A}|^{2}\right).\label{eq:JFuncsinH}
\eea
These expressions are valid beyond our approximations, 
where the $H_{i,V/A}$ form factors are all real.  We used the notation 
$\lambda\equiv \lambda(1,r,s)$ and 
recall that $g_{V,A}$ implicitly contain $d=6$ corrections,  
see Eq.~(\ref{eq:Zll}).
At order $\mathcal{O}(1/\Lambda^2)$, $H_{2,V/A}$ and $H_{3,V/A}$
contribute only through interference with the SM part of the form
factors $H_{1,V/A}$. We drop the $\mathcal{O}(1/\Lambda^4)$ 
terms still contained in Eq.~(\ref{eq:JFuncsinH}).

Only six of the functions $J_i$ in Eq.~(\ref{eq:JFuncsinH}) are
independent. The following relations hold:
\bea
J_5 &=& \frac{\kappa}{4\sqrt{rs}} J_8, \nn \\
J_7 &=& \frac{\sqrt{rs}}{2\kappa} \left(\frac{\kappa^2}{2rs}J_1+ J_2 \right),
\nn\\[0.1cm]
J_9&=&J_1.
\eea
Three of the $J_i$ functions, namely $J_4$, $J_5$, and $J_8$, are CP-odd 
and vanish in the SM  at tree level. 
From the two independent functions among these three, one could determine 
the CP-odd effective couplings $\wh\alpha_{A\widetilde Z}$ and 
$\wh\alpha_{Z\widetilde Z}$.  From the remaining four CP-even 
angular functions, one obtains information on the anomalous couplings 
$\wh\alpha_{1,2}^{\rm eff}$, $\wh\alpha_{AZ}$, and $\wh\alpha_{ZZ}$.
The explicit expressions for the $J$ functions in terms of the
effective couplings are collected in App.~{\ref{app:JFunctions}}. 
We will use them to get analytic insight into the numerical 
analysis presented below.

\subsection{Observables}

Integrating over the three angles in Eq.~(\ref{eq:DiffDecayRate}), the 
differential decay rate as a function of the di-lepton invariant mass is 
given by
 \beq
\frac{d \Gamma}{dq^2} =
 \frac{32\pi}{9} \frac{1}{m_H} \, \mathcal{N}(q^2) \, (4J_1+J_2).
\label{eq:dGammadq2}
\eeq
This observable has been explored recently in
Refs.~\cite{IMT13,GMP13}.  Here, instead, the main focus is on the 
angular asymmetries from which individual $J$ functions can be extracted. Some
of these asymmetries  have already been discussed in~Ref.\cite{BCD13}.
We define the following angular asymmetries normalized to 
$d\Gamma/dq^2$:
\begin{eqnarray}
\mathcal{A}_{\rm  \theta_1} &=& \frac{1}{d\Gamma /dq^2} \, \int_{-1}^1  \,  
d \cos \theta_1 \,\, {\rm sgn}(\cos (2\theta_1)) \, 
\frac{d^2 \Gamma}{d q^2 d\cos \theta_1} \nn \\
 &=& 1-\frac{5}{2\sqrt{2}}+\frac{3J_1}{\sqrt{2}(4J_1+J_2)} \,,
\label{eq:Asm2}\nn\\
\mathcal{A}_{\rm  \phi}^{(1)} &=& \frac{1}{d\Gamma /dq^2} \, 
\int_{0}^{2 \pi}  d \phi  \,\, {\rm sgn}(\sin \phi) \,
\frac{d^2 \Gamma}{d q^2 d \phi} = \frac{9 \, \pi}{32} \,  
\frac{J_4}{4J_1+J_2} , \label{eq:Asm5} \nn\\
\mathcal{A}_{\rm  \phi}^{(2)} &=& \frac{1}{d\Gamma /dq^2} \, 
\int_{0}^{2 \pi}   d \phi  \,\, {\rm sgn}(\sin (2 \phi) ) \,
\frac{d^2 \Gamma}{d q^2 d \phi} = \frac{2}{\pi} \, \frac{J_8}{4J_1+J_2} ,
\label{eq:Asm6}\nn  \\
\AsymThree &=& \frac{1}{d\Gamma /dq^2} \, \int_{0}^{2 \pi}    
d \phi  \,\, {\rm sgn}(\cos \phi) \,
\frac{d^2 \Gamma}{d q^2 d \phi} = \frac{9 \, \pi}{32} \,  
\frac{J_6}{4J_1+J_2} ,  \label{eq:Asm7} \nn \\
\mathcal{A}_{\rm  \phi}^{(4)} &=& \frac{1}{d\Gamma /dq^2} \, 
\int_{0}^{2 \pi}   d \phi  \,\, {\rm sgn}(\cos (2 \phi) ) \,
\frac{d^2 \Gamma}{d q^2 d \phi} = \frac{2}{\pi} \,  
\frac{J_{9}}{4J_1+J_2} .
 \label{eq:Asm8}
\end{eqnarray}
The sign function is $\sgn(\pm |x|) = \pm 1$. We further define the 
 double forward-backward asymmetry
\bea
\AsymCoCt &=&\frac{1}{d\Gamma /dq^2} \int_{-1}^{1} 
d\cos\theta_1\, \sgn(\cos \theta_1)\int_{-1}^{1} 
d\cos\theta_2\, \sgn(\cos \theta_2) 
\frac{d^3 \Gamma}{  dq^2 d \cos \theta_1 d \cos \theta_2 }\nn \\
&=& \frac{9}{16} \, \frac{J_{3}}{4J_1+J_2}. \label{eq:Asm9}
\eea

The single forward-backward asymmetry in the angle $\theta_2$ 
(see App.~\ref{app:KineHZll}), i.e.,
\beq
\frac{1}{d\Gamma /dq^2} \, \int_{-1}^1  \,\,   d \cos \theta_2 \,\, {\rm sgn}(\cos (\theta_2)) \, \frac{d^2 \Gamma}{d q^2 d\cos \theta_2} ,
\eeq
vanishes for $H\to Z(\to \lplm) \lplm$ in the  present 
approximation, 
as already noticed in Ref.~\cite{BCD13}. This is different from the 
analogous forward-backward asymmetry in the electroweak penguin 
decay $B \to K^{\ast} \lplm$, despite a very similar form factor structure. 
To understand this difference, we  note the explicit expression
for the $B \to K^{\ast} \lplm$  decay amplitude in the factorization 
approximation (sufficient for the purpose of explanation),
\begin{eqnarray}
 {\cal M} (B \to K^{\ast} \, \ell^+\ell^-)  & \propto &  \bar u (q_2)
\bigg[ \gamma^\mu \left(C_9^{\rm eff} + C_{10}\,\gamma_5\right) \bigg] 
v(q_1) \,
\langle K^{\ast}(p) |\bar{s} \gamma_\mu  (1-\gamma_5) b |  B(p+q)\rangle     
\nonumber \\
&\propto& \bar u (q_2) \bigg [ \gamma^\mu \left(C_9^{\rm eff} + C_{10}\,
\gamma_5\right) \bigg] v (q_1)  \,
\bigg \{ \frac{2 V(q^2)}{m_B+ m_{K^{\ast}}}\, 
i \varepsilon_{\mu \nu \rho \sigma} \, \epsilon_{K^{\ast}}^{\nu}  
\, p^{\rho} \, q^{\sigma} \nonumber \\
&& + (m_B +m _{K^{\ast}})\, A_1(q^2)  \, \left [ {\epsilon_{K^{\ast}}}_{\mu} 
- \frac{ \epsilon_{K^{\ast}} \cdot q }{  q^2 } \, q_{\mu}  \right ]
+ ...  \bigg \}  \,,
\label{B to Kstar ll amplitude}
\end{eqnarray}
where $V$ and $A_1$ denote $B\to K^\ast$ form factors. 
The forward-backward asymmetry is determined by \cite{Altmannshofer:2008dz}
\begin{eqnarray}
A_{FB}(B \to K^{\ast} \, \ell^+\ell^-) \propto
{\rm Re} \left (A_{\parallel}^L \, {A_{\perp}^L}^{\ast}\right)
-\left(L \rightarrow  R \right),
\end{eqnarray}
where the transversity amplitudes within the current approximation 
are given by
\begin{eqnarray}
A_{\perp}^{L,R} \propto  (C_9^{\rm eff} \mp C_{10}) \, 
\frac{V(q^2)}{m_B+ m_{K^{\ast}}} \,,
\qquad
A_{\parallel}^{L,R} \propto  (C_9^{\rm eff} \mp C_{10}) \, 
\frac{A_1(q^2)}{m_B - m_{K^{\ast}}} \,.
\end{eqnarray}
The single forward-backward asymmetry in the angle $\theta_2$ is
generated by the CP-even part of the interference of transversity
amplitudes, and is proportional to $\mbox{Re}\,(C_9^{\rm eff}
C_{10}^\ast)$.  Comparing Eq.~(\ref{B to Kstar ll amplitude}) and
Eq.~(\ref{eq:MHZll}), and noting the different factors of $i$ in front
of the epsilon symbols, we see that the transversity amplitudes
$A_{\perp}^{L,R}$ in $H\to Z \lplm$ decays are CP-odd at tree level
(when $H_{3,V/A}$ is real), hence the interference of
$A_{\perp}^{L,R}$ and $A_{\parallel}^{L,R}$ cannot induce a CP-even
observable. This implies the vanishing of the single forward-backward
asymmetry in $H\to Z \lplm$ decay at tree level.\footnote{
Beyond the narrow-width approximation, complex form factors 
and a forward-backward asymmetry 
can also be generated by the imaginary part of the $Z$ boson 
propagator~\cite{Chenetal2014}, at the cost of an additional 
$\Gamma_Z/m_Z$ suppression.}

Due to the vanishing of CP-odd Higgs couplings to gauge bosons in the
SM, the angular functions $J_4$ and $J_8$ and hence the 
asymmetries $\mathcal{A}_{\rm \phi}^{(1)}$ and $\mathcal{A}_{\rm
  \phi}^{(2)}$ are generated only by the anomalous couplings 
from the $d=6$ operators. In principle, the asymmetries defined in
Eqs.~(\ref{eq:Asm2}) and (\ref{eq:Asm9}) can determine the
six anomalous couplings appearing in $HZ\ell\ell$ form factors
unambiguously.

\subsection{Higgs couplings in angular asymmetries 
of \boldmath $\HZll$}
\label{sec:AsymHZll}

In this section we discuss the impact of anomalous Higgs
couplings on the angular asymmetries in Eqs.~(\ref{eq:Asm2})
and~(\ref{eq:Asm9}) and, for comparison, the di-lepton invariant 
mass distribution $d\Gamma/dq^2$,
Eq.~(\ref{eq:dGammadq2}).  Some of these asymmetries and their
sensitivity to new physics have already been explored in
Ref.~\cite{BCD13}. We comment on their results in the remainder.
The analysis can be split into the CP-conserving and CP non-conserving 
parts.  At order $\mathcal{O}(1/\Lambda^2)$, the 
CP-odd couplings, $\wh\alpha_{A\widetilde Z}$ and $\wh\alpha_{Z\widetilde Z}$, 
contribute only to $J_4$, $J_5$, and $J_8$ and therefore
do not contribute to the decay rate $d\Gamma/ds$.  The CP-even 
couplings $\wh\alpha_{AZ}$, $\wh\alpha_{ZZ}$, as well as the
combinations $\wh\alpha_1^{\rm eff}$ and $\wh\alpha_2^{\rm eff}$, that
contain the contact $HZ\ell  \ell$ interactions, enter the
remaining angular functions $J_1$, $J_2$, $J_3$, $J_6$, $J_7$, $J_9$. 

Most of the distinctive phenomenology of the angular asymmetries stems 
from the suppression of the vector $Z\ell  \ell$ coupling $g_V$ 
compared to the axial coupling $g_A$. With the conventions of 
Eq.~(\ref{eq:Zll}), $g_V\simeq 0.012$ and $g_A \simeq 15 g_V$.  
Inspecting the
explicit expressions of $J_1$ and $J_2$ in App.~\ref{app:JFunctions},
one sees that $\wh\alpha_{AZ}$ and $\wh\alpha_1^{\rm eff}$ contributions
come with suppression factors of $g_V$ and $g_V^2$, respectively, and
therefore have little effect on $d\Gamma/dq^2 \propto (4J_1 +J_2)$.
In contrast, in $J_3$ and $J_6$, the contributions from 
$\wh\alpha_{AZ}$ and $\aV$ are $1/g_V$ enhanced in comparison with the
other $d=6$ couplings.  The asymmetries that probe these coefficient
functions, $\AsymCoCt$ and $\AsymThree$, are
therefore good candidates to reveal effects that would not be visible in the
 di-lepton mass spectrum.  This pattern motivates our 
focus on two main scenarios for the CP-even sector below. In 
the first we allow only for non-vanishing $\haVA$ and in
the second we consider non-zero $\ha_{AZ}$. In both
cases we set all other anomalous couplings to zero.

The contribution of $\wh\alpha_{ZZ}$ to $J_3$ and $J_6$ is
$g_V$ suppressed compared to $\ha_{AZ}$. We therefore anticipate 
a small effect in the
asymmetries from this coupling.  Nevertheless, it 
can compete with $\haA$ and $\ha_{AZ}$ in
$d\Gamma/dq^2$ 
and in the total cross section $\sigma(s)$ of $\eeHZ$.
Finally, the coupling $\ha_{ZZ}^{(1)}$ amounts to a global shift of the SM
$H\to ZZ$ vertex. Its effect on
asymmetries is again small since it is not enhanced with respect to
the SM contribution by numerical factors or $1/g_V$ terms. 
Since
$\ha_{ZZ}$ and $\ha_{ZZ}^{(1)}$ have essentially no impact on the
angular asymmetries we do not consider specific scenarios for them
but we will comment on their contributions to $d\Gamma/dq^2$ and
$\sigma(s)$.

The $d=6$ corrections to the couplings $g_A$ and $g_V$ are taken
into account and are shown explicitly in expressions in this section. 
To make a clear distinction we define
\begin{equation}
g_A \equiv \barg_A\left(1+\delta g_A\right), \qquad
g_V \equiv \barg_V \left(1- \frac{\barg_A}{\barg_V}\delta g_V\right),
\end{equation}
where $\barg_{V,A}$ are the following combinations of input parameters (free
of $d=6$ corrections):
\begin{equation}
\barg_A \equiv \frac{m_Z}{2}(\sqrt{2} G_F)^{1/2}   , \qquad
\barg_V \equiv \barg_A\,(1 -4 s_W^2).
\end{equation}
Throughout this section  the  $d=6$ corrections to the electromagnetic 
vertex  can be neglected because they always appear in 
combination with the $d=6$ $HZ\gamma$ coupling,
which is already $\mathcal{O}(1/\Lambda^2)$.

In the plots in the remainder of this section we show as a
shaded band the region $0\leq q^2 \leq (12\, \, \mbox{GeV})^2$ (or $s
\lesssim 0.0091$) where the decay $(Z^*,\gamma^*)\to \lplm$ is 
dominated by intermediate $q\bar q$ hadronic resonances and our calculation 
is not valid.\footnote{In experimental studies this region
is removed by means of a kinematic cut on the value of 
$q^2$~\cite{CMS_HiggsQN,ATLAS_HVV}.} We refer to Ref.~\cite{GI14} 
for a discussion of the low-$q^2$ part of the spectrum.

\subsubsection{\boldmath Contact $HZ\ell\ell$ interactions}

First, we concentrate on the observability of the contact $HZ\ell \ell$ 
interaction, setting the other anomalous couplings to zero. 
The relevant couplings  $\haV$
and $\haA$ are defined in Eq.~(\ref{eq:alphaVandA}) in terms of the 
$d=6$ Lagrangian couplings. They enter
the form factors $H_{1,V/A}$, which are non-vanishing 
already in the SM, through the combinations $\wh\alpha_{1,2}^{\rm eff}$, 
and implicitly through the $Z\ell\ell$ couplings $g_{V,A}$ according
to~Eq.~(\ref{eq:cA6}). 

\begin{figure}[t]
\begin{center}
\subfigure[$d\Gamma/ds$ (in $10^{-6}$~GeV)]
{\includegraphics[width=.32\columnwidth,angle=0]{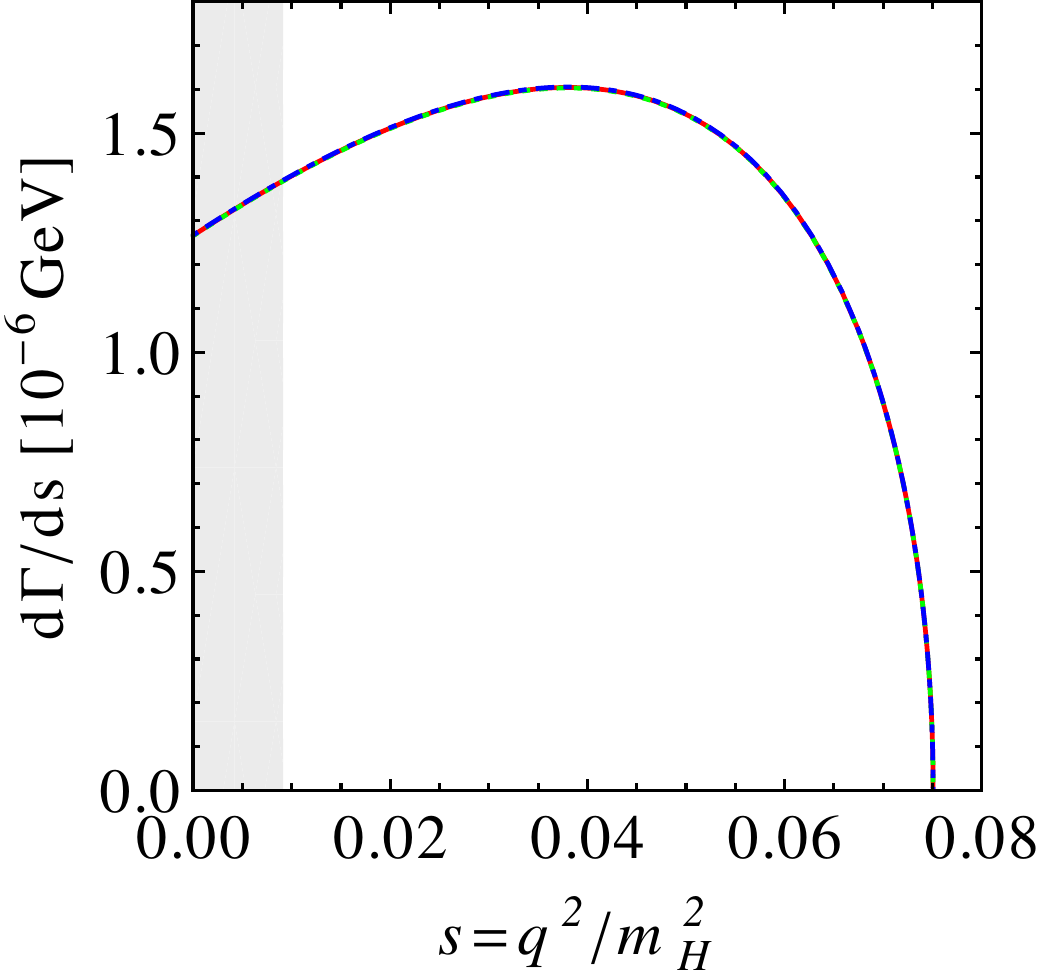}
\label{fig:ContactVectora}}
\subfigure[$-\AsymThree$]
{\includegraphics[width=.31\columnwidth,angle=0]{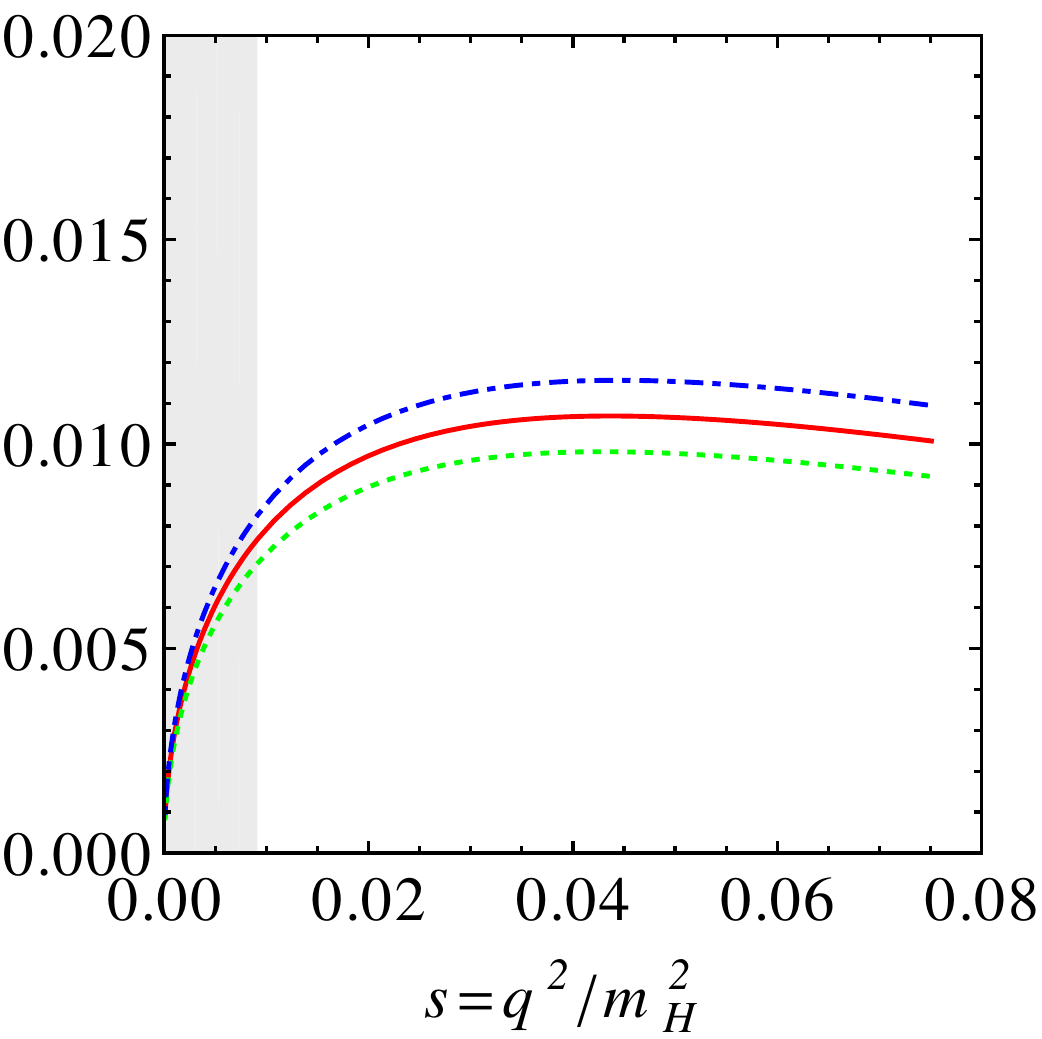}
\label{fig:ContactVectorb}}
\subfigure[$-\AsymCoCt$]
{\includegraphics[width=.31\columnwidth,angle=0]{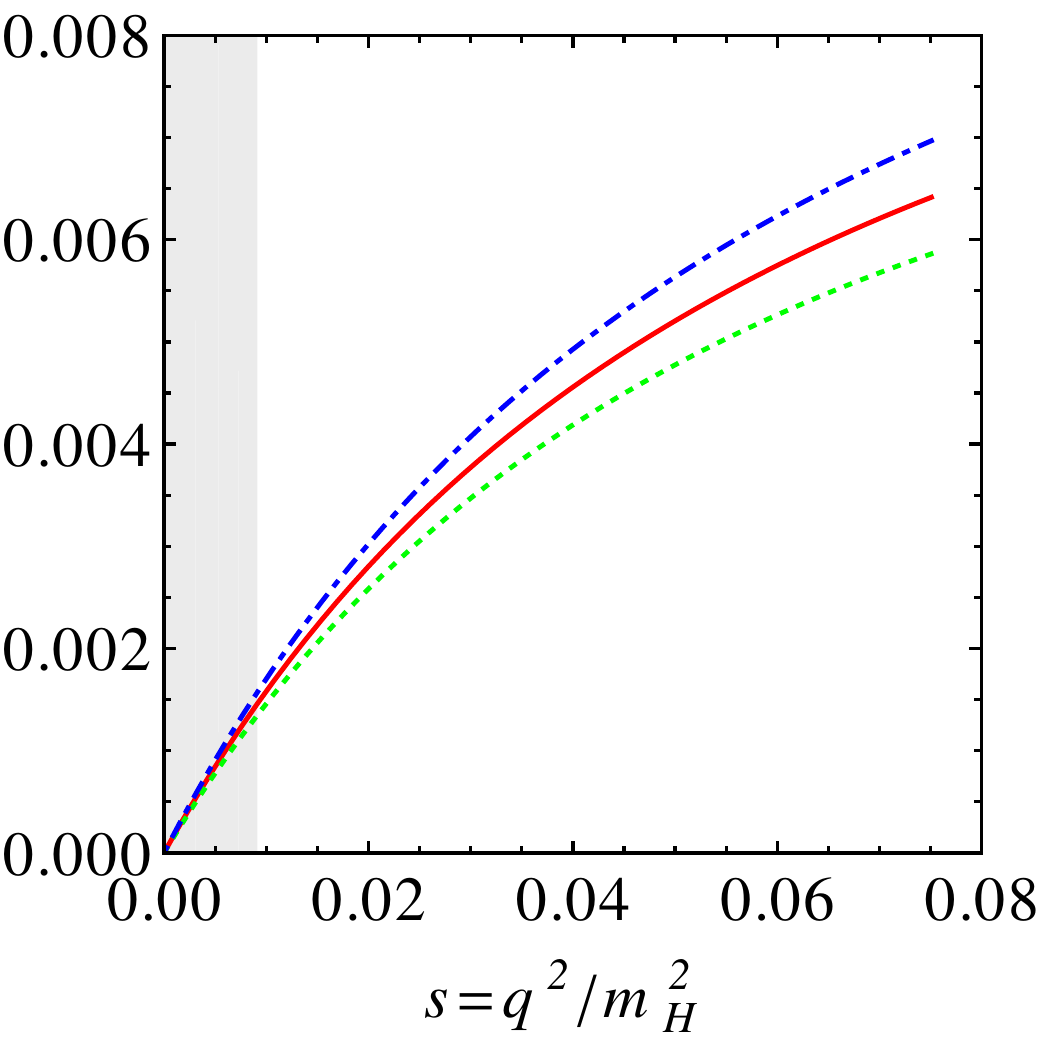}
\label{fig:ContactVectorc}}
\caption{(a) $d\Gamma/ds$, (b)  $-\AsymThree$, (c) $-\AsymCoCt$.
Three scenarios are considered. The red solid-line is the SM case. The  
dotted green line corresponds to $(\wh\alpha^V_{\Phi \ell},
\wh\alpha^A_{\Phi \ell})=(-5,0)\times 10^{-3}$, and the dot-dashed blue 
line to $(\wh\alpha^V_{\Phi \ell},\wh\alpha^A_{\Phi \ell})=(5,0)
\times 10^{-3}$.  The shaded bands exclude $\sqrt{q^2} < 12$ GeV, 
where hadronic resonances dominate. 
}
\label{fig:ContactVector}
\end{center}
\end{figure}

We begin our discussion by focusing on the vector contact interaction, 
that is, we put $\wh\alpha^A_{\Phi\ell}=0$ for the moment,
which in our operator basis amounts to $\wh\alpha_{\Phi e }=
(\wh\alpha_{\Phi\ell }^{(1)}+\wh\alpha_{\Phi\ell }^{(3)})$. 
Due to the $g_V$ suppression, the impact of the vector
contact interaction in $J_1$ and $J_2$ and hence $d\Gamma/ds$  is small. 
This is confirmed in Fig.~\ref{fig:ContactVectora} which, besides 
the SM result, shows two (barely visible) curves that describe the 
modifications for the maximally and minimally allowed values in the 
range of Eq.~(\ref{contactrange}). To understand 
this analytically we exploit here and below the hierarchy 
$g_V \ll g_A$ to write simplified expressions\footnote{In the 
figures, however, we always use the exact expressions, not the 
simplified versions.}  for the 
angular functions that exhibit
the dominant effects in $d\Gamma/ds$ and the asymmetries. 
We also employ the approximation $r\approx 1/2$, which is correct up to
$5\%$, and make use of the fact --- appropriate for $\HZll$ --- that $s\ll 1$.
Within these approximations, the combination $4J_1 + J_2$ that enters 
$d\Gamma/ds$ in Eq.~(\ref{eq:dGammadq2})
can be written as
\begin{align}
&4J_1 +J_2 \simeq \sqrt{2}\,m_H^2\,G_F\,  \barg_A^4\,(1+16s)   
\times \nn \\[0.15cm]
& \left[ 1 +2\ha_{ZZ}^{(1)}-\frac{48  s}{1+16s} \,\ha_{ZZ} + 
4 \left(\delta g_A - \frac{\barg_V}{\barg_A} \delta g_V  \right) +
2(1-2s)\left(\haA -\frac{\barg_V}{\barg_A}\haV\right)\right].
\label{eq:J1J2approx1}
\end{align}
This expression is valid including terms  of order $\mathcal{O}(s)$.  
In the scenario considered here where $\ha_{ZZ}=\ha_{ZZ}^{(1)}=0$,  the
corrections to the SM result are very small. This is due to
the $g_V$ suppression of the $\aV$ terms in Eq.~(\ref{eq:J1J2approx1}) (both
the direct contribution and the indirect one due to $\delta g_V$). 
In fact, the simplified formula shows that in a generic situation  
$\ha_{ZZ}, \ha_{ZZ}^{(1)}$ and the axial contact interaction are expected 
to be more important than the vector contact interaction. However, 
none of the anomalous couplings is enhanced relative to the SM 
contribution.

In contrast, the asymmetries $\AsymThree$, $\AsymCoCt$ proportional 
to $J_3$ and $J_6$ are sensitive to the vector contact
coupling since this 
and only this contribution is enhanced by $1/g_V$. 
The results of Fig.~\ref{fig:ContactVectorb}
and~\ref{fig:ContactVectorc} display the corresponding larger sensitivity 
to $\aV$. However, although larger than in $d\Gamma/ds$, the 
contact interaction is still a small correction 
of ${\cal O}(10)\%$ to the SM result. Larger asymmetries were 
obtained in Ref.~\cite{BCD13}, since they allowed larger values of $\haV$. 
While we formally agree with their results, the estimate 
Eq.~(\ref{contactrange}) excludes these values of $\haV$.

The above observations can be easily understood from 
the simplified expressions for the angular functions $J_3$ and $J_6$, 
which we can write  as 
\bea
J_3 &\simeq&-64\sqrt{2}\,m_H^2\,G_F  \,\barg_A^2\, \barg_V^2  s
\left( 1  -  \haA +\frac{\barg_A}{\barg_V}\haV \right),\nn\\
J_6 &\simeq&- 32\,m_H^2\,G_F   \, \barg_A^2 \barg_V^2 \sqrt{s} 
\left( 1 -  \haA +\frac{\barg_A}{\barg_V}\haV      \right).
\label{eq:J3J6approx}
\eea
In the last expressions we put $\ha_{ZZ}=\ha_{ZZ}^{(1)}=0$ and 
used Eq.~(\ref{eq:cA6}) to fix $\delta g_{V,A} = -\haVA$.
Including the contributions from the denominator in their definition, 
the asymmetries  $\AsymThree$ and $\AsymCoCt$ are approximated by
\bea
- \AsymThree &\simeq& \frac{9\pi \sqrt{2}}{2}  \frac{\barg_V^2}{\barg_A^2} 
\frac{\sqrt{s}}{1+16s} \left(1 +\haA +\frac{\barg_A}{\barg_V} \haV\right), 
\nn \\
  -\AsymCoCt &\simeq&\frac{36\,\barg_V^2}{\barg_A^2}\frac{s}{1+16s} \left( 1  +\haA +\frac{\barg_A}{\barg_V} \haV  \right) .
\label{eq:A3AcoctApprox1}
\eea
These asymmetries are largely dominated by the vector contact 
interaction enhanced by the factor $\barg_A/\barg_V\approx 15$. 
All other effects are subleading (including those arising from 
the  denominator, given in Eq.~(\ref{eq:J1J2approx1})).
Unfortunately, the asymmetries proportional to these functions are 
intrinsically small, because they contain a global $g_V^2$ factor.
Note that in the denominators of the last expressions the term $16s$
is of order one and cannot be expanded.  This term is largely
responsible for the shape of the curves in
Figs.~\ref{fig:ContactVectorb} and~\ref{fig:ContactVectorc}.
Interestingly, in the absence of other anomalous couplings the ratio
between these two asymmetries, 
\begin{equation}
\frac{\AsymCoCt}{\AsymThree} \simeq \frac{4\sqrt{2s}}{\pi},
\end{equation}
is given by pure kinematics, independent of $d=6$ corrections.

\begin{figure}[t]
\begin{center}
\subfigure[$d\Gamma/ds$ (in $10^{-6}$~GeV)]
{\includegraphics[width=.32\columnwidth,angle=0]{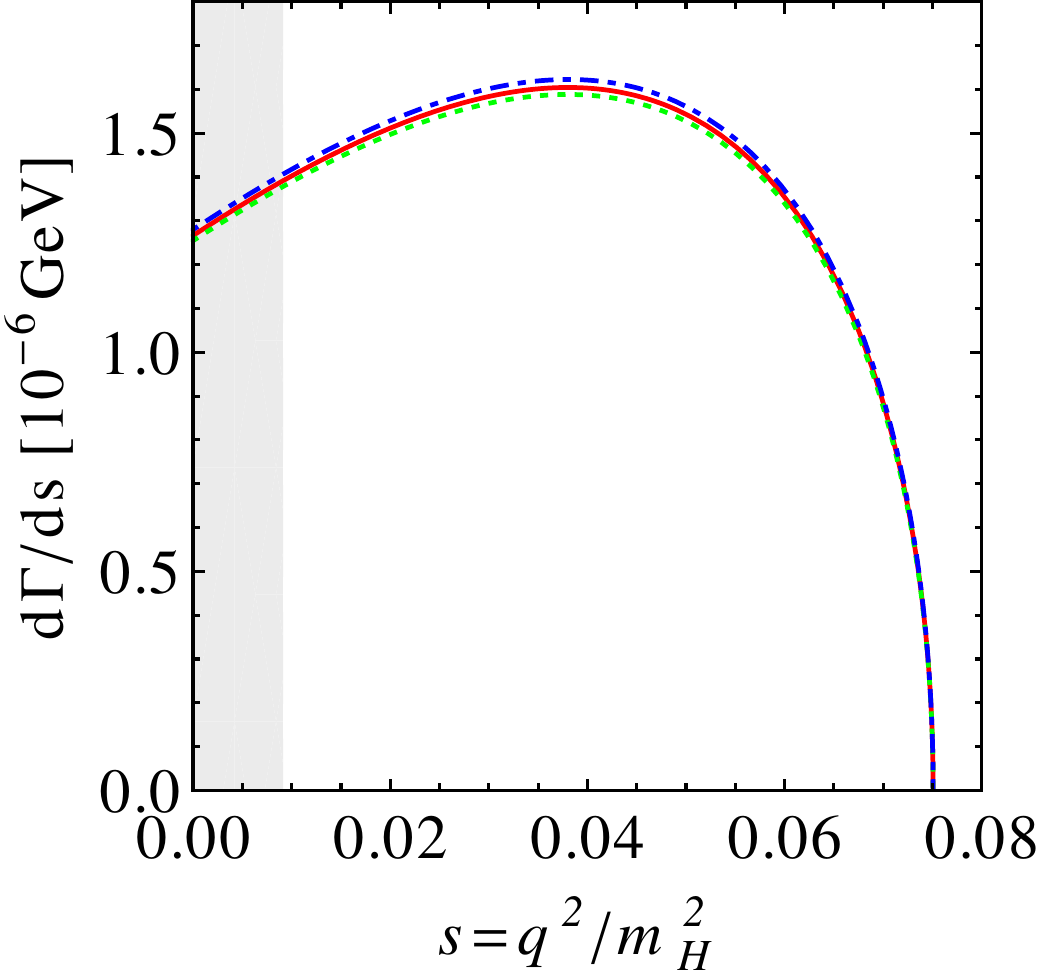}
\label{fig:ContactAxiala}}
\subfigure[$-\AsymThree$]
{\includegraphics[width=.31\columnwidth,angle=0]{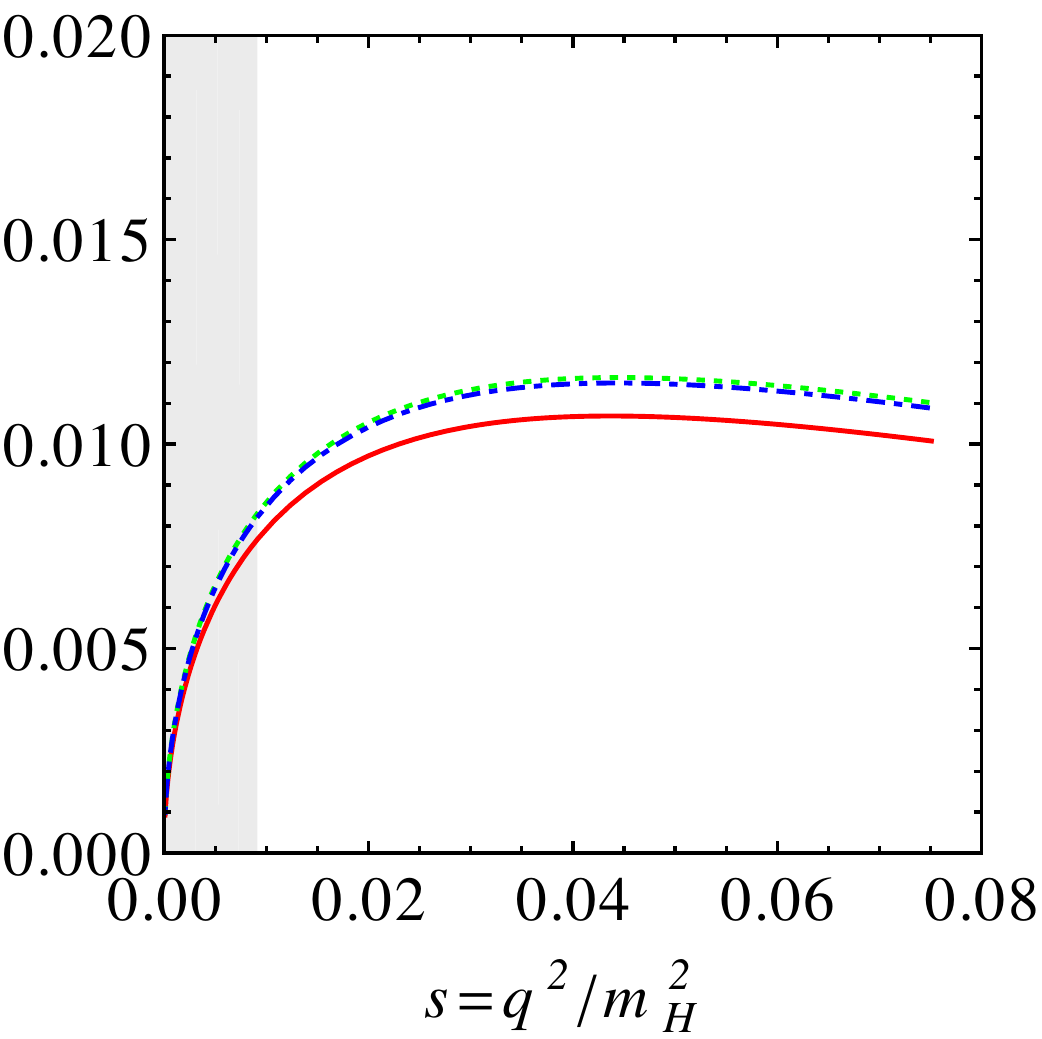}
\label{fig:ContactAxialb}}
\subfigure[$-\AsymCoCt$]
{\includegraphics[width=.32\columnwidth,angle=0]{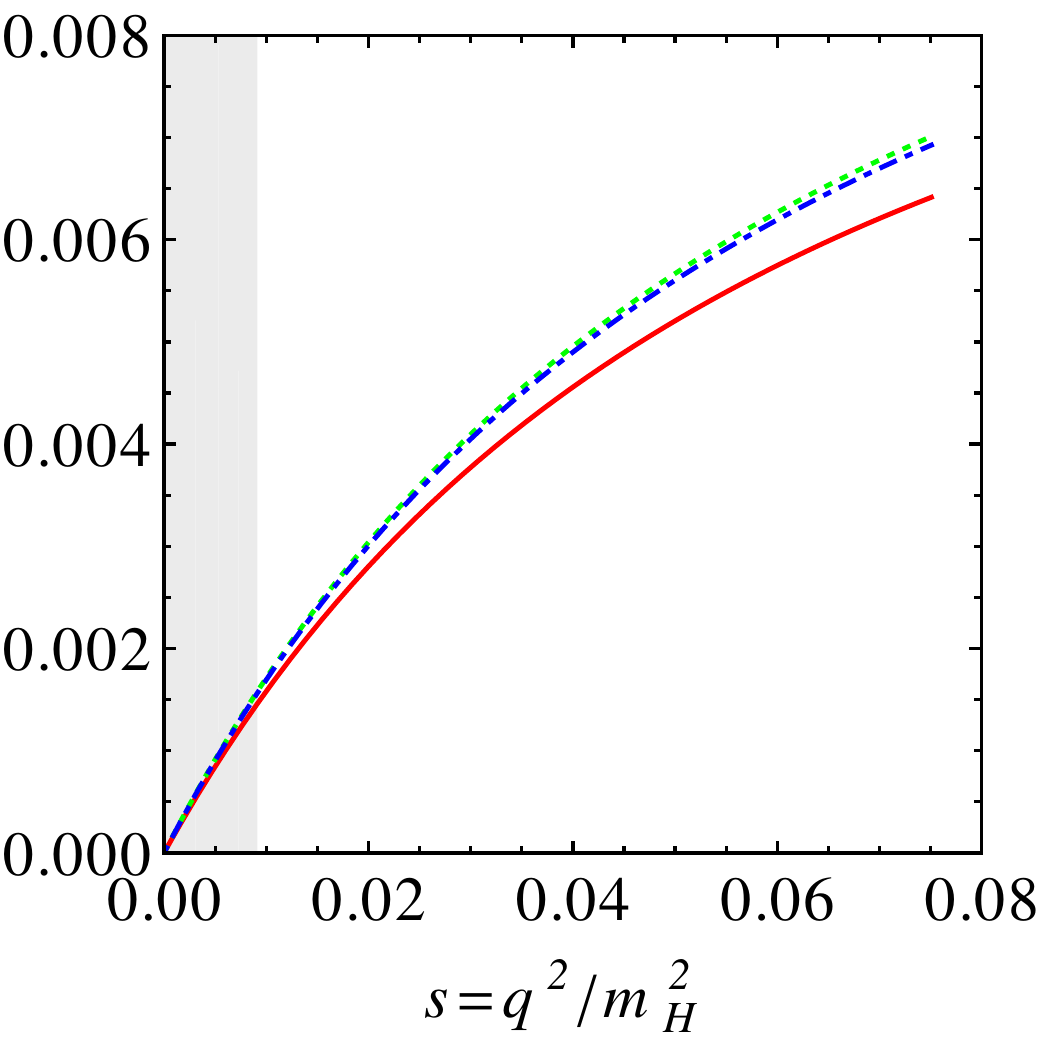}
\label{fig:ContactAxialc}}
\caption{(a) $d\Gamma/ds$, (b)  $-\AsymThree$, (c) $-\AsymCoCt$.
The red solid-line is the SM case. The  dotted green line corresponds 
to $(\wh\alpha^V_{\Phi \ell},\wh\alpha^A_{\Phi \ell})=(5,5)\times 10^{-3}$, 
whereas the dot-dashed blue line to  $(\wh\alpha^V_{\Phi \ell},
\wh\alpha^A_{\Phi \ell})=(5,-5)\times 10^{-3}$. 
}
\label{fig:ContactAxial}
\end{center}
\end{figure}

We now turn to the more general case where both the vector and axial-vector
$HZ\ell  \ell$ couplings contribute.  
Fig.~\ref{fig:ContactAxial} shows $d\Gamma/ds$ and the same
asymmetries as Fig.~\ref{fig:ContactVector} for two values 
of the axial coupling $\wh\alpha_{\Phi \ell}^A$ and fixed 
$\wh\alpha^V_{\Phi \ell} = 0.005$. The essential
features can be easily understood with the help of the approximate 
expressions of Eqs.~(\ref{eq:J1J2approx1}),~(\ref{eq:J3J6approx}), 
and~(\ref{eq:A3AcoctApprox1}). From Eq.~(\ref{eq:J1J2approx1}) we see 
that $\haA$ gives the dominant contribution in the $d=6$ corrections to  
$d\Gamma/ds$, but the fact that $\haA$ cannot exceed a few permille makes 
the modifications to SM result very small (Fig.~\ref{fig:ContactAxiala}).
In the asymmetries,
Figs.~\ref{fig:ContactAxialb} and~\ref{fig:ContactAxialc}, the
deviations from the SM are essentially due to $1/g_V$ enhanced contribution
from $\wh\alpha^V_{\Phi \ell}$. The inclusion of $\haA$ for fixed $\haV$ 
barely alters this result.

\subsubsection{\boldmath Anomalous $HZ\gamma$ coupling}

\begin{figure}[t]
\begin{center}
\subfigure[$d\Gamma/ds$ (in $10^{-6}$~GeV)]
{\includegraphics[width=.32\columnwidth,angle=0]{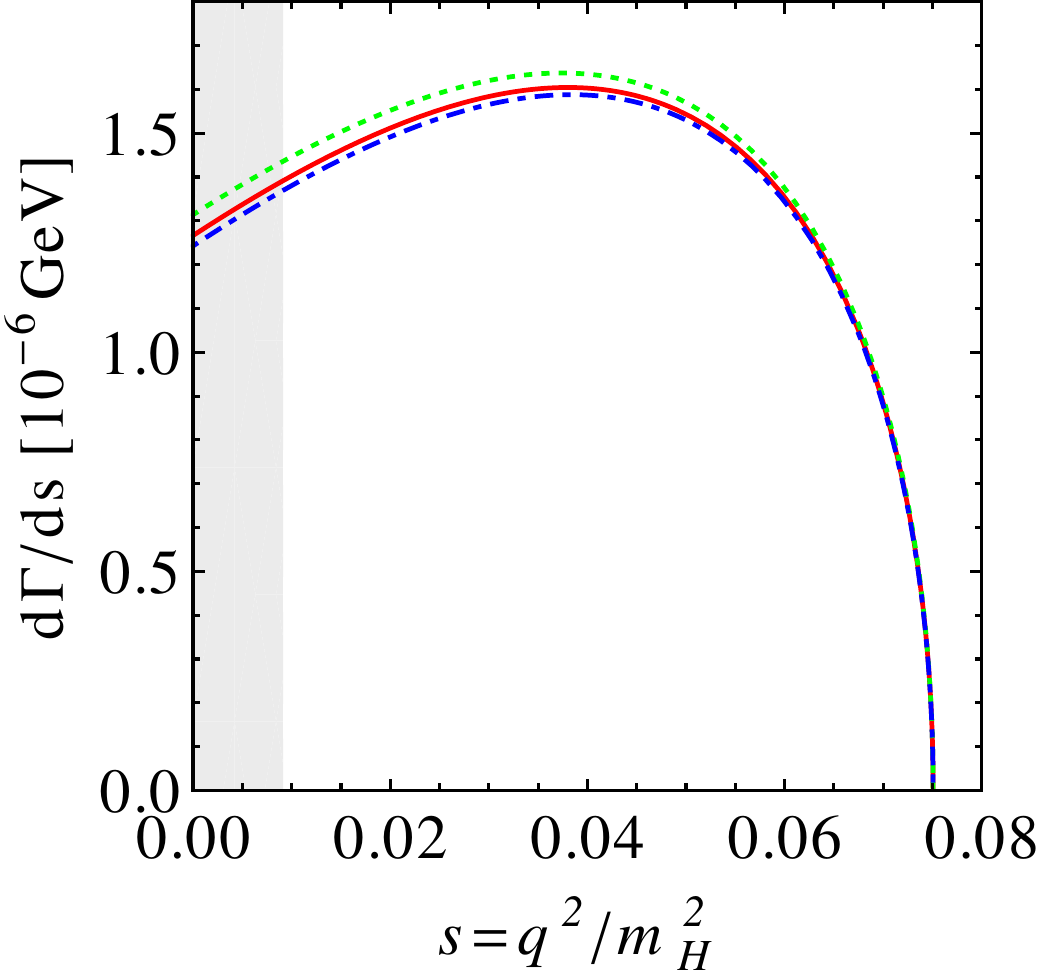}
\label{fig:alphaAZa}}
\subfigure[$-\AsymThree$]
{\includegraphics[width=.31\columnwidth,angle=0]{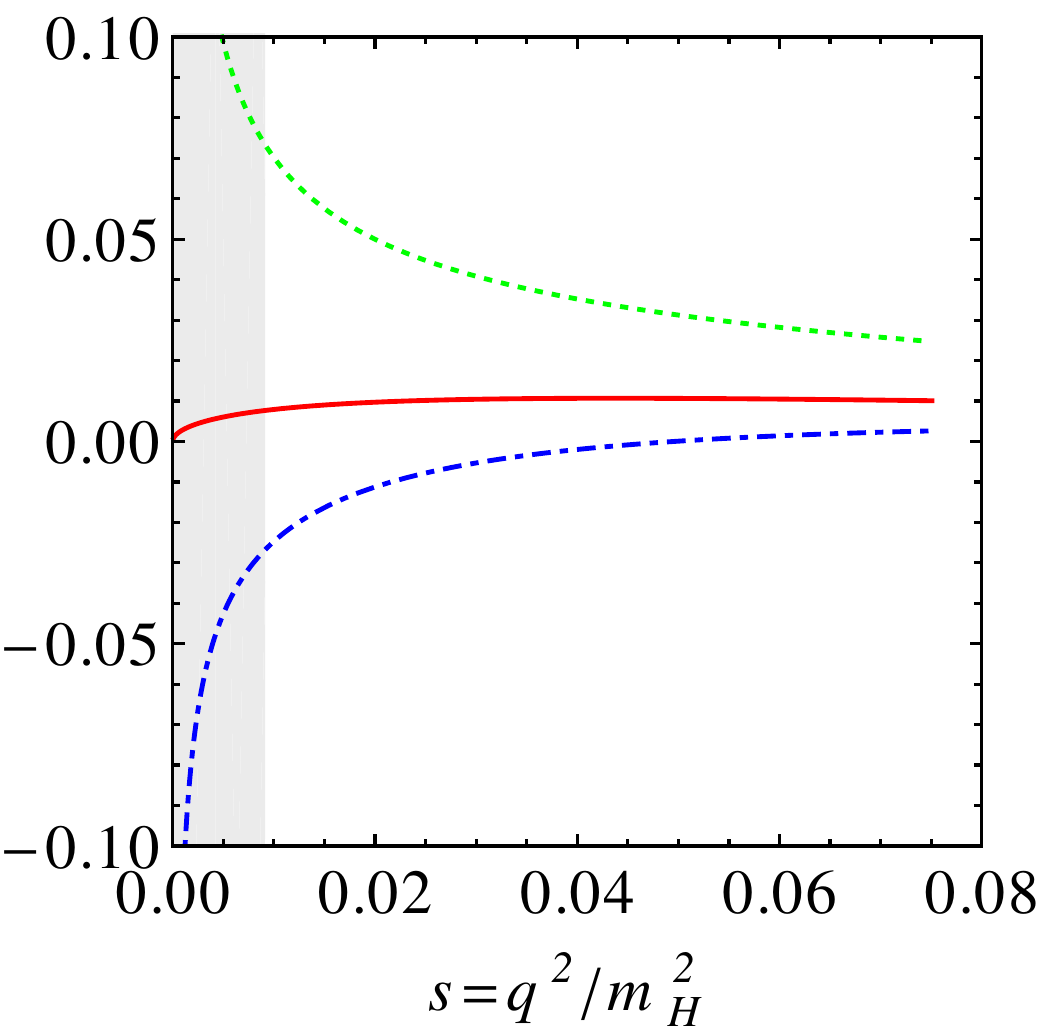}
\label{fig:alphaAZb}}
\subfigure[$-\AsymCoCt$]
{\includegraphics[width=.31\columnwidth,angle=0]{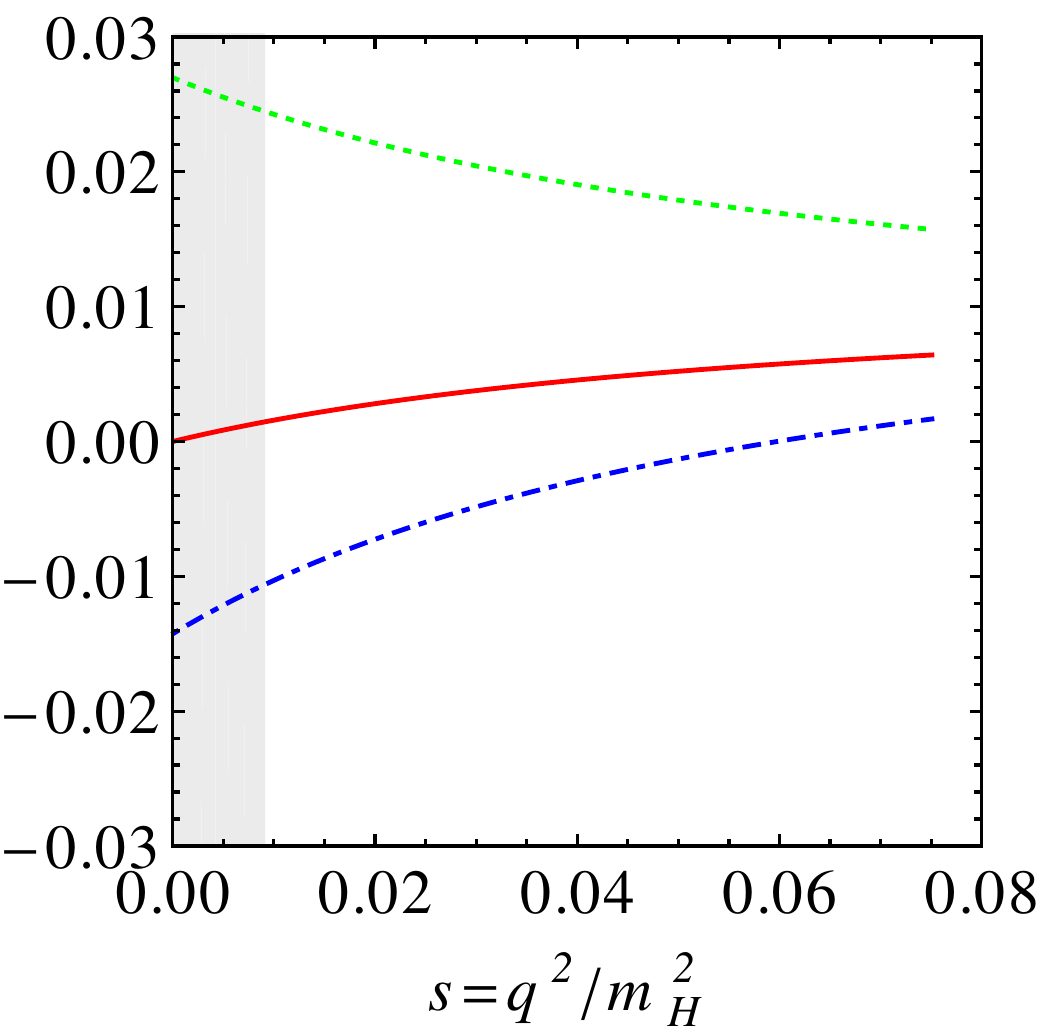}
\label{fig:alphaAZc}}
\caption{(a) $d\Gamma/ds$, (b)  $-\AsymThree$, (c) $-\AsymCoCt$.
Three scenarios are considered. The red solid-line is the SM case. 
The dot-dashed blue line corresponds to $\wh\alpha_{AZ}=-1.3\times 10^{-2}$,
 whereas the  dotted green line corresponds to 
$\wh\alpha_{AZ}=2.6\times 10^{-2}$.
}
\label{fig:alphaAZ}
\end{center}
\end{figure}

We next investigate the effect of the anomalous $HZ\gamma$ coupling,
$\wh\alpha_{AZ}$, and display it for the two values $\wh\alpha_{AZ} = -0.013$ 
and $\wh\alpha_{AZ}=0.026$, 
which limit the allowed range 
Eq.~(\ref{hzarange}) as discussed in Sec.~\ref{sec:Dyn}.

The simplified expression relevant to $d\Gamma/ds$ is 
given by 
\begin{equation}
4J_1+J_2 \simeq \sqrt{2}\,m_H^2\,G_F \, \barg_A^4\, (1+16s)
  \left( 1  - \frac{  12\, \barg_V g_{\rm em}\,Q_\ell}{\barg_A^2(1+16s)} 
\, \ha_{AZ} \right),
\label{eq:J1J2approxAZ}
\end{equation}
from which we immediately deduce that the overall effect on $d\Gamma/ds$ 
is very small due to the $\barg_V$ suppression of the $\ha_{AZ}$ term 
(see Fig.~\ref{fig:alphaAZa}). 
Remarkably however, despite a $g_V^2$ suppression, 
the asymmetry $\AsymThree$,  Fig.~\ref{fig:alphaAZb},  
can reach $5\%$ for values of $\wh\alpha_{AZ}$ close to the upper bound.
The effect is less pronounced in the asymmetry $\AsymCoCt$, but 
can reach the percent level, see Fig.~\ref{fig:alphaAZc}. 
The larger effect in the asymmetries is due to the fact that in 
$J_3$ and $J_6$ the $\wh\alpha_{AZ}$ contribution is $1/g_V$ enhanced.
Moreover, the photon pole in $\AsymThree$ is only partially  cancelled 
by the $\sqrt{s}$ factor. These features become evident from the 
approximate expressions
\bea
- \AsymThree &\simeq& \frac{9\pi \sqrt{2}}{2}  \frac{\barg_V^2}{\barg_A^2} 
\frac{\sqrt{s}}{1+16s} \left(1  -\frac{g_{\rm em}\,Q_\ell}{8\,\barg_V s }
\ha_{AZ}  \right),\nn\\
-\AsymCoCt &\simeq&  \frac{36\,\barg_V^2}{\barg_A^2}\frac{s}{1+16s} 
\left( 1  -\frac{g_{\rm em} \,Q_\ell}{4\,\barg_V s}\ha_{AZ}  \right).
\label{eq:A3AcoctApprox2}
\eea
 The double enhancement by the factor 
$1/(g_V s) \sim {\cal O}(10^3)$ implies that these asymmetries can 
exceed their SM expectation, even when the anomalous couplings are 
generated by BSM physics in the multi-TeV range.
Note that in the presence of the anomalous $\ha_{AZ}$ coupling the 
ratio of $\AsymThree$ and $\AsymCoCt$ 
is no longer free of $d=6$ corrections.

\begin{figure}[t]
\begin{center}
\includegraphics[width=0.45\columnwidth]{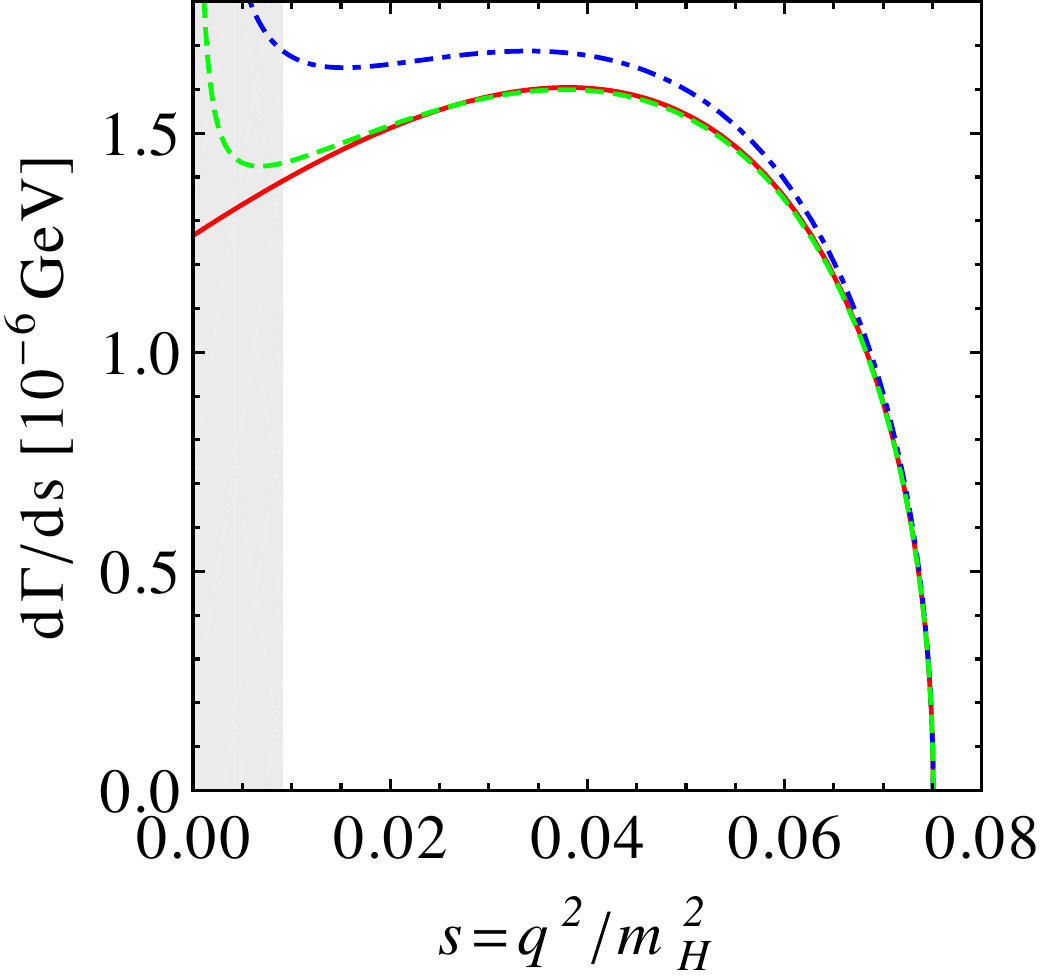}
\vspace*{0.2cm}
\caption{ $d\Gamma/ds$ including  terms of order  $\mathcal{O}(1/\Lambda^4)$
in the squared amplitude. The three scenarios of Fig. 4 are considered: the red solid-line is the SM case, the dot-dashed blue line corresponds to $\wh\alpha_{AZ}=-1.3\times 10^{-2}$,
 whereas the  dotted green line corresponds to 
$\wh\alpha_{AZ}=2.6\times 10^{-2}$.}
\label{fig:dGammads1overLambda4}
\end{center}
\end{figure}

In the  case of the  anomalous $\alpha_{AZ}$ coupling, the square of the 
form factors $H_{i,V/A}$ contains terms proportional to
$\alpha_{AZ}^2/\Lambda^4$, which are enhanced by the photon pole for 
small $s$. Being formally of higher order in the $1/\Lambda^2$ expansion, 
they have been consistently neglected up to this point. However, the 
$1/\Lambda^4$ photon-pole enhanced terms may give the dominant 
contribution from anomalous couplings, also when the effective Lagrangian 
is extended to $d=8$ operators,  
because they are enhanced by $1/(s\,g_V)$ with respect to the terms of 
order $\mathcal{O}(1/\Lambda^2)$. In the specific case of $\alpha_{AZ}$, 
it is therefore mandatory to investigate the photon-pole enhanced 
$1/\Lambda^4$ terms in the expressions for the $J$ functions. 
The effect on the di-lepton mass distribution $d\Gamma/ds$ is indeed 
sizeable as can be seen by comparing Fig.~\ref{fig:alphaAZa} to 
Fig.~\ref{fig:dGammads1overLambda4}, which includes the enhanced 
$\mathcal{O}(1/\Lambda^4)$ terms. In the low-$s$ region, $d\Gamma/ds$ 
can now be enhanced by up to $20\%$ above the SM for the larger value of 
$\alpha_{AZ}$. The effects on asymmetries are less relevant and affect 
chiefly the part inside the shaded region 
$q^2 \leq  (12\, \, \mbox{GeV})^2$. Therefore, we refrain from 
displaying the asymmetries in the presence of $\mathcal{O}(1/\Lambda^4)$
 terms.

\subsubsection{\boldmath  CP-odd couplings}

\begin{figure}[t]
\begin{center}
\subfigure[$\mathcal{A}_{\phi}^{(1)}$]
{\includegraphics[width=.315\columnwidth,angle=0]{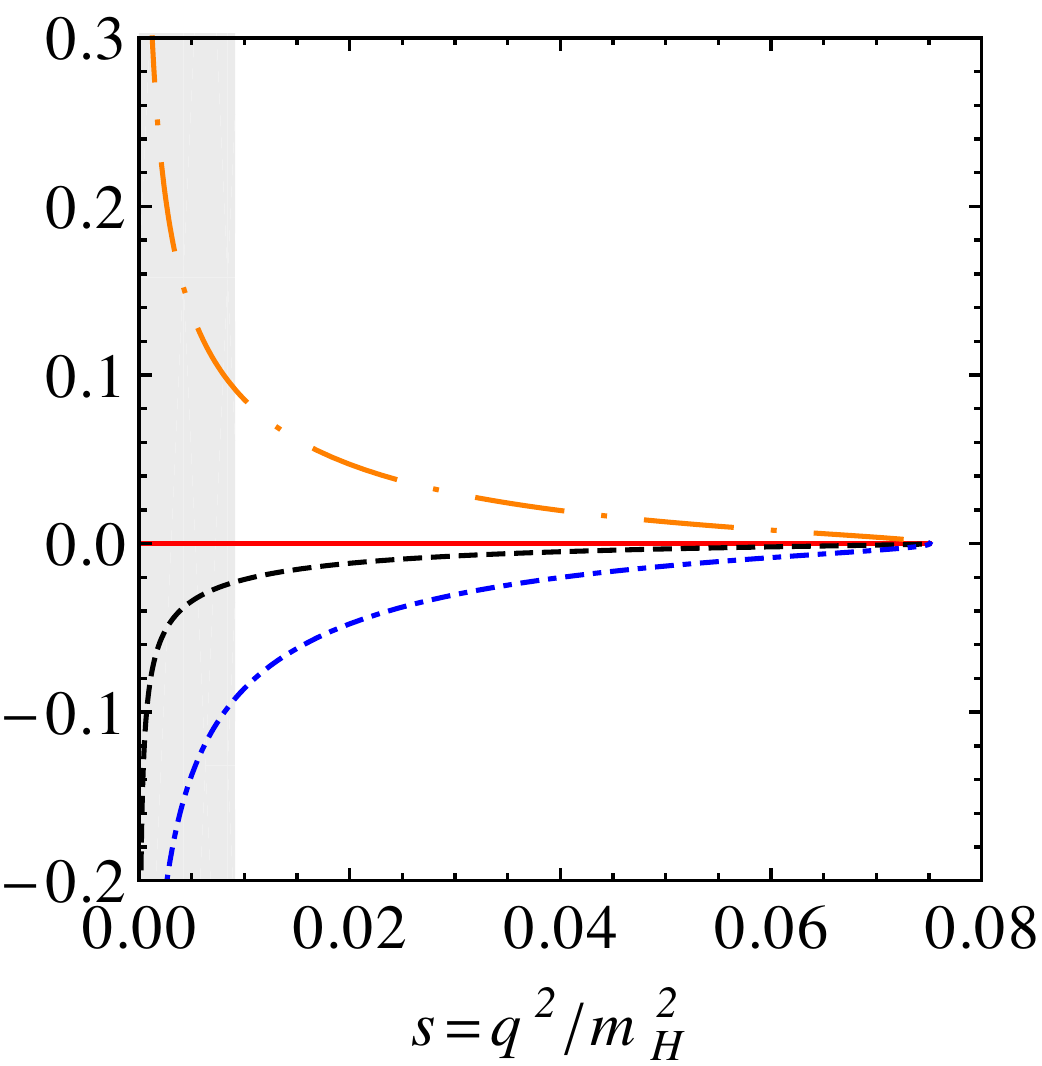}
\label{fig:CPoddA1}}
\subfigure[$\mathcal{A}_{\phi}^{(2)}$]
{\includegraphics[width=.33\columnwidth,angle=0]{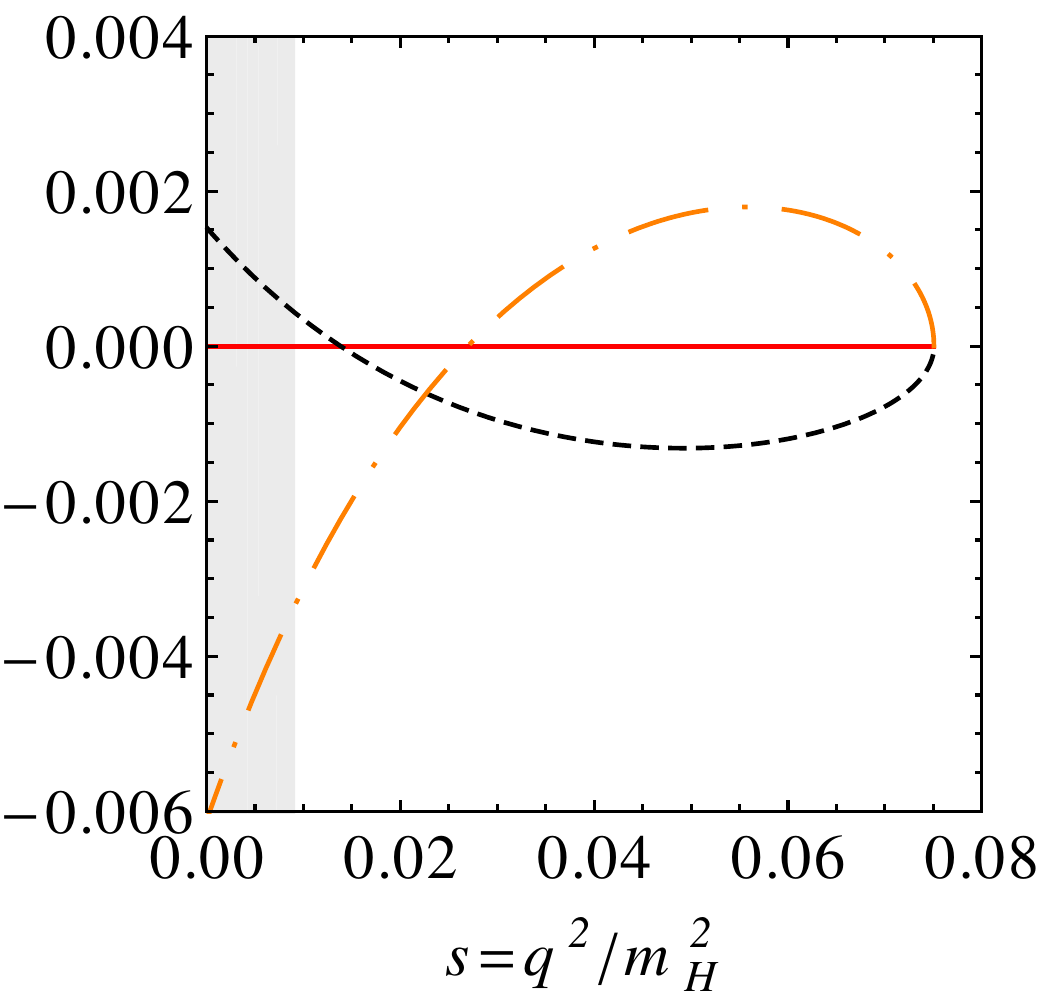}
\label{fig:CPoddb}}
\subfigure[$\mathcal{A}_{\phi}^{(2)}$]
{\includegraphics[width=.31\columnwidth,angle=0]{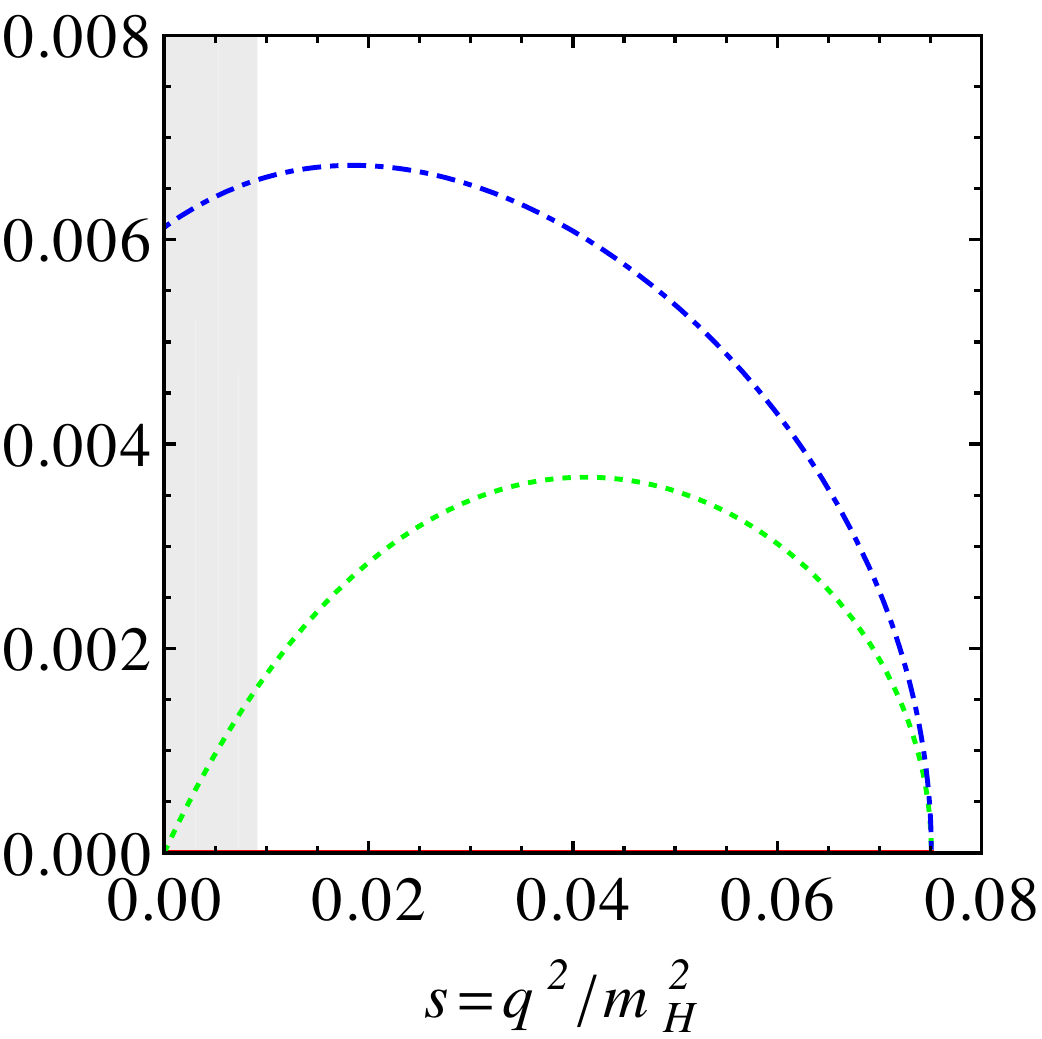}
\label{fig:CPodda}}
\caption{ Asymmetries  $\mathcal{A}_{\phi}^{(1,2)}$ in four different 
scenarios.   The dot-long-dashed orange line corresponds  to  
$(\wh\alpha_{Z\widetilde Z},\wh \alpha_{A\widetilde Z})=(4,4)\times 10^{-2}$,  
the dashed black line to  $(\wh\alpha_{Z\widetilde Z}, 
\wh\alpha_{A\widetilde Z})
=(-2,-1)\times 10^{-2}$, the dot-dashed blue line corresponds to 
$(\wh\alpha_{Z\widetilde Z}, \wh\alpha_{A\widetilde Z})=(4,-4)\times 10^{-2}$, 
and the dotted green line corresponds to $(\wh\alpha_{Z\widetilde Z}, 
\wh\alpha_{A\widetilde Z})=(4,0)\times 10^{-2}$. The solid red line is 
the vanishing SM result.
}
\label{fig:CPoddDecay}
\end{center}
\end{figure}

There are two asymmetries that are sensitive to the CP-odd couplings
$\wh\alpha_{A \widetilde Z}$ and $\wh\alpha_{Z\widetilde Z}$,
$\mathcal{A}_{\phi}^{(1)}$ and $\mathcal{A}_{\phi}^{(2)}$, as defined in 
Eq.~(\ref{eq:Asm8}).  The asymmetry
$\mathcal{A}_\phi^{(1)}$, proportional to $J_4$, is enhanced by a
prefactor $1/{\sqrt{s}}$ at small $s$. This asymmetry is
largely dominated by the coupling $\wh\alpha_{A\widetilde Z}$ due
to  a suppression by $g_V s$ of the $\wh\alpha_{Z\widetilde Z}$ term.
On the other hand, the asymmetry $\mathcal{A}_\phi^{(2)}$, 
proportional to $J_8$, 
receives contributions from both  $\wh\alpha_{A\widetilde Z}$ and 
$\wh\alpha_{Z\widetilde Z}$. Although the $\wh\alpha_{A\widetilde Z}$ term is 
multiplied by $g_V$, the small $s$ factor in front of 
$\wh\alpha_{Z\widetilde Z}$ 
renders both contribution to be of the same order. These features
can be seen from the approximate expressions 
\bea
\mathcal{A}^{(1)}_\phi &\simeq& - \frac{9\pi\sqrt{2}}{16} 
\frac{\sqrt{1-12s}}{\sqrt{s}(1+16s)} \frac{\barg_V\,g_{\rm em}
\,Q_\ell}{\barg_A^2} \ha_{A\widetilde Z},\nn  \\
\mathcal{A}^{(2)}_\phi &\simeq& \frac{16\,\sqrt{1-12s}}{\pi(1+16s)} 
\left( s\, \alpha_{Z\widetilde Z}  + 
\frac{\barg_V\,g_{\rm em}\,Q_\ell}{4\barg_A^2}
\wh\alpha_{A\widetilde Z}\right).
\label{eq:A2Approx}
\eea
The interplay between the two terms can generate an asymmetry-zero in 
$\mathcal{A}^{(2)}_\phi$, provided both CP-odd couplings have the same 
sign (recall $Q_\ell=-1$). Its approximate location is at 
\beq
s_0 =-\frac{\barg_V\,g_{\rm em}\,Q_\ell}{4\barg_A^2}
\frac{\wh\alpha_{A\widetilde Z}}{\ha_{Z\widetilde Z}}, 
\label{eq:ApproxZero}
\eeq
the measurement of which would establish a relation between 
the two CP-odd effective anomalous couplings. We illustrate these 
results in Fig.~\ref{fig:CPoddDecay}. 
For want of stringent experimental bounds on these
couplings we assume that they will not exceed a few times $10^{-2}$, 
as is the case for the other couplings previously studied. 
In Fig~\ref{fig:CPoddA1},
$\mathcal{A}_\phi^{(1)}$ and three different scenarios for
$\wh\alpha_{Z\widetilde Z}$ and $\wh\alpha_{A\widetilde Z}$ are displayed.  
For lower values of
$s$ the asymmetry can be of the order of $10$\%, but it goes to
zero rather quickly at the kinematic end point.  This
asymmetry is essentially independent of $\wh\alpha_{Z\widetilde Z}$.  In
Fig.~\ref{fig:CPoddb} and~\ref{fig:CPodda} we show 
$\mathcal{A}_\phi^{(2)}$ for four different choices of
$\wh\alpha_{Z\widetilde Z}$ and $\wh\alpha_{A\widetilde Z}$.  Fig.~\ref{fig:CPoddb}
shows two cases where the couplings have same signs and the asymmetry
has a zero.  The position of the zero can be estimated from the
approximate expression Eq.~(\ref{eq:ApproxZero}). For
$\wh\alpha_{A\widetilde Z} =\wh \alpha_{Z\widetilde Z}$ the zero predicted 
at $s_0\simeq 0.028$, in good agreement with Fig.~\ref{fig:CPoddb}
(dot-long-dashed orange curve).  The significance of the asymmetry-zero is 
somewhat limited in practice, since the asymmetry itself is
only at the permille level. The asymmetry, however, could reach a few percent
for values of the anomalous couplings one order of magnitude larger, which are 
not ruled out experimentally~\cite{CMS_HiggsQN}. 
In the next section we show that in
$\eeHll$ the CP-odd couplings can generate the asymmetry
$\mathcal{A}_\phi^{(2)}$ at the percent level for anomalous couplings of 
a few times $10^{-2}$.

Similar to the CP-even $HZ\gamma$ coupling, the anomalous 
coupling $\alpha_{A\widetilde
    Z}$ also generates $\mathcal{O}(1/\Lambda^4)$ terms that 
are photon-pole enhanced. The effect on  $d\Gamma/ds$ is
similar to the CP-even case shown in
  Fig.~\ref{fig:dGammads1overLambda4} 
and we do not show the CP-odd case here explicitly.
The asymmetries are again less affected by the $1/\Lambda^4$, and 
change mostly in  the shaded region $q^2 \leq  (12\, \, \mbox{GeV})^2$.


\section{\boldmath Angular asymmetries of  
$e^+e^- \to H Z (\to \lplm)$}
\label{sec:eeHZ}

From the result for the decay $\HZll$ it is straightforward to
calculate the cross section for the crossing-symmetric process
$\eeHll$. In order to fully exploit crossing symmetry we define the
kinematics for $\eeHll$ as discussed in App.~\ref{app:KineeeHZ}.
In particular, the angles $\theta_1$, $\theta_2$ and $\phi$ are now
defined as in Fig.~\ref{fig:KineeeHZ}. According to these definitions,
the cross section can be written in terms of the same function
$\mathcal{J}(q^2,\theta_1,\theta_2,\phi)$.  The process is described by
the same set of form factors $H_{i,V/A}$ and angular functions 
$J_i$, see Eq.~(\ref{eq:FullJ}), analytically continued
in the energy $s$ to describe the different kinematic regime.
The main difference between the two processes is that the di-lepton
invariant mass $q^2=s\,m_H^2$ is now given by the CM energy of the 
initial-state $e^+e^-$
pair. The differential cross section for $\eeHll$ is therefore expressed 
as before as 
\beq
\frac{d\sigma}{ d\cos\theta_1 \,d\cos\theta_2  d\phi} = \frac{1}{m_H^2} \, 
\mathcal{N_\sigma}(q^2) \, \mathcal{J}(q^2,\theta_1,\theta_2,\phi) ,
\eeq
where the new normalisation reads
\beq
\mathcal{N_\sigma}(q^2) = \frac{1}{2^{10}(2\pi)^3}\frac{1}{\sqrt{r}\, 
\gamma_Z}\frac{\sqrt{\lambda(1,s,r)}}{s^2}.
\eeq
Note that we still use the
Higgs mass to construct the dimensionless variables $s$, $r$, and
$\gamma_Z$, as in Eq.~(\ref{eq:AdVar}). 

The threshold energy for the reaction is given by $\sqrt{q_{\rm th}^2}
= (m_H+m_Z)\approx 217$~GeV which gives, in units  of $m_H^2$, the 
minimal $s$ value
\beq
s_{\rm th}= q_{\rm th}^2/m_H^2 \approx 2.98.
\eeq
The form factors are therefore probed at much higher energies, which 
leads to non-trivial phenomenological consequences in comparison 
with $\HZll$. We limit our numerical analysis to intermediate energies 
accessible to a first-stage high-energy $e^+ e^-$ collider, and study 
the range
\beq
s_{\rm th}\leq s \leq 7.0,
\eeq
which translates into $q_{\rm th}^2  
\leq q ^2\lesssim (332\, \mbox{GeV})^2$. Depending on the 
value of the BSM scale $\Lambda$, the effective Lagrangian description 
ceases to be valid for very high values of $s$. In the theoretical 
expressions for the production process, this is seen from the 
fact that the $d=6$ corrections relative to the SM generally 
contain terms of order $s\alpha_k$. The above chosen range for 
$s$ guarantees that the EFT description is  valid when $\Lambda$ 
is above $1\,$TeV.

The total $\eeHZ$ cross section is given by
\beq
\sigma(s) = \frac{32\pi}{9}\frac{1}{m_H^2} \mathcal{N}_\sigma (4J_1 +J_2).
\eeq
We define angular asymmetries analogous to those of \Eqs{eq:Asm8}{eq:Asm9},
normalizing them by the total cross section. Since the normalization
$\mathcal{N}_\sigma$ drops out in the ratios, the final expression for
the asymmetries in terms of $J$ functions are identical to those of $\HZll$.

The SM cross section can be used to estimate the number of produced events. 
At $\sqrt{q^2} = 250$~GeV and with an integrated luminosity of 
$250$~fb$^{-1}$ one expects around $2300$ events, of which up to $1900$ 
could be reconstructed~\cite{Andersonetal14} (assuming $H\to b\bar b$).  
This number decreases to around
$1400$ for $\sqrt{q^2} = 350$~GeV and integrated luminosity of
$350$~fb$^{-1}$ due to a decrease in the cross section 
(see Fig.~\ref{fig:eeHZContactVectora}).

In the remainder of this section we study the total cross section and 
asymmetries to assess their sensitivity to $d=6$ effective Higgs couplings 
in analogy with the decay $\HZll$. For the purpose of comparison we 
consider the same scenarios as in the previous section. Note that 
in $e^+ e^-$ collisions, due to the clean environment, one could 
also consider $Z$ decay to quarks. The vector and axial-vector couplings 
should then be replaced by the appropriate values. A detailed 
anomalous coupling analysis of this possibility is, however, beyond 
the scope of this paper.

\subsection{\boldmath Contact $HZ\ell \ell$ interactions}

\begin{figure}[t]
\begin{center}
\subfigure[$\sigma(s)$]
{\includegraphics[width=.31\columnwidth,angle=0]{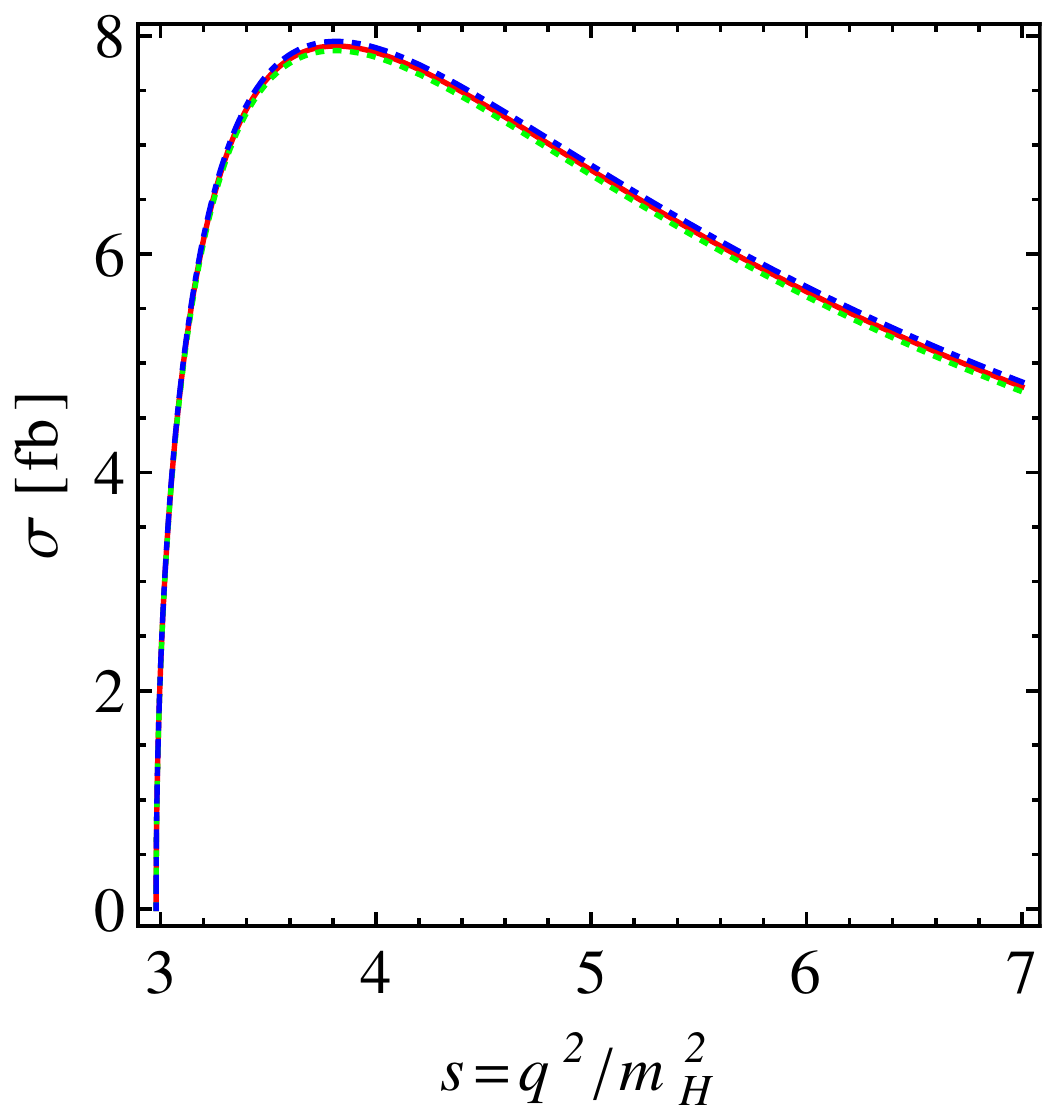}
\label{fig:eeHZContactVectora}}
\subfigure[$-\AsymThree$]{\includegraphics[width=.32\columnwidth,angle=0]{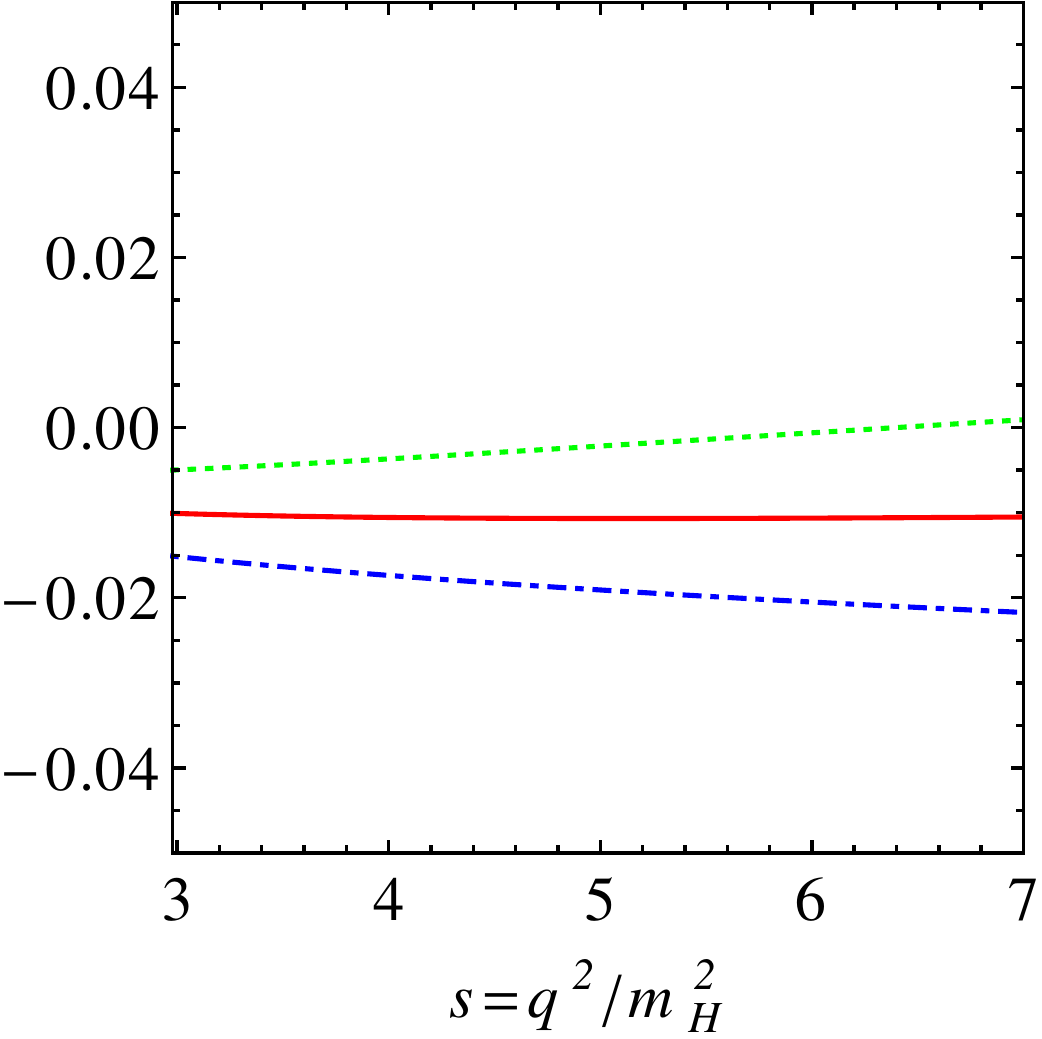}
\label{fig:eeHZContactVectorb}}
\subfigure[$-\AsymCoCt$]
{\includegraphics[width=.32\columnwidth,angle=0]{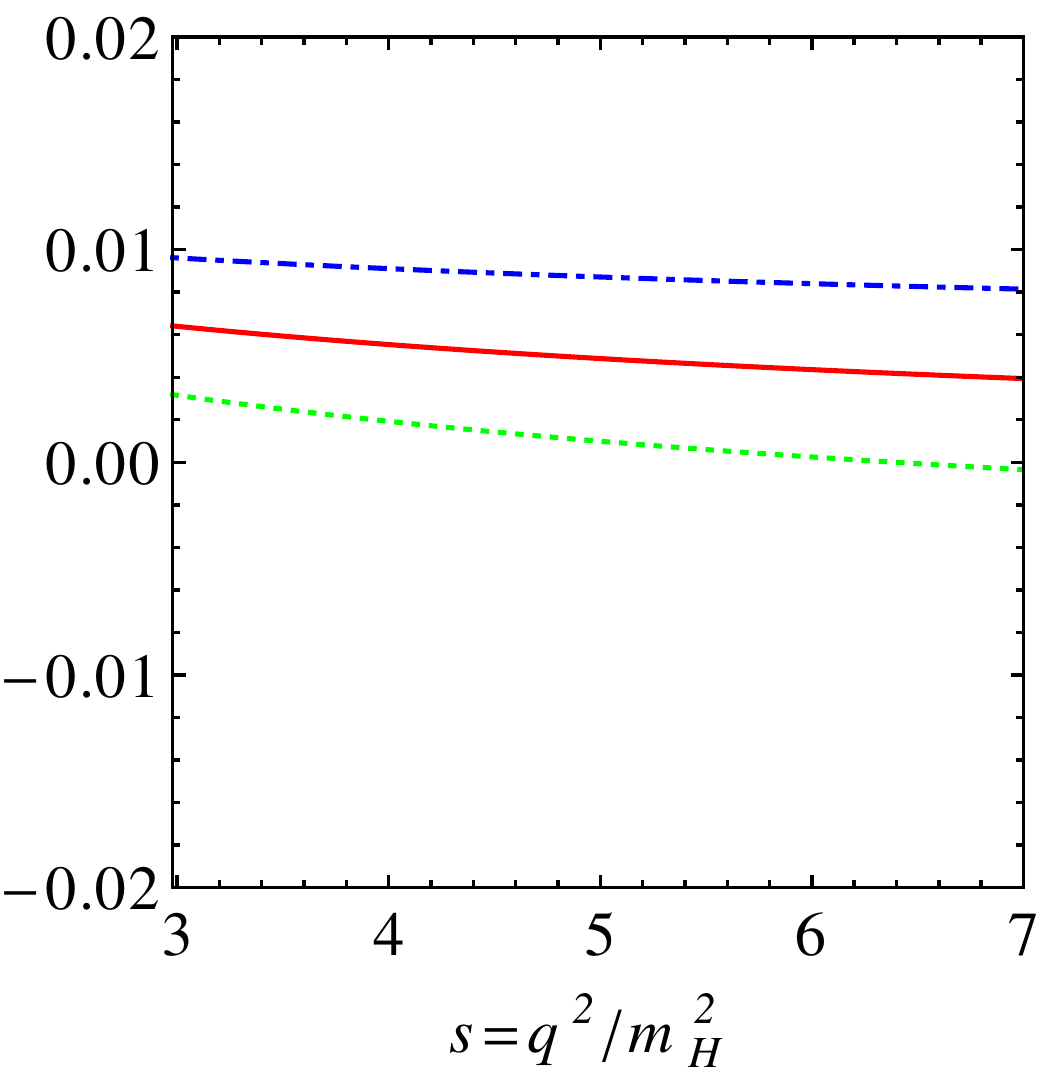}
\label{fig:eeHZContactVectorc}}
\caption{(a) $\sigma(s)$, (b)  $-\AsymThree$, (c) $-\AsymCoCt$.
Three scenarios are considered. The red solid-line is the SM case. The  
dotted green line corresponds to $(\wh\alpha^V_{\Phi \ell},
\wh\alpha^A_{\Phi \ell})=(-5,0)\times 10^{-3}$, and the dot-dashed blue 
line to $(\wh\alpha^V_{\Phi \ell},\wh\alpha^A_{\Phi \ell})=
(5,0)\times 10^{-3}$. 
}
\label{fig:eeHZContactVector}
\end{center}
\end{figure}

We again begin with the case where the
axial-vector $HZ\ell  \ell$ interactions are set to zero.  In
Fig.~\ref{fig:eeHZContactVector} we show results for the same
observables and coupling parameter choices that we investigated for 
$\HZll$.  In the total cross section
$\aV$ effects remain $g_V$ suppressed and therefore insignificant, 
as shown in Fig.~\ref{fig:eeHZContactVectora}. In the asymmetries, 
the modification of the SM value due to $\wh\alpha^V_{\Phi \ell}$ is 
more pronounced in $\eeHll$ than in the decay $\HZll$ 
due to higher values of~$s$, but the effect is still not dramatic, 
as shown in Figs.~\ref{fig:eeHZContactVectorb}. The asymmetries can be 
at most  at the level of 1~to~2\%.

\begin{figure}[t]
\begin{center}
\subfigure[$\sigma$]
{\includegraphics[width=.31\columnwidth,angle=0]{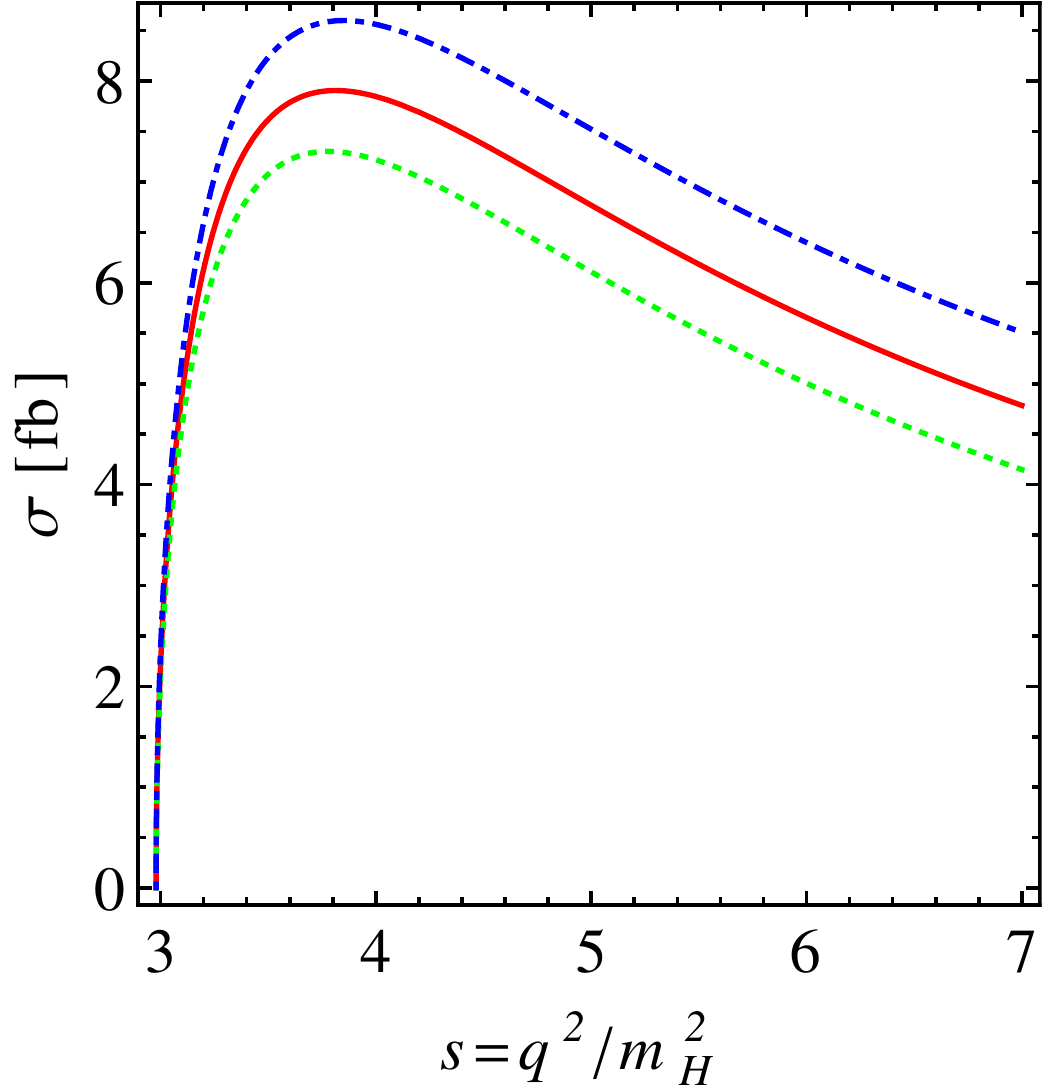}
\label{fig:eeContactAxiala}}
\subfigure[$-\AsymThree$]
{\includegraphics[width=.32\columnwidth,angle=0]{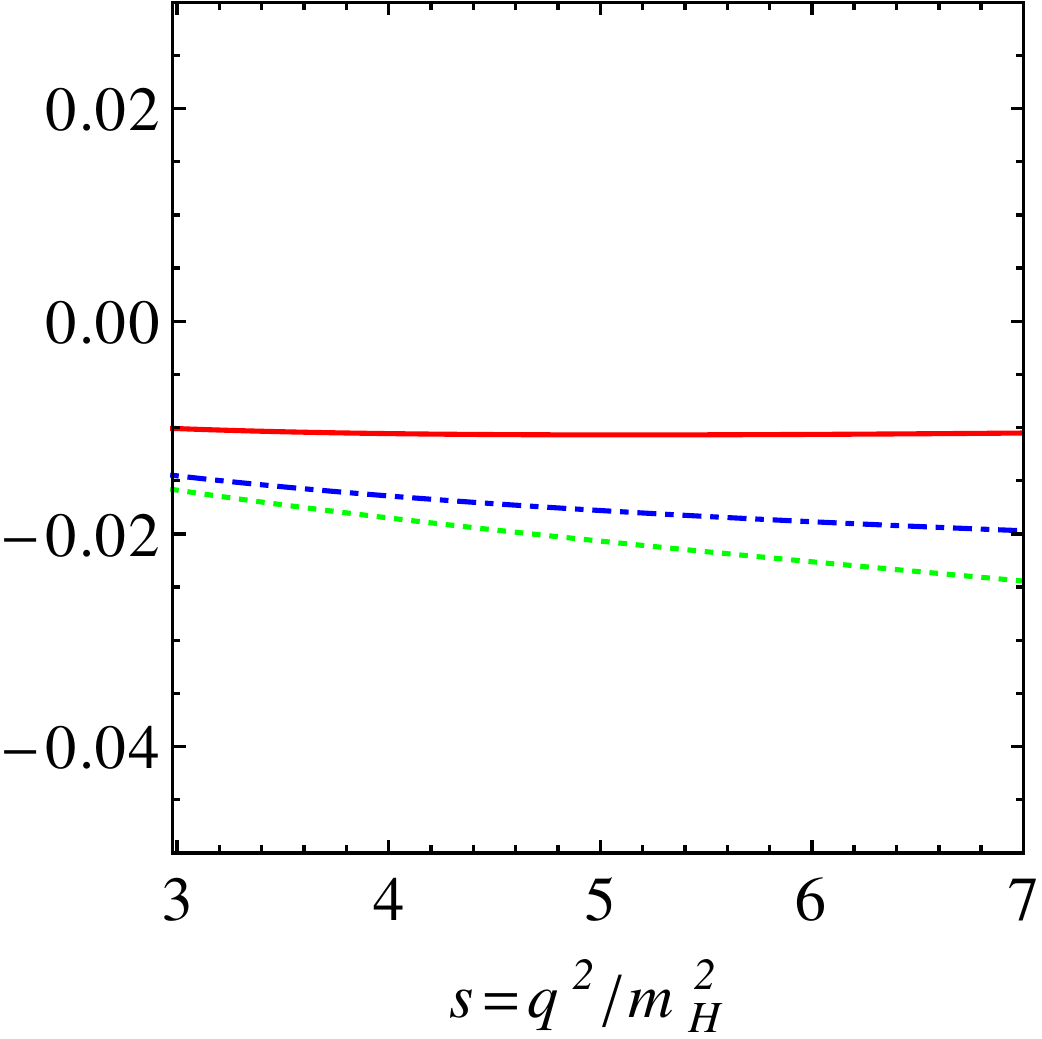}
\label{fig:eeContactAxialb}}
\subfigure[$-\AsymCoCt$]
{\includegraphics[width=.32\columnwidth,angle=0]{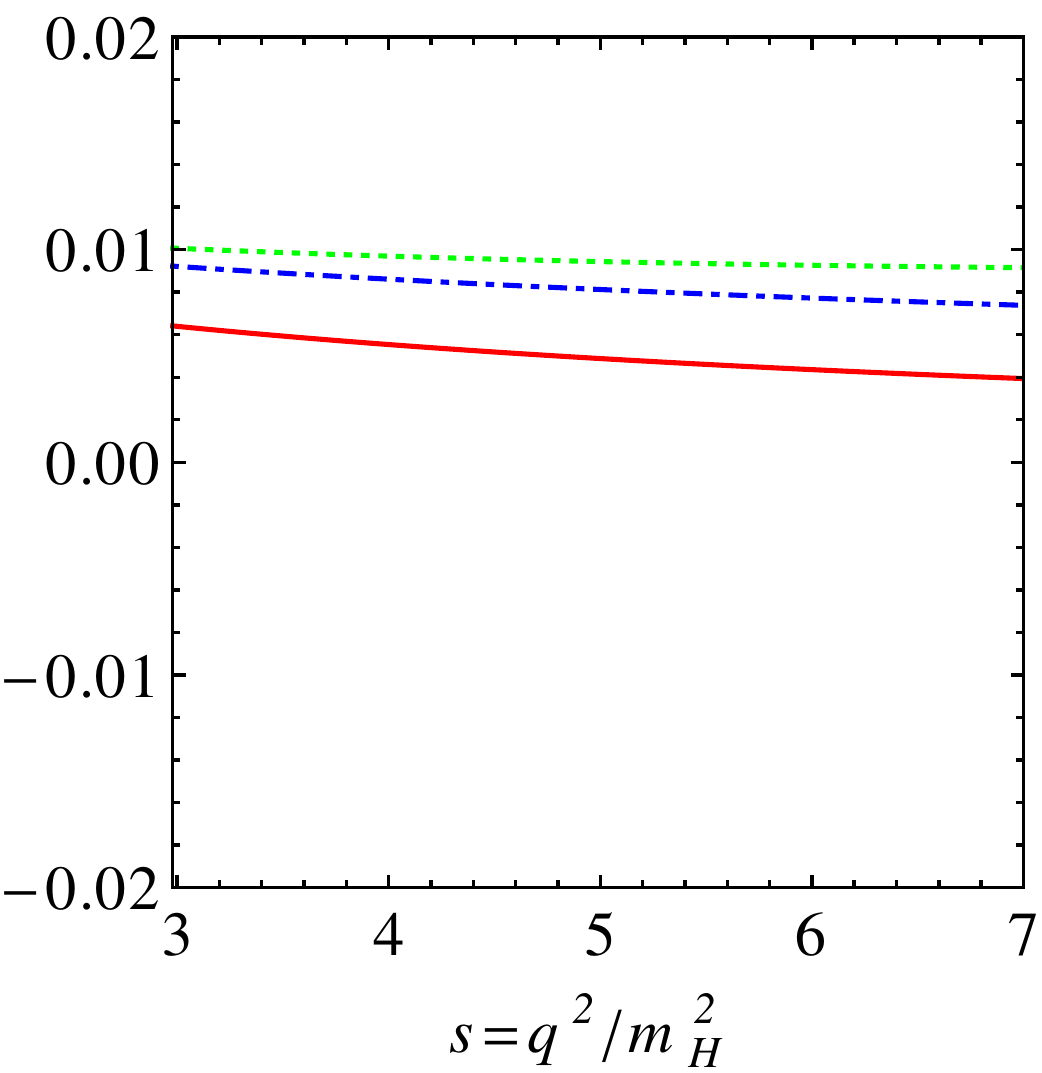}
\label{fig:eeContactAxialc}}
\caption{(a) $\sigma(s)$, (b)  $-\AsymThree$, (c) $-\AsymCoCt$.
Four scenarios with the same $\wh\alpha^V_{\Phi \ell}$ coupling are 
considered. The red solid-line is the SM case. The  dotted green line 
corresponds to $(\wh\alpha^V_{\Phi \ell},\wh\alpha^A_{\Phi \ell})=
(5,5)\times 10^{-3}$, whereas the dot-dashed blue line to  
$(\wh\alpha^V_{\Phi \ell},\wh\alpha^A_{\Phi \ell})=(5,-5)\times 10^{-3}$.
}
\label{fig:eeContactAxial}
\end{center}
\end{figure}

The situation is more interesting, and different from 
Higgs decay, when the axial-vector contact 
interaction is also present. Fig~\ref{fig:eeContactAxiala} shows that 
the total cross section is quite sensitive to the 
axial-vector contact coupling. 
This can be understood with the help of the approximate expression 
for the combination $4J_1+J_2$. As before, we exploit $g_V\ll g_A$ 
and approximate $r=1/2$, but we can no longer use that $s$ is small. 
We then~find
\begin{eqnarray}
4J_1+J_2 &\simeq &\sqrt{2}\,m_H^2\,G_F\,  \barg_A^4\,\frac{s+3}{s-1}  \, 
\times\Bigg[ 1+2\ha_{ZZ}^{(1)} +\frac{12(2s-1)}{s+3}\,\ha_{ZZ} \nn \\
&& 
-2(2s-1) \left( \haA - \frac{\barg_V}{\barg_A}\haV   \right)
+4\left(  \delta g_A - \frac{\barg_V}{\barg_A} \delta g_V  \right) \Bigg] 
\nn\\
&\simeq& \sqrt{2}\,m_H^2\,G_F\,  \barg_A^4  \, \frac{s+3}{s-1}\, 
\left[ 1 -2(1+2s)\,\haA \right]. 
\label{eq:J1J2ee}
\end{eqnarray}
In the last equation we neglected the contributions from $\haV$ that are  
suppressed by $\barg_V$ and we used that in the adopted scenario  
$\ha_{ZZ}^{(1)}=\ha_{ZZ}=0$.
Since now $4s\sim O(10)$, the contribution 
from $\haA$ is significantly larger than in the invariant mass 
distribution $d\Gamma/ds$ of $\HZll$. For $\haA = 5\times 10^{-3}$
the modification of the SM cross section 
reaches $15\%$ as shown in Fig.~\ref{fig:eeContactAxiala}.

The anomalous contributions to the asymmetries  $\AsymThree$, 
$\AsymCoCt$ shown in Figs.~\ref{fig:eeContactAxialb}
 and~\ref{fig:eeContactAxialc} are still largely determined 
by $1/\barg_V$ enhanced $\haV$ contributions. The main
dependence on $\haA$ comes from the denominator of the asymmetries,
but is subleading compared to the $\haV$ terms from the numerator.
For non-vanishing contact couplings the asymmetries are well approximated by
\begin{align}
-\AsymThree &\simeq - \frac{9\pi\sqrt{2}}{2} \,\frac{\barg_V^2}{\barg_A^2}\,  
\frac{s-1}{2s-1} \frac{\sqrt{s}}{s+3}\left[ 1 -(1+2s)\left(\haA-
\frac{\barg_A}{\barg_V} \haV \right)  \right], \nn\\[0.1cm]
-\AsymCoCt &\simeq  9 \,\frac{\barg_V^2}{\barg_A^2} \, 
\frac{1}{s+3}\left[ 1 -  (1+2s)\left( \haA-\frac{\barg_A}{\barg_V} 
\haV \right)  \right].
\label{eq:A3AcoctApproxeeAV}
\end{align}
The asymmetries can reach $2$\% for allowed values of $\haVA$.  
 Relative to the SM value of the asymmetry, the correction 
from anomalous couplings can 
still be $100\%$. 
The ratio of the asymmetries is determined by kinematics as for 
$\HZll$.

Thus we conclude that the total cross section $\sigma(s)$ can be 
significantly modified by $\wh\alpha^A_{\Phi \ell}$ but is insensitive 
to $\wh\alpha^V_{\Phi \ell}$ in comparison, while 
the situation is opposite for (some of) the angular asymmetries.
Eq.~(\ref{eq:J1J2ee}) shows that the cross section 
of $\eeHZ$ is also quite sensitive to $\ha_{ZZ}$ 
due to the  factor $12(2s-1)/(s+3)\sim \mathcal{O}(20)$. Overall, $\eeHZ$ therefore 
seems to be better 
suited to discover contact interactions than $\HZll$.

\subsection{\boldmath Anomalous $HZ\gamma$ coupling}

\begin{figure}[t]
\begin{center}
\subfigure[$\sigma $]
{\includegraphics[width=.31\columnwidth,angle=0]{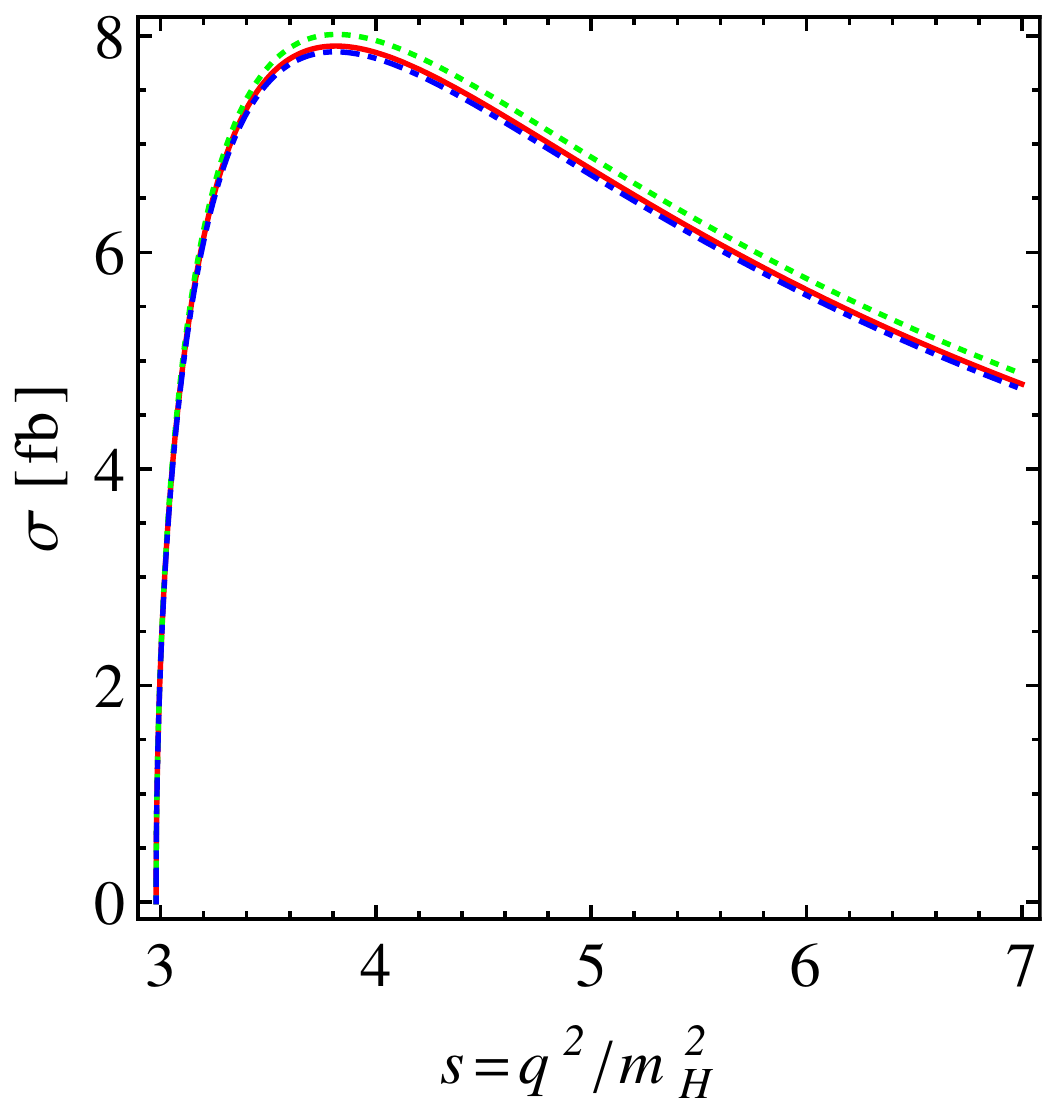}
\label{fig:ContactGammaa}}
\subfigure[$-\AsymThree$]
{\includegraphics[width=.31\columnwidth,angle=0]{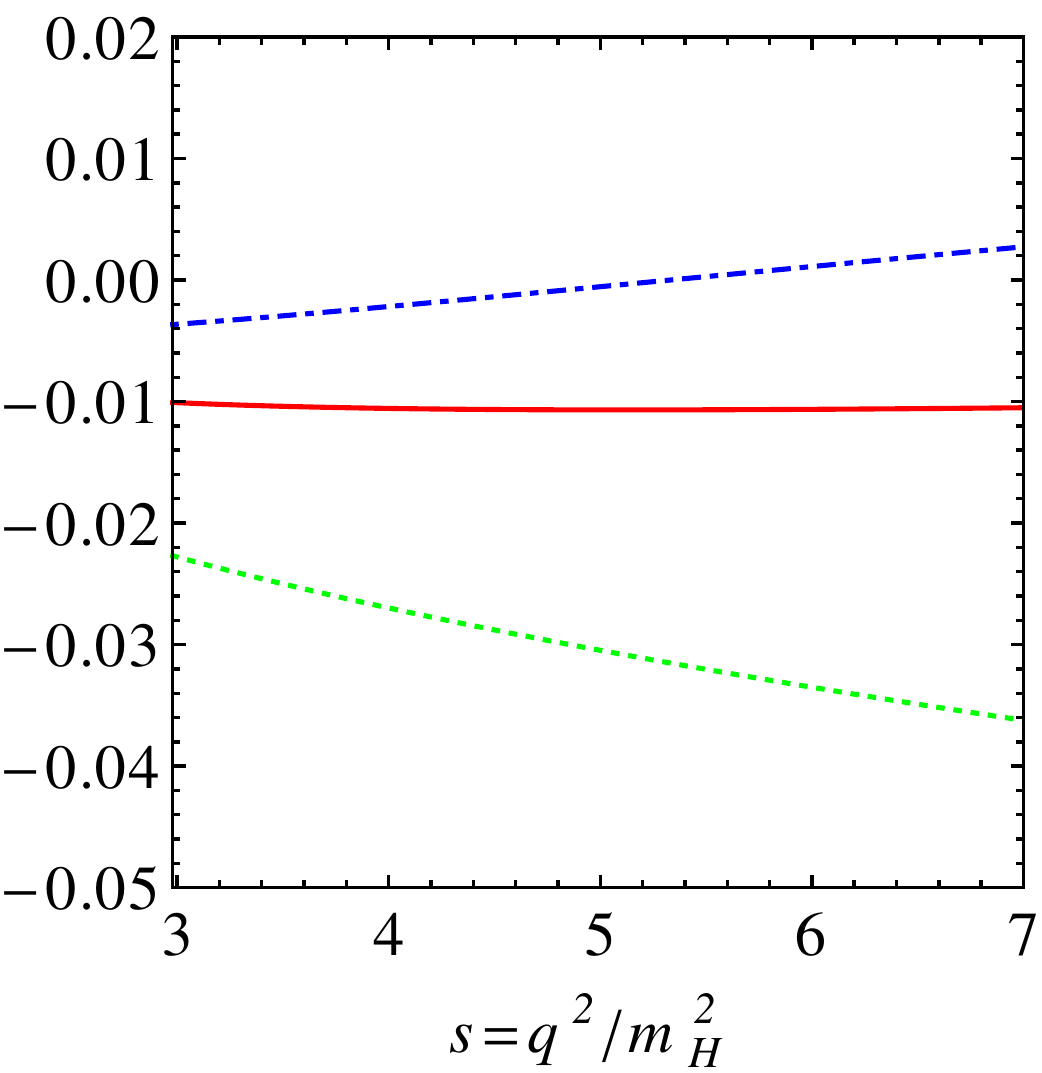}
\label{fig:ContactGammab}}
\subfigure[$-\AsymCoCt$]
{\includegraphics[width=.32\columnwidth,angle=0]{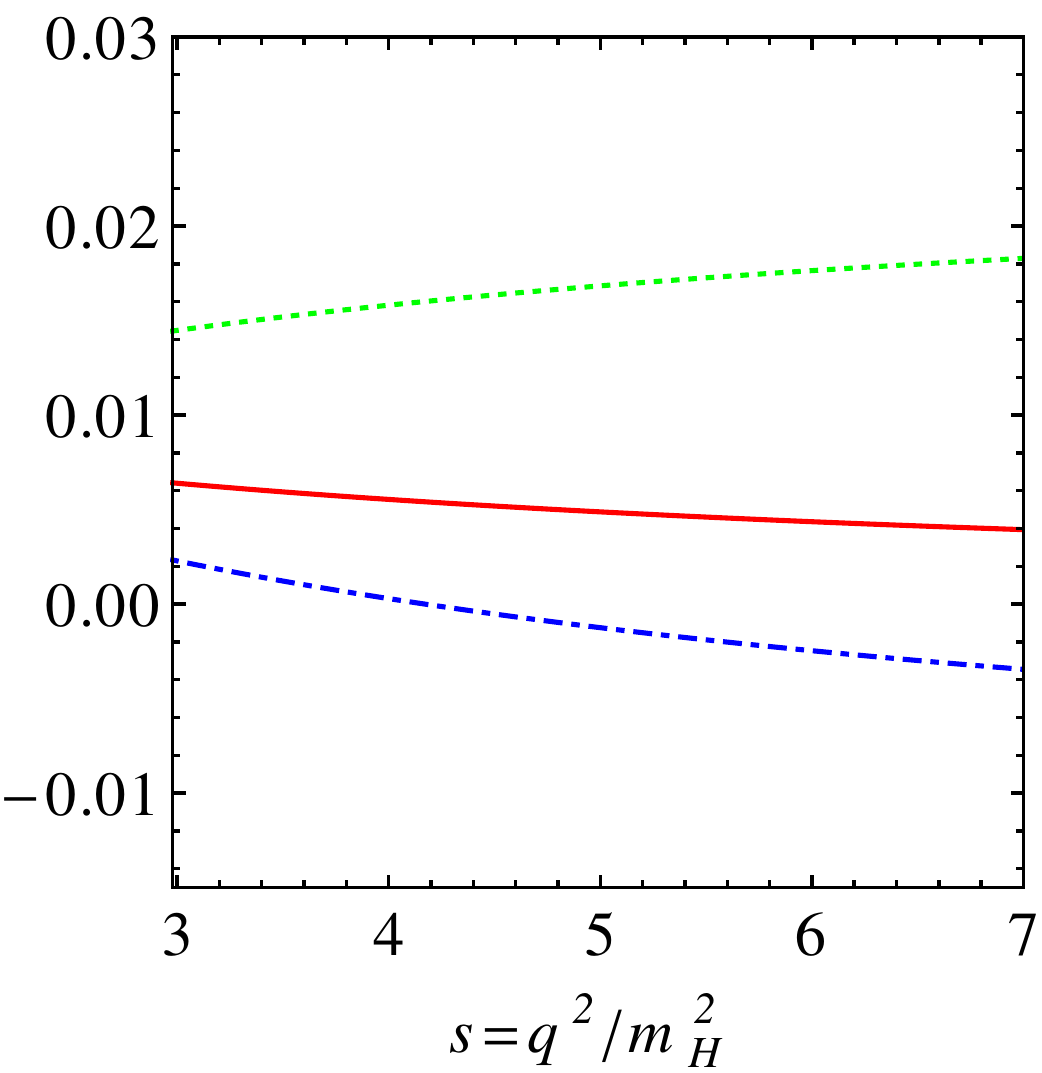}
\label{fig:ContactGammac}}
\caption{(a) $\sigma(s)$, (b)  $-\AsymThree$, (c) $-\AsymCoCt$.
Three scenarios are considered: The red solid-line is the SM result. The dot-dashed blue line corresponds to $\wh\alpha_{AZ}=-1.3\times 10^{-2}$, the  dotted green line  to $\wh\alpha_{AZ}=2.6\times 10^{-2}$.
}
\label{fig:ContactGamma}
\end{center}
\end{figure}

Turning to the anomalous $HZ\gamma$ coupling, we find 
the approximate expression 
\beq
4J_1+J_2 \simeq  \sqrt{2}\,m_H^2\,G_F\, \barg_A^4  
\,\frac{s+3}{s-1}\left( 1 -\frac{s-1}{s+3}\,
\frac{12 \,\barg_V\,g_{\rm em} Q_\ell }{\barg_A^2}\ha_{AZ}  \right)
\eeq
for the combination of angular functions that determines the 
cross section. Similar to the case of $\HZll$ the correction 
is $g_V$ suppressed and has little influence on  $\sigma(s)$ 
as shown in Fig.~\ref{fig:ContactGammaa}. 

The asymmetries can reach a
few percent (Figs.~\ref{fig:ContactGammab} and~\ref{fig:ContactGammac})  
for the largest allowed values of the $HZ\gamma$ coupling. This again is 
due to the $1/g_V$ enhancement of the correction.
Assuming other couplings to vanish, approximate expressions for the 
asymmetries in the presence of the $\wh\alpha_{AZ}$ coupling are
\begin{align}
-\AsymThree &\simeq - \frac{9\pi\sqrt{2}}{2} \, 
\frac{\barg_V^2}{\barg_A^2}\, \frac{s-1}{2s-1} \frac{\sqrt{s}}{s+3} 
\left[ 1 -\frac{g_{\rm em} Q_\ell\, (s+1)}{2\barg_V}\,\ha_{AZ}    \right], 
\nn\\[0.15cm]
-\AsymCoCt &\simeq  9\,\frac{  \barg_V^2}{\barg_A^2} \, \frac{1}{s+3}
\left[ 1 -\frac{g_{\rm em} Q_\ell \, (s-1)}{\barg_V} \, \ha_{AZ}  \right] .
\label{eq:A3AcoctApproxAZ}
\end{align}
There is no photon-pole enhancement in this case. 
Nevertheless, relative to the SM value of the asymmetry, the correction 
from the anomalous coupling can 
still be $100\%$. 

\subsection{\boldmath CP-odd couplings}

\begin{figure}[t]
\begin{center}
\subfigure[$\mathcal{A}_{\phi}^{(1)}$]
{\includegraphics[width=.32\columnwidth,angle=0]{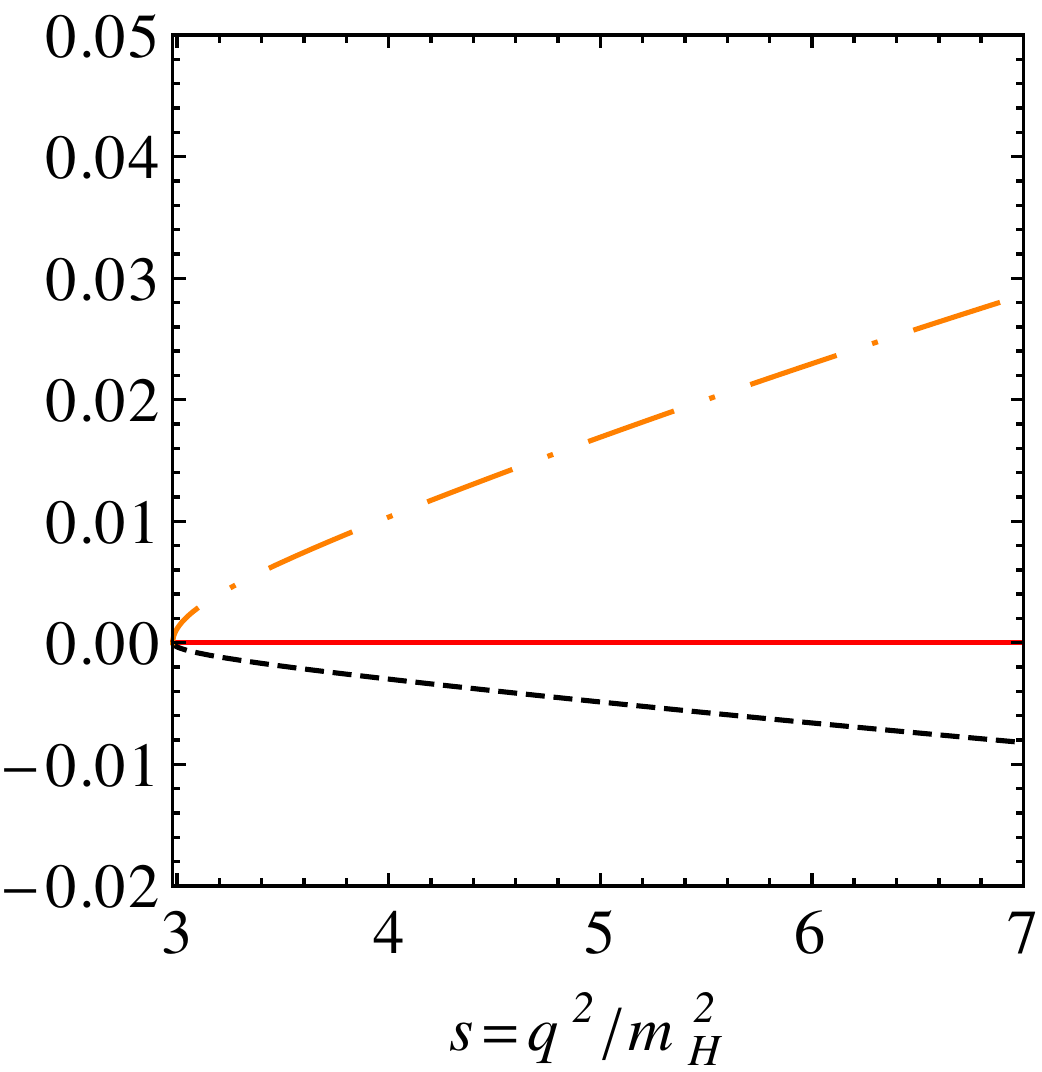}
\label{fig:CPoddeea}}
\subfigure[$\mathcal{A}_{\phi}^{(2)}$]
{\includegraphics[width=.32\columnwidth,angle=0]{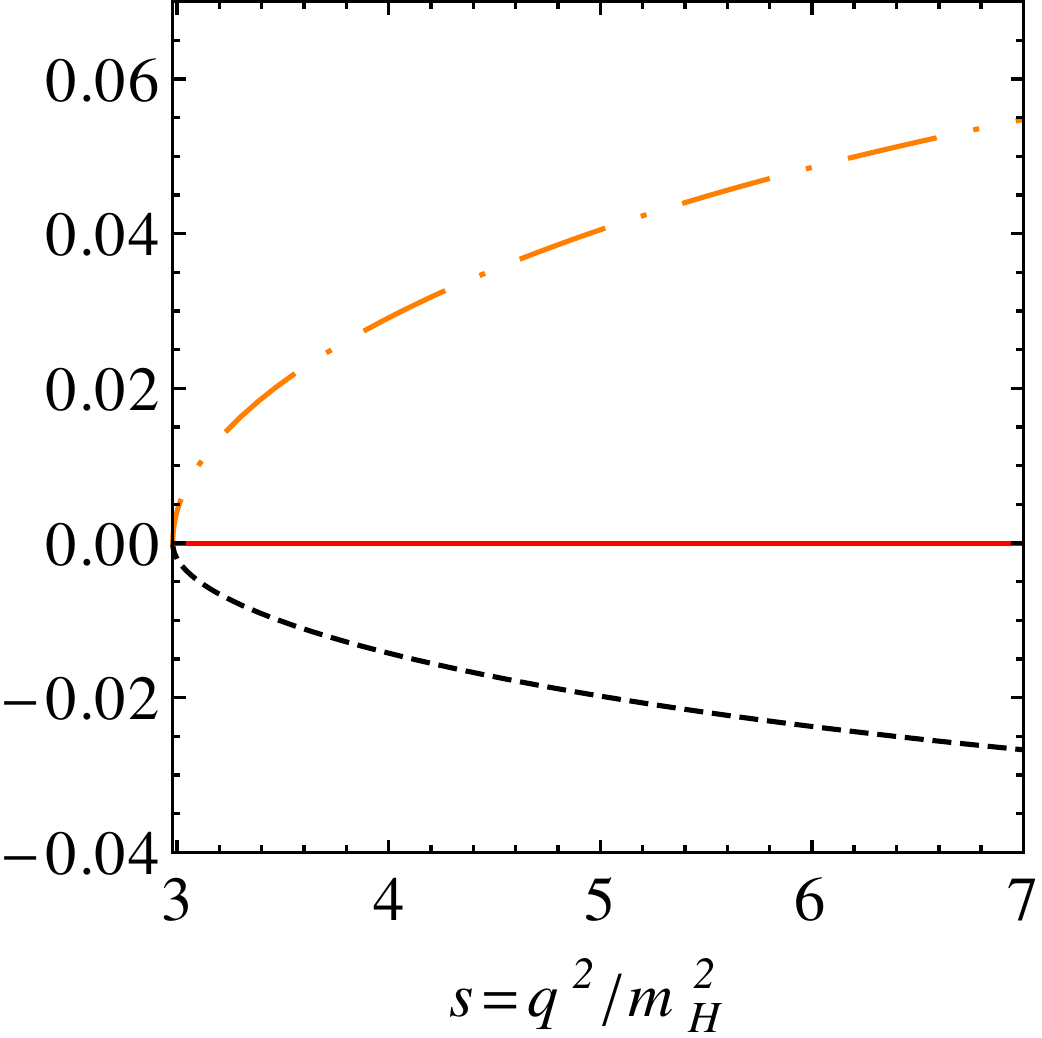}
\label{fig:CPoddeeb}}
\caption{ Asymmetries  $\mathcal{A}_{\phi}^{(1,2)}$ in two different 
scenarios.   The dot-dashed orange line corresponds  to  
$(\wh\alpha_{Z\widetilde Z},\wh \alpha_{A\widetilde Z})=(4,4)\times 10^{-2}$, and  
the dashed black line to  $(\wh\alpha_{Z\widetilde Z}, 
\wh\alpha_{A\widetilde Z})=(-2,-1)\times 10^{-2}$. The red solid line is
the vanishing SM result.
}
\label{fig:CPoddee}
\end{center}
\end{figure}

In the $\eeHll$ case, the asymmetry $\mathcal{A}_{\phi}^{(1)}$ is
again dominated by $\wh\alpha_{A\widetilde Z}$, but the contribution from
$\wh\alpha_{Z\widetilde Z}$ is less suppressed than in $\HZll$ 
due to the larger values
of $s$. The situation is opposite for 
$\mathcal{A}_{\phi}^{(2)}$, where $\wh\alpha_{Z\widetilde Z}$ dominates
and $\wh\alpha_{A\widetilde Z}$ gives a small $g_V$ suppressed
contribution. Although a zero may appear in both asymmetries due to the 
interplay of the two CP-odd couplings, whenever a zero occurs the 
strong cancellation between the two contributions keeps the asymmetry 
below the permille level. We therefore provide approximate expressions
that contain only the dominant effects:
\bea
\mathcal{A}_{\phi}^{(1)}&\simeq&-\frac{9\pi \sqrt{2}}{8} \, 
\frac{\barg_V g_{\rm em}Q_\ell}{\barg_A^2} \, 
\frac{\sqrt{\lambda}\,(s-1)}{\sqrt{s}\,(s+3)}  
\,\ha_{A\widetilde Z},\nn \\
\mathcal{A}_{\phi}^{(2)}&\simeq& \frac{8\sqrt{\lambda}}{\pi (s+3)} 
\,\wh \alpha_{Z\widetilde Z}.
\eea
From  these expressions one sees that the asymmetries can be at the percent 
level for CP-odd couplings $\mathcal{O}(10^{-2})$. 
The exact results for the 
asymmetries $\mathcal{A}_{\phi}^{(1,2)}$ are shown in  
Fig.~\ref{fig:CPoddee} for two coupling value sets.


\section{Estimate of SM loop effects}
\label{sec:SM-loop}

Electroweak one-loop contributions to the processes studied here can
be of similar size as the tree-level $d=6$ corrections discussed in 
the previous sections.  For example, they are around $2$\% percent 
for the $H\to 4\ell$ decay rate~\cite{BDDW06}.  In this section we 
perform a rough estimate of SM loop contributions and compare them 
to the effect from the anomalous $HZ\gamma$ coupling. A full analysis 
is beyond the scope of the present work. 

Let us consider the SM one-loop $HZ\gamma^*$ and $HZZ^*$ amplitudes, whose   
explicit analytical expressions can be found 
in Refs.~\cite{Kniehl90,Kniehl91,Kniehl94}.
The  amplitude for the transition $H\to ZV$ (with $V=Z^*,\gamma^*$) 
involves five form factors in general. However, when the particles are on-shell 
or  coupled to conserved currents, which is the case of interest here, 
only two form factors contribute. We therefore write the 
amplitudes in the form
\begin{equation}
\mathcal{M}^{\mn}_{HZV}(H\to Z(p)V(q)) = 
{2 m_Z^2\,(\sqrt{2} G_F)^{1/2}}\, 
\left [\frac{q^{\mu} p^{\nu}}{m_H^2} \,   D_{V} (q^2)  +  
g^{\mu \nu}\,E_{V}(q^2)  \right ], 
\label{eq:HZVLoop} 
\end{equation}
where the loop functions $D_{Z,\gamma}$ and $E_{Z,\gamma}$ are functions of
$q^2$. The tree-level $HZZ$ vertex is treated separately and already included in
Eq.~(\ref{eq:effHVVCouplings}).\footnote{The $d=6$ corrections from the
redefinition of the Lagrangian input parameters in the last equation
can be neglected since they generate terms that are loop and
$1/\Lambda^2$ suppressed. It also needs to pointed out that the form
factors $D_{Z,\gamma}$ and $E_{Z,\gamma}$ are gauge invariant only if
both external states are on their mass shells. We use the expressions
from Refs.~\cite{Kniehl90,Kniehl91,Kniehl94} where the 't Hooft-Feynman gauge
is adopted, and drop the (presumably small) box-diagram contributions
to the $\HZll$ and $\eeHZ$ processes, which would be required to 
restore gauge invariance. In the $q^2$ range relevant to the decay $\HZll$
the gauge dependence of the one-loop expressions for the $HZ\gamma^\ast$ 
amplitude is expected to be small because the photon is nearly on the 
mass shell in relation to $m_H^2$.}  

In the previous sections we discussed the modifications of the 
form factors $H_{1,V}$ and $H_{2,V}$ due to the anomalous $HZ\gamma$ 
coupling $\ha_{AZ}$. Including the one-loop $H\to ZV$ amplitudes of
Eq.~(\ref{eq:HZVLoop}) into the defining expression Eq.~(\ref{eq:MHZll}), 
we find 
\bea
H_{1,V}&=&\frac{2m_H (\sqrt{2} G_F)^{1/2} \,r}{s-r} g_V
\left[1 + E_Z(q^2)
+ \frac{Q_\ell \,g_{\rm em}\,\kappa\, (s-r)}{  2 r s\,g_V}   
\left(    \wh\alpha_{AZ}  -\frac{2\,r}{\kappa}E_\gamma(q^2)   
\right) \right],  
\nonumber \\[0.1cm]
H_{2,V}&=& \frac{2m_H (\sqrt{2} G_F)^{1/2}}{s-r}\, g_V 
\left[ \,r\,D_Z(q^2) -   \frac{Q_\ell \,g_{\rm em} (s-r)}{  s\,g_V}
\left(  \wh\alpha_{AZ}  + r\, D_\gamma(q^2) \right)\right],
\label{eq:H2VLoop}
\eea
which should be compared to Eq.~(\ref{eq:ExplicitH}). For the 
present purpose we have kept only the anomalous $HZ\gamma$ 
interaction, setting all other  $d=6$ couplings to zero. 
The terms with an intermediate photon are $1/g_V$ enhanced with
respect to the terms with an intermediate $Z$. Since the one-loop
$H\to ZZ$ amplitude is of the same order as the $H\to
Z\gamma$ amplitude, we can neglect the contributions from 
$D_Z$ and $E_Z$ in the further discussion.

\begin{figure}[t]
\begin{center}
\subfigure[$\delta H_{1,V}$ in $\HZll$]
{\includegraphics[width=.40\columnwidth,angle=0]{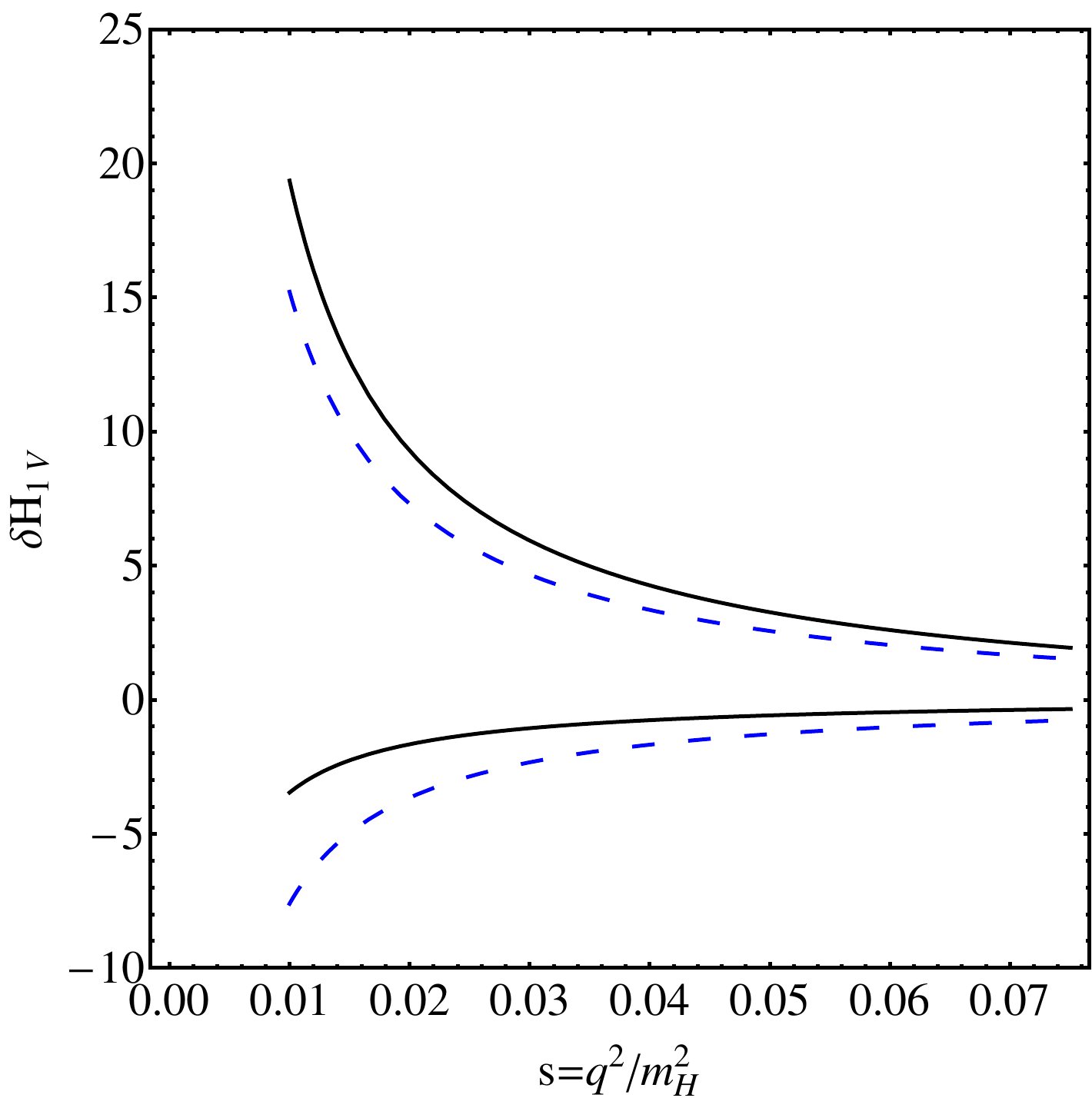}
\label{fig:H1VLoopDecay}}
\subfigure[$ H_{2,V}$ in $\HZll$]
{\includegraphics[width=.40\columnwidth,angle=0]{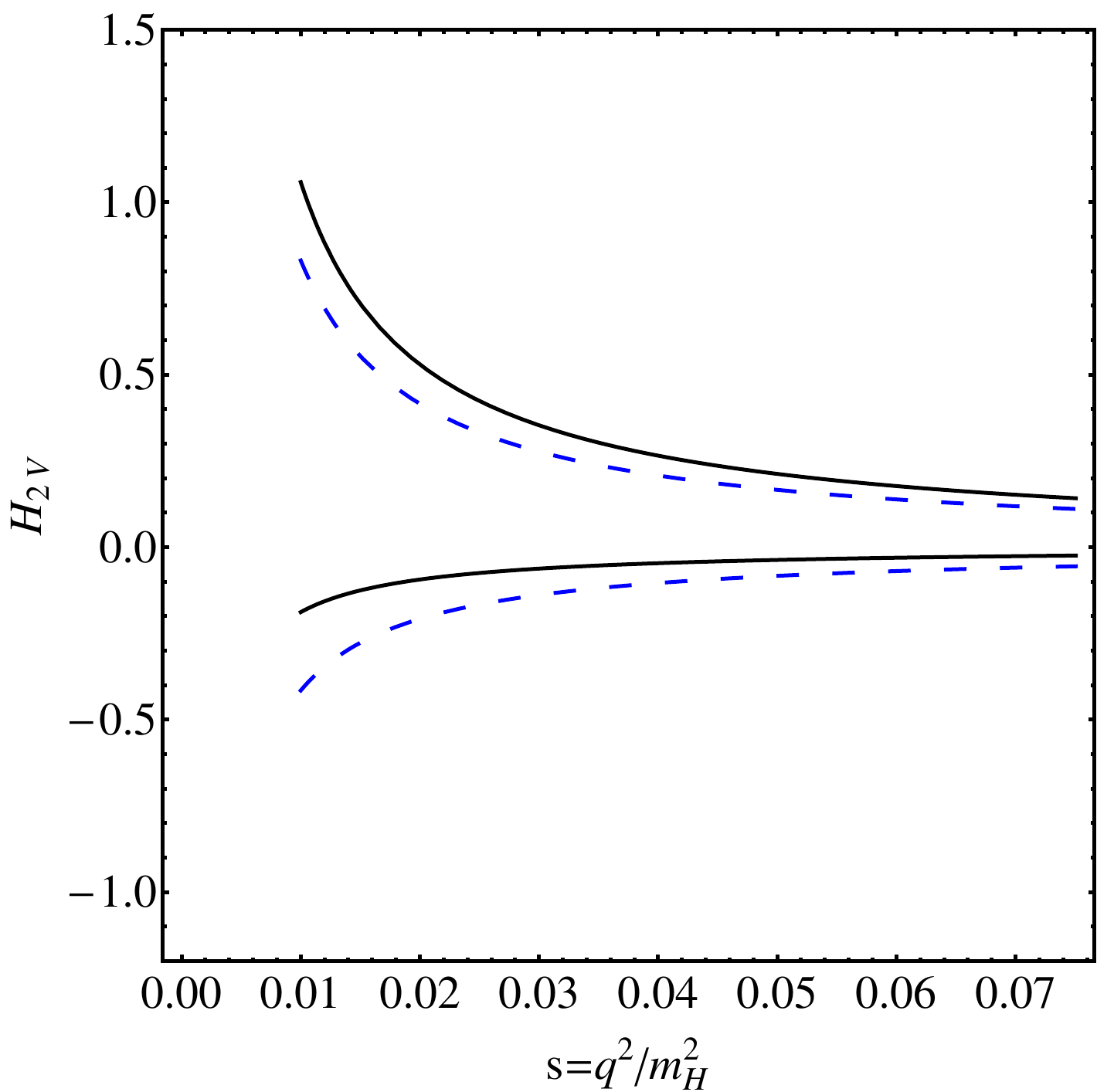}
\label{fig:H2VLoopDecay}}
\caption{Dominant effects due to the $H\to ZV$ one loop amplitude in the 
form factors $H_{1,V}$ and $H_{2,V}$ in decays $\HZll$.  In (a) we 
show $\delta H_{1,V}$,  Eq.~(\ref{eq:deltaH1V}), and  (b) $H_{2,V}$. 
The results within the solid lines include the dominant loop contribution 
and $\ha_{AZ}\in [-1.3,2.6]\times 10^{-2}$. Results within
the  dashed line include solely the effects of  $\ha_{AZ}$.  }
\label{fig:H1VLoopCorrec}
\end{center}
\end{figure}

We start by discussing the modifications to $H_{1,V}$ and $H_{2,V}$ 
in the decay $\HZll$. The first important observation is that, since
the Higgs boson cannot decay into $WW$ or $t\bar t$, the loop contribution is
real and does not generate an imaginary part of the form factors. 
Therefore, the angular structures in the presence of these
loop contributions remain the same as discussed in the previous
sections.  Second, the $s$ dependence of the loop contribution is
small, since $s\ll 1$. Therefore, the inclusion of the $HZV$ amplitude 
at one loop amounts, essentially, to shifting the value of 
$\ha_{AZ}$ by an amount given by the expressions in round brackets in 
Eq.~(\ref{eq:H2VLoop}). To estimate the size of this 
shift in  $H_{1,V}$ and  $H_{2,V}$, 
we compare the allowed range for the anomalous  $HZ\gamma$ coupling, 
\beq
 \wh\alpha_{AZ} \in [-1.3,2.6]\times 10^{-2},
\label{range2}
\eeq 
to the quantities 
\beq
\frac{2\,r}{\kappa}E_\gamma (s=0.01) = -7.1\times 10^{-3},
\eeq
and 
\beq
rD_\gamma(s=0.01)=   7.1\times 10^{-3},
\eeq
respectively. (The energy dependence of this function is small in the $s$ 
range relevant to $\HZll$.) The shift is therefore small 
relative to the allowed values of $\wh\alpha_{AZ}$. This is shown 
explicitly in Fig.~\ref{fig:H1VLoopDecay} for 
\beq
\delta H_{1,V} =   \frac{Q_\ell \,g_{\rm em}\, \kappa\, (s-r)}
{  2\,r\,s\,g_V}   \left(    \wh\alpha_{AZ}  -\frac{2\,r}{\kappa}
E_\gamma(q^2)   \right) ,\label{eq:deltaH1V}
\eeq
and in Fig.~\ref{fig:H2VLoopDecay} for $ H_{2,V}$. We therefore conclude 
that the previously discussed asymmetries are not 
affected dramatically by loop effects, at least in the study of the 
anomalous $HZ\gamma$ interaction. In any case, SM loop effects 
are calculable and should simply be included in a definitive 
analysis, when sufficient experimental data are available.

\begin{figure}[t]
\begin{center}
\subfigure[${\rm Re}\left(\delta H_{1,V}\right)$ in $\eeHZ$]
{\hspace{0.4cm}\includegraphics[width=.37\columnwidth,angle=0]{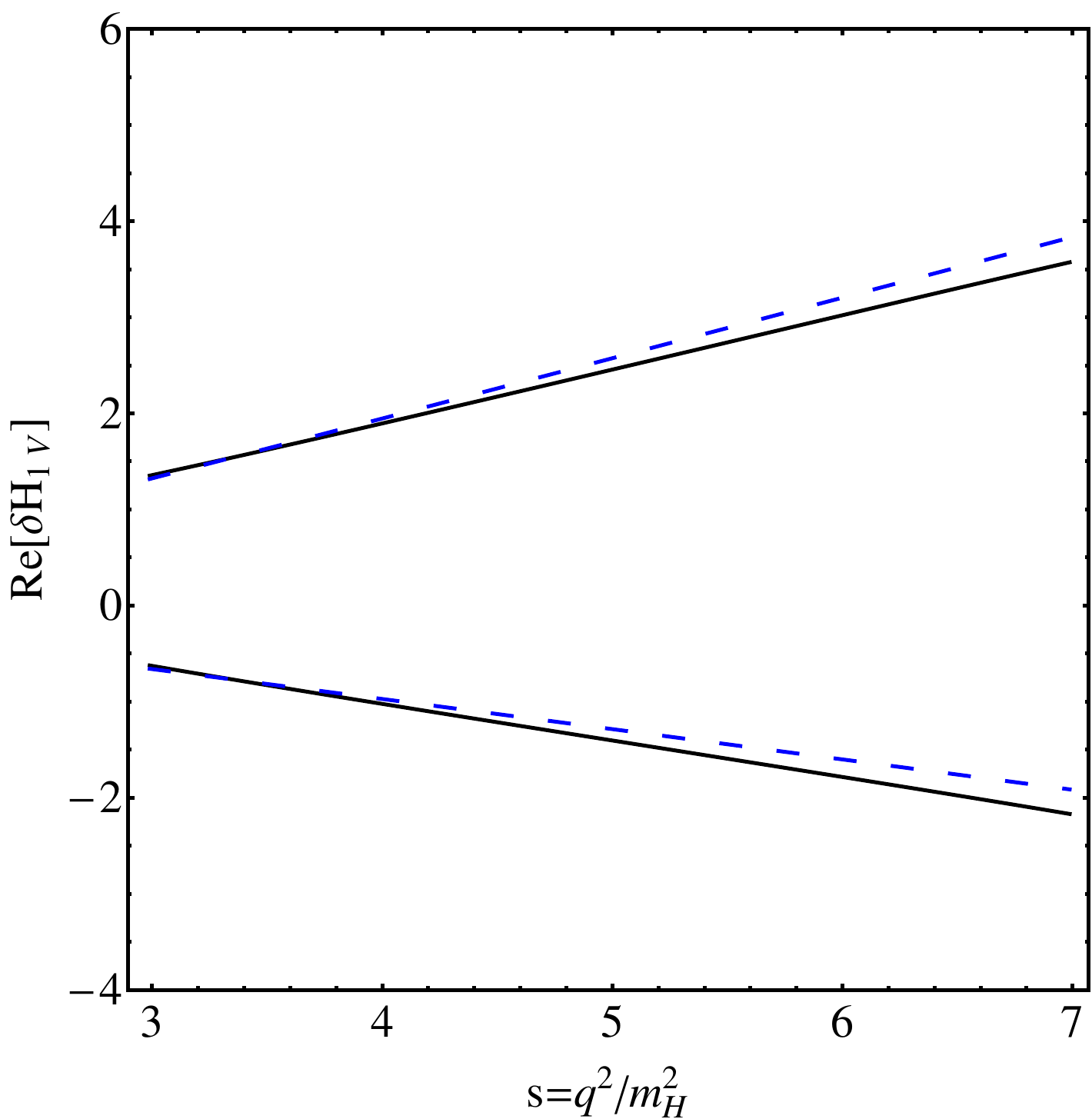}
\label{fig:ReH1VLoopScatt}}
\subfigure[${\rm Im}\left(\delta H_{1,V}\right)$ in $\eeHZ$]
{\includegraphics[width=.37\columnwidth,angle=0]{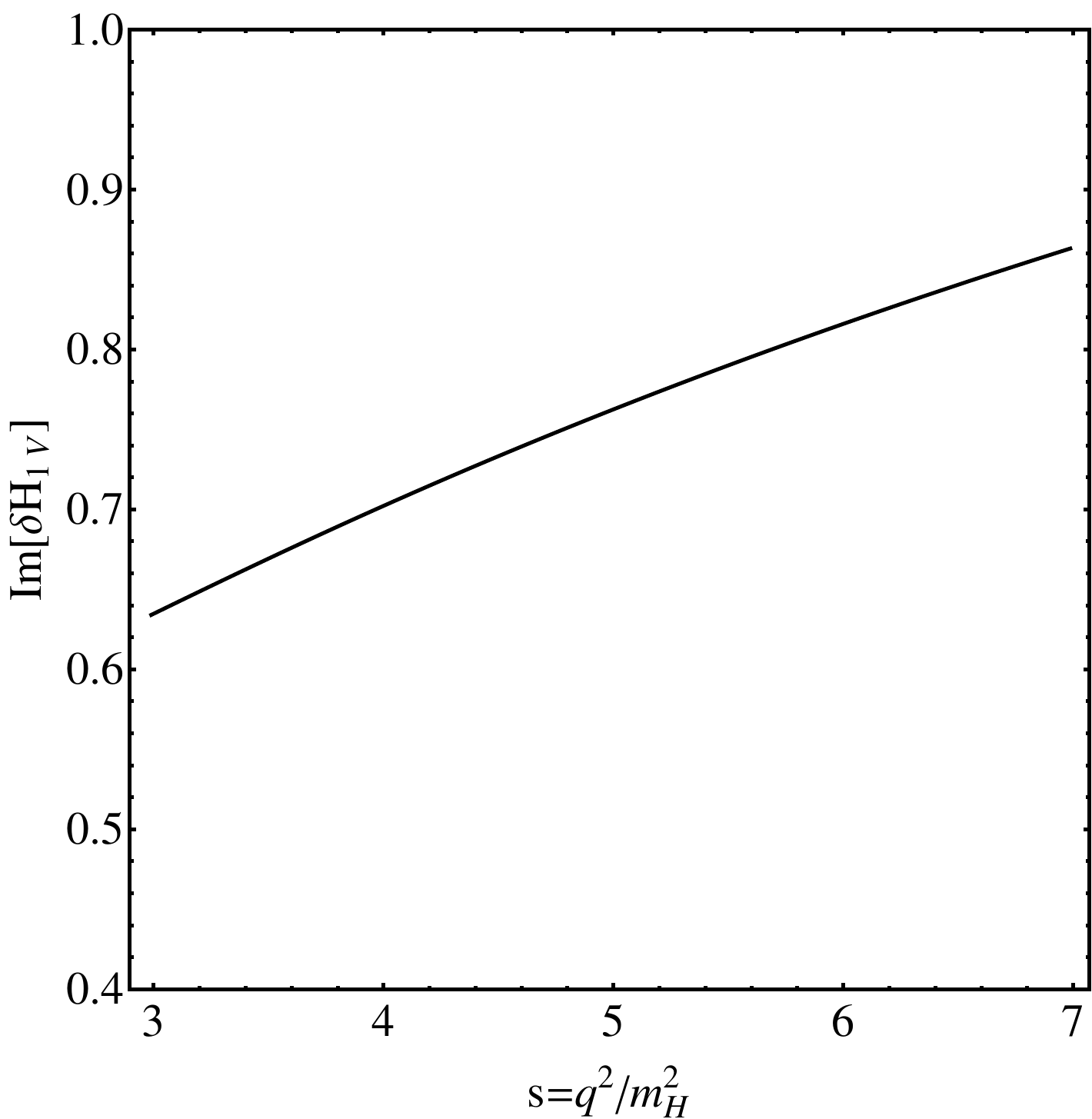}
\label{fig:ImH1VLoopScatt}}
\subfigure[${\rm Re}\left( H_{2,V}\right)$ in $\eeHZ$]
{\includegraphics[width=.40\columnwidth,angle=0]{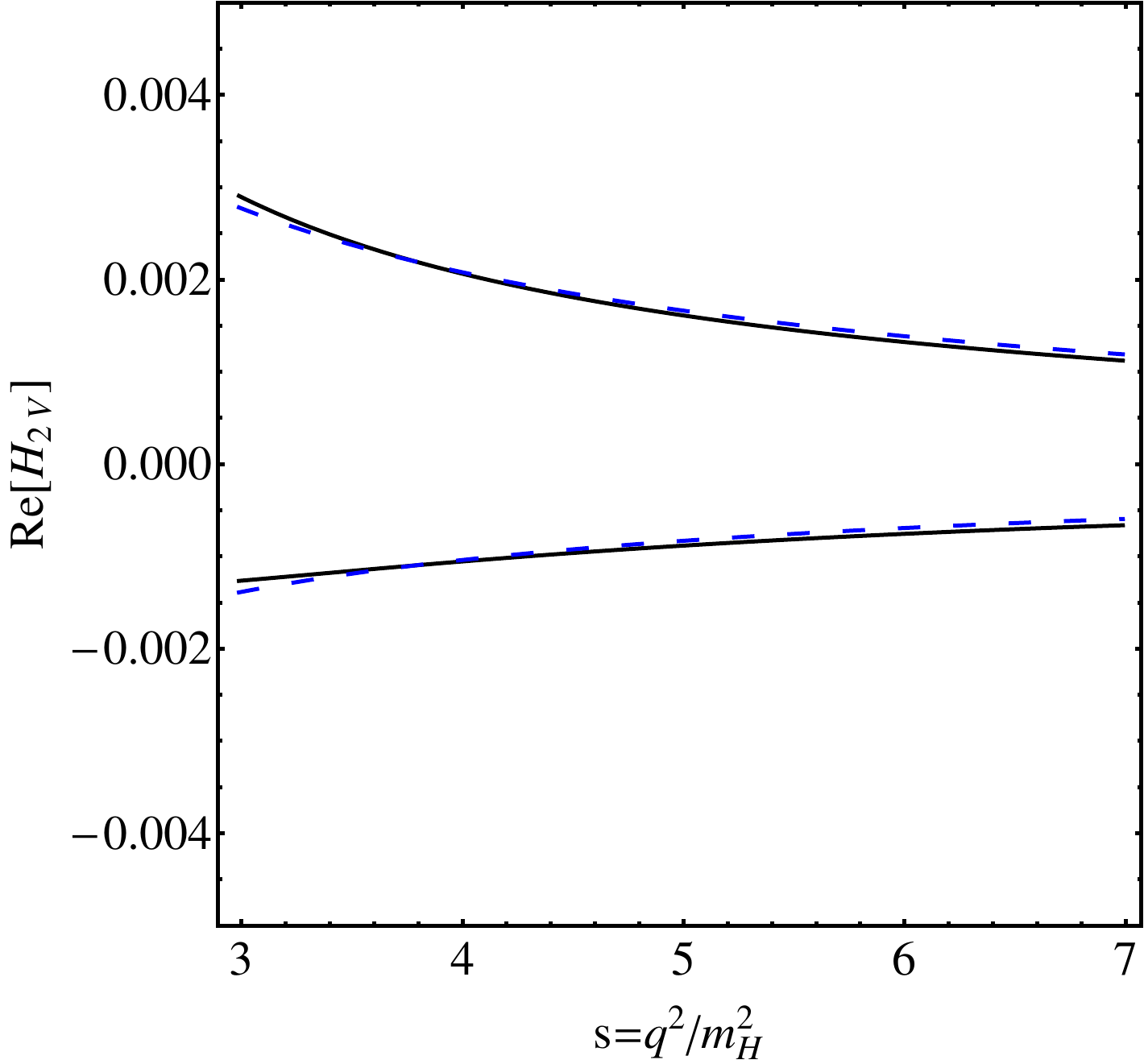}
\label{fig:ReH2VLoopScatt}}
\subfigure[${\rm Im}\left( H_{2,V}\right)$ in $\eeHZ$]
{\includegraphics[width=.40\columnwidth,angle=0]{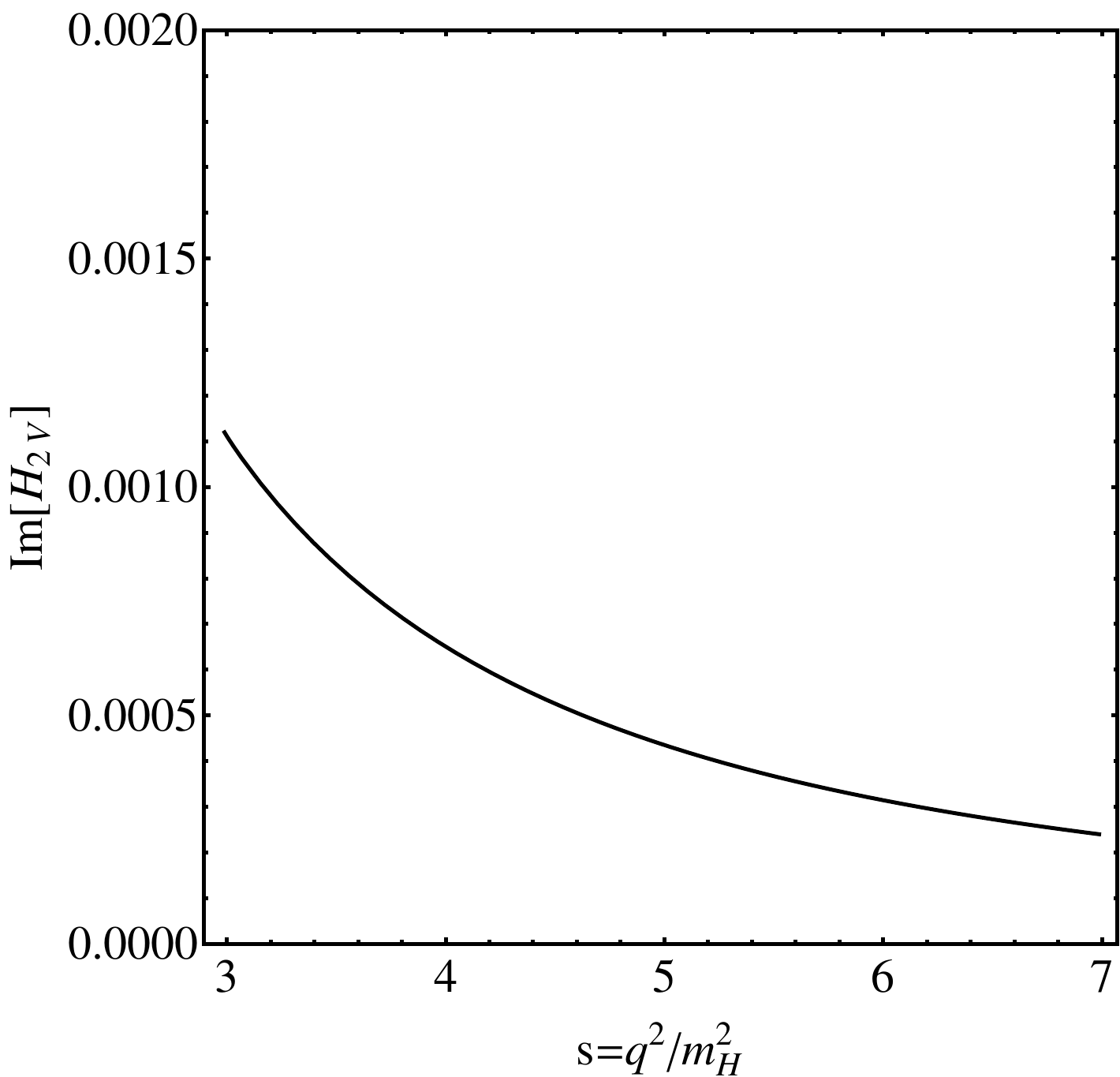}
\label{fig:ImH2VLoopScatt}}
\caption{Dominant effects due to the $H\to ZV$ one loop amplitude in 
the form factors $H_{1,V}$ and $H_{2,V}$ in $\eeHll$. In (a) and (b) 
we show ${\rm Re}(\delta H_{1,V})$
and ${\rm Im}(\delta H_{1,V})$, respectively. In (c) and (d) we show 
${\rm Re}(H_{2,V})$ and ${\rm Im}(H_{2,V})$ respectively. 
Results within the solid lines include the dominant loop contribution 
and $\ha_{AZ}\in [-1.3,2.6]\times 10^{-2}$. Results within
the  dashed lines include solely the effects of  $\ha_{AZ}$.  }
\label{fig:H2VLoopCorrec}
\end{center}
\end{figure}

Turning to $\eeHZ$, we note that the form factors are now probed in 
the kinematic range, where the off-shell momentum $q^2\geq (m_H+m_Z)^2$.
The loop functions $D_V$ and $E_V$ develop imaginary parts and therefore the
form factors $H_{1,V}$ and $H_{2,V}$ are complex, which generates 
additional angular structures in Eq.~(\ref{eq:FullJ}).
However, while the imaginary parts are sizable, as shown in 
the two right panels Fig.~\ref{fig:ImH1VLoopScatt} and 
Fig.~\ref{fig:ImH2VLoopScatt}, the real parts of the form factors 
$H_{1,V}$ and $H_{2,V}$ relevant to the asymmetries discussed in the 
previous sections, are not dramatically altered and even
smaller than for $\HZll$. The corresponding 
results for  $ \delta H_{1,V}$ and $H_{2,V}$ are displayed 
in the two left panels of 
Fig.~\ref{fig:H2VLoopCorrec}. Numerically, the contribution from
$\frac{2\,r}{\kappa} E_\gamma(q^2)$ to $\delta H_{1,V}$ now ranges from  
$(-0.66 -i\,12) \times 10^{-3}$ near threshold ($s=3$) 
to $(1.7 -i\,5.9) \times 10^{-3}$ at $s=7$. Similarly $rD_\gamma(q^2)$, 
which affects $H_{2,V}$ varies from $(1.4 + i\, 11) \times 10^{-3}$ 
at $s=3$ to $  (-1.5 + i\, 5.8) \times 10^{-3}$ at $s=7$. The real 
part of these numbers should again be compared to the range 
given in Eq.~(\ref{range2}). 

\section{Summary}
\label{sec:Conc}

In this work we studied the observability of anomalous $d=6$ Higgs
couplings in $\Hllll$ decay and in the crossing-symmetric process
$\eeHll$.  We computed the differential decay width $d\Gamma/dq^2$ of
$\HZll$, the total cross section of $\eeHll$, $\sigma(s)$, as well
as angular asymmetries in both processes. Our particular 
interest regarded the question, see also Ref.~\cite{BCD13}, 
whether angular asymmetries have the 
potential to reveal BSM physics that would be hidden in 
$d\Gamma/dq^2$ and $\sigma(s)$. In some of these asymmetries, the
anomalous $HZ\gamma$ coupling, $\ha_{AZ}$, and the vector contact
$HZ\ell \ell$ interaction parametrized by $\haV$, are
enhanced with respect to the SM contribution by a factor of $1/{g_V}$. 
These two types of interactions are therefore the prime targets 
of the asymmetry analysis.
Our main conclusions can be summarized as follows:
\begin{itemize}
\item We identify several angular asymmetries, which are indeed 
very sensitive to anomalous couplings.
\item Within the presently allowed range of the anomalous 
$HZ\gamma$ interaction strength,  $\ha_{AZ}$, 
modifications of angular asymmetries 
of ${\cal O}(1)$ and even larger relative to the SM value 
are still possible indicating sensitivity to multi-TeV scales.
\item Anomalous $HZ\ell\ell$ contact interactions have smaller 
effects. This is  mainly because we find that their size is already 
tightly constrained by existing data, in agreement with 
the constraints derived in Ref.~\cite{PR13} (although this
refers to another operators basis). The effects of the contact 
$HZ\ell \ell$ interactions in the angular asymmetries of $\HZll$ 
were previously investigated in Ref.~\cite{BCD13}. While we 
formally agree with their results, we find significantly smaller 
asymmetries, since the typical values of 
$\haV$ adopted in that paper are about a factor of four 
larger than those allowed in the present analysis.
\item At present, the CP-odd $d=6$ couplings are not strongly 
constrained by data. We showed that CP-odd asymmetry 
$\mathcal{A}_\phi^{(1)}$ can reach 
the few percent level in both in $\HZll$ decay and $\eeHZ$ 
Higgs production. In $\HZll$ an asymmetry-zero may
occur. However, for allowed values of the CP-odd couplings 
the asymmetry that can display this zero is never large.
\item Most interesting asymmetries are small in absolute 
terms, reaching at most 10\%, 
and often much less, because they are suppressed by the small 
vector $Z\ell\ell$ coupling.
\item Overall, the process $\eeHZ$ seems better suited 
than $\HZll$ for the study of anomalous $HZ\ell\ell$ 
contact interactions due to 
the higher di-lepton invariant masses.  This is particularly 
true for the contributions of $\haA$ (as well as of $\ha_{ZZ}$) 
to the total cross section, where 
$15$\% percent modifications are possible. On the other hand, 
$\HZll$ provides better sensitivity to the anomalous $HZ\gamma$ 
coupling due to the photon-pole enhancement.
\end{itemize}

We further provided a rough estimate of SM loop contributions
to the processes discussed here. These loop contributions have been
calculated in the past and our estimate suggests that loop effects 
are small compared to the presently allowed $d=6$ effects. 
Once sufficient data is available to attempt constraining $d=6$ 
couplings from angular asymmetries, SM loop effects should 
simply be included. However, the
experimental detection of angular asymmetries will be challenging
even with the planned higher statistics up-grades of the LHC.  

\section*{Acknowledgements}

We thank G. Buchalla, O. Cat\`a, and G. D'Ambrosio for
correspondence regarding Ref.~\cite{BCD13}. We would like to thank
M.~Ramon, F.~Riva, M. Spira and M. Trott for illuminating discussions. 
This work is supported in part by the Gottfried Wilhelm Leibniz programme 
of the Deutsche Forschungsgemeinschaft (DFG). The
work of DB was supported by the Alexander von Humboldt
Foundation. YMW is supported by the 
DFG Sonder\-forschungs\-bereich/Trans\-regio~9 
``Computer\-gest\"{u}tzte Theoretische Teilchen\-physik".

\appendix
\section{Kinematics}
\label{app:Kine}
\subsection{\boldmath $H\to Z \ell^+ \ell^-$}
\label{app:KineHZll}

\begin{figure}[t]
\begin{center}
\includegraphics[width=0.9\columnwidth]{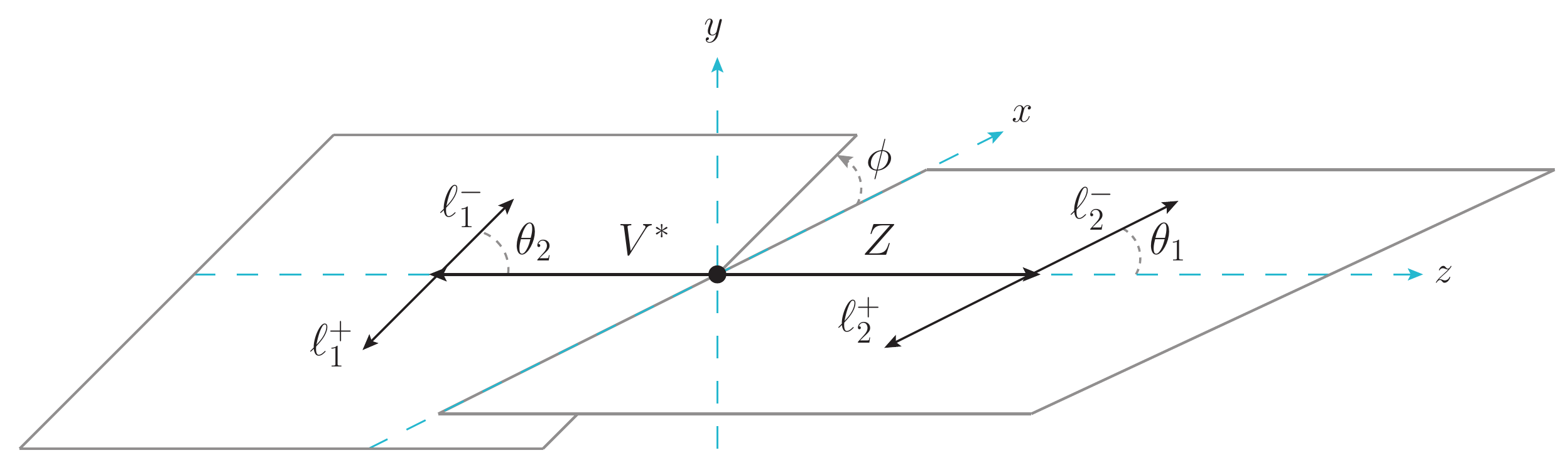}
\vspace*{0.2cm}
\caption{Kinematics of the  four-body decay $\Hllll$. 
}
\label{fig:KineHZll}
\end{center}
\end{figure}

Here we describe the kinematics and the angle conventions used in
our results. The reaction is labelled as 
$H(p_H) \to  \ell_1^-(p_3) \ell_1^+(p_4) 
Z(p)(\to \ell_2^- (p_1) \ell_2^+(p_2))$, where we labelled the two pairs 
of leptons to distinguish the pair $\ell_2^- \ell_2^+$ that arises 
from the decay of the on-shell $Z$ boson from the other. We have
\begin{equation}
p= p_1+p_2, \qquad 
q = p_3+p_4, 
\end{equation}
and $p^2=m_Z^2$.
We denote momenta in the $\ell_2^+ \ell_2^-$ rest frame by an upper
bar ($\bar p$), whereas momenta in the $\ell_1^+ \ell_1^-$ rest frame
are denoted by an asterisk ($p^*$).

Using the conventions for the axes given in Fig.~\ref{fig:KineHZll},
we define the positive $z$ direction to be that of the on-shell $Z$
three-momentum $\bp$ in the Higgs rest frame. The angle
$\theta_1$ is the angle between the momentum $\bp_1$ of $\ell_2^-$ and
the $z$ axis, in the $\ell_2^+ \ell_2^-$ rest frame. Accordingly, in
the massless limit  the momenta $p_{1,2}$ are written in the $Z$ rest frame as
\bea
\bar p_1 &=& \frac{m_Z}{2}(1,  \sin\theta_1,0, \cos\theta_1),\\
\bar p_2 &=& \frac{m_Z}{2}(1, - \sin\theta_1,0,- \cos\theta_1).
\eea

The angle $\theta_2$ is the angle between the momentum $\bp_3$ of
$\ell_1^-$ in the $\ell_1^+ \ell_1^-$ rest frame and the $z$ axis.
The momenta $p_{3,4}$ in the rest frame of the lepton pair are written
as
\bea
p^*_3 &=&\frac{\sqrt{q^2}}{2} (1,  \sin\theta_2 \cos\phi,\sin\theta_2 
\sin\phi , \cos\theta_2),\\
p^*_4 &=& \frac{\sqrt{q^2}}{2}(1, - \sin\theta_2 \cos\phi,-\sin\theta_2 
\sin\phi ,- \cos\theta_2),\label{eq:p3andp4inHZll}
\eea
where $\phi$ is the angle between the normal of the planes defined
by the $z$ direction and the momenta $p_1$ and $p_3$. It is measured
positively from the $\ell_2^+\ell_2^-$ plane to the $\ell_1^+\ell_1^-$
plane.

\subsection{\boldmath $\eeHll$}
\label{app:KineeeHZ}

The momenta are labelled as $e^-(p_{-})e^+(p_+)\to 
H(p_H)Z(p)(\to \ell^-(p_1)\ell^+(p_2))$,
where in the final state we kept the conventions used in the $\HZll$.
We choose the $z$ direction to be defined by the momentum of
the on-shell $Z$ boson in the initial state rest frame, here the incoming 
$e^+e^-$ rest frame.
The $xz$ plane coincides with the plane defined by $\bp$ and $\bp_1$, 
which complies with the previous definition. 
For the final state leptons, in the dilepton rest frame and with $m_\ell=0$, 
the expressions of the momenta are formally the same as in the $\HZll$ case
\bea
\bar p_1 &=& \frac{m_Z}{2}(1,  \sin\theta_1,0, \cos\theta_1),\\
\bar p_2 &=& \frac{m_Z}{2}(1, - \sin\theta_1,0,- \cos\theta_1),
\eea
where again $\theta_1$ is the angle between $\bp_1$, the momentum of 
$\ell^-$, and the $z$ axis.

With these definitions, the incoming momenta in the  $e^+e^-$ rest frame 
(denoted with an asterisk) are given by
\bea
p_-^* &=& \frac{\sqrt{q^2}}{2} (1,  \sin\theta_2^- \cos\phi,\sin\theta_2^- 
\sin\phi , \cos\theta_2^-),\\
p_+^*&=& \frac{\sqrt{q^2}}{2} (1, - \sin\theta_2^- \cos\phi,-\sin\theta_2^- 
\sin\phi ,- \cos\theta_2^-),
\eea
where, to make a clear distinction, the angle $\theta_2^-$ is  the angle 
between  the direction of flight of the $e^-$ and the $z$ axis in the 
$e^+e^-$ rest frame. 
To best exploit the crossing symmetry of the two processes, one should 
describe the reaction using the angle $\theta_2^+$ measured from the $z$ 
axis to the direction of flight of the $e^+$, since in $\HZll$ we 
chose to use the angle between the direction of flight of $\ell_1^-$ and 
the $z$ axis. Our results in Sec.~\ref{sec:eeHZ} are therefore written in 
terms of the angle
\beq
\theta_2^+ \equiv \theta_2 =\pi-\theta_2^-,
\eeq
which makes the expressions for the squared amplitude in decay and 
scattering formally identical.

\begin{figure}[t]
\begin{center}
\includegraphics[width=0.9\columnwidth]{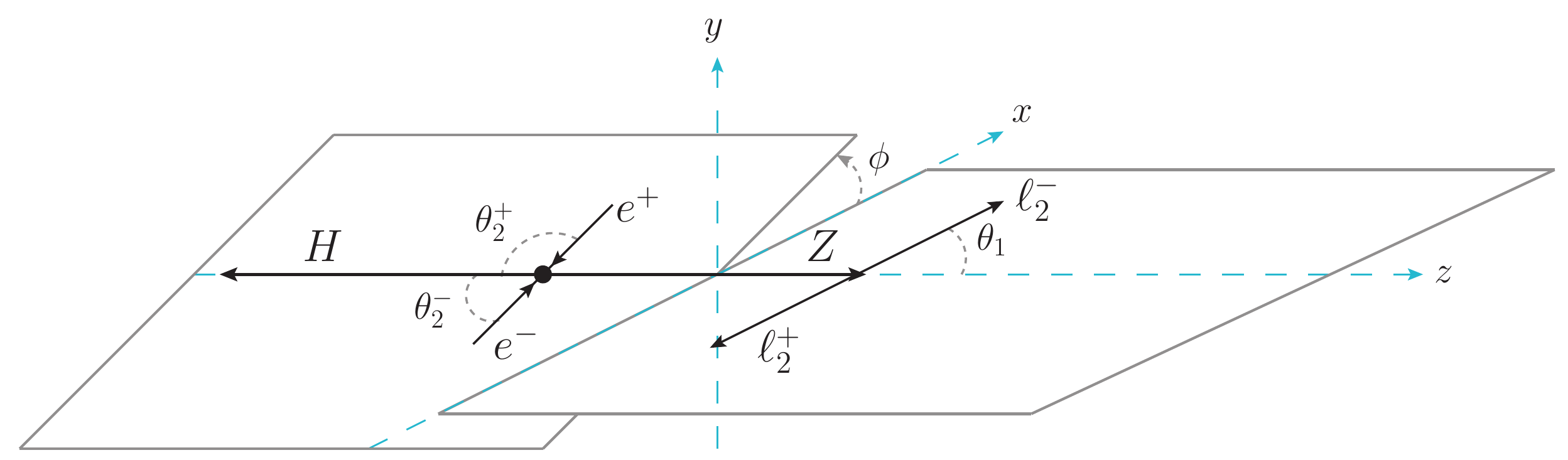}
\vspace*{0.2cm}
\caption{Kinematics for the scattering $\eeHll$.
}
\label{fig:KineeeHZ}
\end{center}
\end{figure}

\section{\boldmath Explicit expressions for the $J$ functions }
\label{app:JFunctions}

Here we give the  expressions of the $J$ functions defined in Eq.~(\ref{eq:JFuncsinH}). In the following results
$\lambda$ stands for
\beq
\lambda\equiv \lambda(1,s,r) =1+s^2+r^2 -2s -2r -2r s,
\eeq
and we recall that $\kappa = 1-s-r$. 
The couplings $g_{V,A}$ are those of Eq.~(\ref{eq:Zll}) and contain the 
$d=6$ corrections. The explicit expressions for the $J$ functions read 
at ${\cal O}(1/\Lambda^2)$
\bea
J_1&=& \frac{8\sqrt{2}\,m_H^2\,G_F}{(s-r)^2}  
  \left(g_A^2+g_V^2\right)^2\,r^3 s \times\nn\\
&& \left ( 1+ \frac{2\,(g_V^2\haone +g_A^2\hatwo)}{g_A^2+g_V^2}
-\frac{2\,\kappa \,\haZZ}{r} +\frac{g_V\, Q_\ell g_{\rm em} (s-r) \kappa\,\haAZ}{(\gAgVsq) \,rs}
  \right) ,\nn \\[0.2cm]
J_2&=&\frac{4\sqrt{2}\,m_H^2\,G_F}{(s-r)^2}   \left(g_A^2+g_V^2\right)^2\kappa^2 \, r^2 \times\nn\\
&& \left ( 1+ \frac{2\,(g_V^2\haone +g_A^2\hatwo)}{g_A^2+g_V^2}
-\frac{8\,s\,\haZZ }{\kappa} +\frac{4\,g_V\, Q_\ell\, g_{\rm em} (s-r) \haAZ}{(\gAgVsq) \kappa}  \right),\nn\\[0.2cm]
J_3&=& - \frac{128\sqrt{2}\,m_H^2\,G_F}{(s-r)^2}    \,g_A^2 \, g_V^2\, r^3 \, s 
 \left ( 1+\haone+\hatwo -\frac{2\,\kappa\,\haZZ}{r} +\frac{ Q_\ell\, g_{\rm em} (s-r)\kappa \haAZ}{2\, g_V r\,s}\right),\nn\\[0.2cm]
J_4&=&-  \frac{16\sqrt{2}\,m_H^2\,G_F}{(s-r)^2}    \, g_A^2 \, g_V^2\, 
\kappa \sqrt{\frac{\lambda \, r^3}{s}}\, \left( 4\,s\, \ha_{Z\widetilde Z}+ 
\frac{ Q_\ell\,g_{\rm em} \, (r-s)\ha_{A\widetilde Z}}{g_V}     
\right),\nn\\
J_5&=& \frac{\kappa}{4\sqrt{r\,s}}\,J_8,\nn\\[0.2cm]
J_6&=& - \frac{64\sqrt{2}\,m_H^2\,G_F}{(s-r)^2}   \,g_A^2\, g_V^2\, \kappa\,\sqrt{s\,r^5} \times\nn \\
&&\left(   1 +  \haone+\hatwo   +\frac{(\lambda-2\kappa^2)\haZZ}{r\,\kappa}
 + \frac{ Q_\ell\, g_{\rm em} (r-s)(\lambda-2\kappa^2)\haAZ}{4 \, g_V\, r\, s\, \kappa }\right), \nn \\
J_7&=& \frac{4\sqrt{2}\,m_H^2\,G_F}{(s-r)^2}   \left(g_A^2+g_V^2\right)^2\,\kappa\,\sqrt{s\,r^5} \times \nn\\
&&\left(1 +\frac{2(g_V^2\haone +g_A^2\hatwo)}{g_A^2+g_V^2} + \frac{(\lambda-2\kappa^2)\haZZ}{r\,\kappa}+ \frac{ g_V\,Q_\ell\, g_{\rm em}  (r-s)(\lambda-2\kappa^2)\haAZ}{2(g_A^2+g_V^2) \, r\, s\,  \kappa}
\right),\nn \\[0.15cm]
J_8&=&\frac{8\sqrt{2}\,m_H^2\,G_F}{(s-r)^2}  \left(g_A^2+g_V^2\right)^2  
\,r^2 \,\sqrt{\lambda} \,\left( 2\,s\, \ha_{Z\widetilde Z} + 
\frac{ g_V\,Q_\ell\,g_{\rm em} \, (r-s)\ha_{A\widetilde Z}}{(g_A^2+g_V^2)}    
\right),\nn \\[0.15cm]
J_9&=&J_1.
\eea


\end{document}